\documentclass[traditabstract]{aa-almost} 
\usepackage[varg]{txfonts}
\usepackage{time}
\usepackage{color}
\usepackage{natbib}
\usepackage[percent]{overpic}
\usepackage{amstext,amsmath,mathtools}
\usepackage{graphicx}
\usepackage{textcomp}
\usepackage{txfonts}
\usepackage{url}
\usepackage{booktabs} 
\begin{document}
\titlerunning{EST FPI systems}
\title{Proposal for the optical design of three robust and highly performing FPI systems for the European Solar Telescope}

\author{G.B. Scharmer\inst{1}
\and
B. Lindberg\inst{2}}

\institute{Institute for Solar Physics, Dept. of Astronomy, Stockholm University,
AlbaNova University Center, SE 106\,91 Stockholm, Sweden \and
Lens Tech AB, Tallbackagatan 11, SE 931\,64 Skellefte\aa, Sweden }
\date{Draft: \now\ \today}
\frenchspacing

\abstract{We describe a proposal for the optical design of three dual Fabry-Perot based narrowband filter systems for for the future European Solar Telescope (EST). These are intended to constitute the core elements of three imaging spectropolarimeters, foreseen to become amongst the most important science instruments for EST. The designs proposed here rely heavily on the heritage of CRISP and CHROMIS, developed for the Swedish 1-m Solar Telescope and described in detail in a companion paper (Scharmer et al. 2025, in prep.). The outstanding performance of these systems, and the simplicity of their designs, provide strong support of our proposal to build similar systems for EST. The design concepts involve i) minimising the FPI clear aperture diameter by means of numerical simulations based on constraints on Strehl and spectral resolution set by the EST Science Advisory Group (SAG); ii) a compact telecentric optical design with an optical path length of less than 4.7~m; iii) a straight-through optical system based on lenses and without any folding mirrors; iv) the combination of a high resolution etalon with high reflectivity and a low reflectivity, low resolution etalon, to mitigate the effects of cavity errors (Scharmer 2006, Scharmer et al. 2025); v) flexibility in terms of image scale by simple replacement of the last lens (the camera lens) of the FPI system. We propose to compensate for the focus curve of ESTs Pier Optical Path (POP) by focusing the camera lenses of the FPI systems. However, to prevent large movement of the focal plane between the FPIs and the large movement FPI camera lens, it is necessary that the focus error at the input side of the FPI system is very small - we suggest that it should be about 5~mm or less. Therefore, a slow focusing mechanism is also needed that involves the entire construction of the FPI system, including its wideband system, filter wheel, polarimeter and cameras.

\par
\vspace{2mm}
 The proposed systems should offer several advantages over other much more complex systems, including manufacture, alignment, stability, flexibility of changes of image scale, and costs. The underlying design concepts also make the proposed FPI systems robust and highly performing in terms of image quality, overall transmission, and fidelity of the spectral transmission profile. }
 
\keywords{ Instrumentation: high angular resolution  -- Instrumentation: polarimeters -- Instrumentation: spectrographs -- Methods: observational -- Techniques: image processing -- Techniques: imaging spectroscopy -- Techniques: polarimetric -- Techniques: high angular resolution 
}

\maketitle

\section{Introduction}
The European Solar Telescope \citep[EST;][]{2022A&A...666A..21Q} will be equipped with several state-of-the-art science instruments, including three Fabry-Perot interferometer-based (FPI-based) imaging spectropolarimeters, in the following referred to as EST-B, EST-V, and EST-R. By collecting data simultaneously at high cadence, these three instruments will provide unprecedented information about the thermodynamics, line-of-sight velocities, and the magnetic field of the solar atmosphere from the deep photosphere to the upper chromosphere. 

The designs proposed here rely heavily on the experience and heritage of three FPI-based spectropolarimeters developed for the Swedish 1-m Solar Telescope \citep[SST; ][]{2003SPIE.4853..341S}, namely CRISP \citep[][referred to as CRISPm after a recent upgrade with new camera lens, modified image scale, and new cameras]{2008ApJ...689L..69S}, CHROMIS \citep[][now being upgraded with new cameras and a polarization modulator]{2017psio.confE..85S}, and CRISP2 (soon to replace CRISPm, and with a much larger field of view). In a companion paper (Scharmer et al. 2025, in prep.), we provide an extensive summary of the rationale and design details of these systems, including a review of the numerical simulations made to establish critical design and performance parameters of the systems. In that paper we also describe in detail the corresponding numerical simulations for the three FPI systems for EST – these are not repeated here. The optical designs of the EST FPI systems are also summarised in that publication, including their relation to the corresponding systems developed for SST, but are described in more detail here. In the companion paper, we also describe procedures of calibration and data processing and image reconstruction, developed for CRISP and CHROMIS. These are crucial for extracting high-quality science data from the spectropolarimetric data of the FPI-based instruments and are of obvious relevance for understanding of the procedures  that will need implementation at EST as well. We also give in Scharmer et al. (2025) examples of data recorded with CRISP and CHROMIS and point to earlier work in which the high image quality of CRISP has been firmly established.

In the companion publication (Scharmer et al. 2025), our emphasis is on the EST-B system, which is the most challenging of the three FPI systems for EST, and the same emphasis is made here. This system is described in much greater detail than is the case for EST-V and EST-R. Some of the aspects investigated for EST-B have not yet been investigated for EST-V and EST-R, partly because circumstances do not make that necessary and partly because of lack of resources. These aspects should be studied as part of a final design study, if and when there is a decision to build these systems. On the other hand, simulations should also be made to investigate the combined effects of tolerances on manufacturing and assembly for EST-B, once a final design is more mature.

The top-level requirements for the three FPI systems have been set by the EST Science Advisory Group (SAG). These requirements include a circular field of view with a diameter of 1 arc minute and other parameters given in Table \ref{table_sect1_1}.

\begin{table}[h]
  \centering
  \small
  \begin{tabular}{|l|c|c|c|c|}
        \hline
    \mathstrut
FPI&Wavelength&Spectral&Strehl&Image scale \\
& range (nm)&resolution&&(arcsec/pixel) \\
\hline
EST-B&380-500&50,000 &>90\% &0.010\\
&&at 396 nm&at 380 nm&\\
\hline
EST-V&500-680&100,000&>90\%&0.013\\
&&at 630 nm&at 500 nm&\\
\hline
EST-R&680-1000&80,000&>90\%&0.017\\
&&at 854 nm& at 680 nm&\\
 \hline
  \end{tabular}
  \vspace{1mm}
  \caption{Top-level requirements for the EST FPI based filter systems}
 \label{table_sect1_1}  
\end{table}

\begin{table}[h]
  \centering
  \small
  \begin{tabular}{|l|c|c|c|c|}
        \hline
    \mathstrut
    Wavelength (nm)& Arc sec/pixel\\
    \hline
380 (EST-B)& 0.0093\\
500 (EST-B/V)& 0.0123\\
680 (EST-V/R)& 0.0167\\
1000 (EST-R)& 0.0246\\
     \hline
  \end{tabular}
  \vspace{1mm}
  \caption{Critical sampling, given by $\lambda/2D$ (radians/pixel), where $D$ is the entrance pupil diameter of the
telescope (4.2 m).}
 \label{table_sect1_2}  
\end{table}

The Strehl values given in Table \ref{table_sect1_1} refer to a combination of apodization effects originating from the
tilted rays corresponding to the F-ratio of the optics at the location of the FPI etalons (that are mounted
in a telecentric setup), and limitations of the optical design of the FPI system, with its integrated re-
imaging optics.

The apodization effects depend primarily on the spectral resolution required and the F-ratio of the
system at the etalons, and were evaluated through calculations that are reported on separately
(Scharmer et al. 2025). Given the spectral resolution required, it was possible to determine the
smallest possible F-ratio that allows a Strehl of at least 95\%, based on these apodization effects.
Choosing the smallest possible F-ratio for a given spectral resolution is of fundamental importance in
maximising the field-of-view of the entire system with a given clear aperture of the etalons, or in
minimising the clear aperture of the etalons when the diameter of the field-of-view is given. The
minimum F-ratio thus obtained from calculations was found to be in the range 140-150 for the EST-V
and EST-R systems, and 100-110 for the EST-B system (thanks to its lower spectral resolution). The
so-called TIS (tunable imaging spectropolarimeters) consortium, while G. Scharmer was still a member, 
therefore decided to agree on an f-number of 147 for the EST-V and EST-R systems
(which would correspond to etalons with clear apertures of 180 mm), and 110 for the EST-B systems,
and these values are adopted in the following.

The present report describes preliminary optical designs of all three EST FPI systems, but emphasis is
on the EST-B system, which covers the 380-500 nm wavelength range. The organisation of this report
is as follows. In Sect. \ref{simplifications_decisions}, we summarise the decisions and simplifications made that are common to
the optical designs of the three EST FPI  systems developed. The EST-B system is described in Sect. \ref{EST-B_FPI_system}. Section \ref{EST-B_config} contains all
information about system parameters and system configurations and Sect. \ref{EST-B_optimisation} describes the
optimisation of its optical design. Section \ref{EST-B_performance} provides discussions and conclusions related to the
performance and alignment of the EST-B system, in particular it has an in-depth discussion about
potential problems and remedies associated with POP’s focus curve and likely thermal drifts, and
image scale variations connected to refocusing. Section \ref{EST-V_FPI_system} summarises design and performance of the EST-V system, and Sect. \ref{EST-R_design3} gives a similar summary of the EST-R system. Section \ref{im_scale_variations} summarises the variation of image scales across the wavelength ranges of the three FPI systems discussed, with emphasis on the difference between the narrowband and wideband systems. In Sect. \ref{comments_manufacture}, we give comments relating to the manufacture, installation and alignment of the FPI systems. In Sect. \ref{tolerances}, we discuss tolerances briefly, and in Sect. \ref{discussion} we discuss and summarise the results of this development.  In parallel, we move a number of tables and spot diagrams to appendices: Appendix \ref{EST-B_design2} contains details of the optical design of EST-B, and tables of Strehl values and spot diagrams related to the performance of EST-B, Appendix \ref{EST-V_design2} gives similar but more condensed information about EST-V, and Appendix \ref{ESTR2} the same condensed information is given for EST-R. Appendix \ref{tolerances2} contains additional information about tolerances and Appendix \ref{files_used}, finally, gives a list of the Zemax files used to construct the designs proposed - these can be obtained upon request. 

We emphasise, that the present proposal of FPI systems for EST is indeed only a proposal that has not yet been evaluated or commented on by the EST Project, but part of this proposal has earlier been presented to the TIS consortium by G. Scharmer.
 
\section{Simplifications and decisions} \label{simplifications_decisions}
\subsection{POP}
\begin{figure}[htbp]
\center
\includegraphics[angle=0, width=0.95\linewidth,clip]{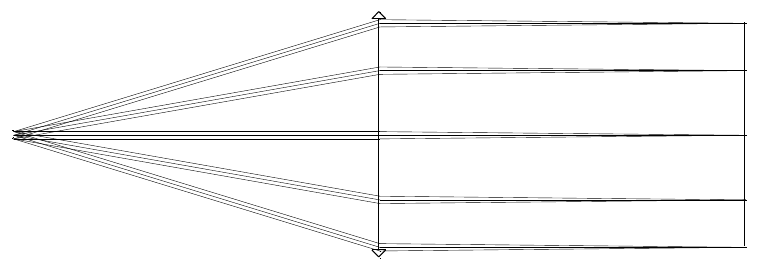}
 \caption{Idealized model of the EST telescope and POP.}
\label{fig:Fig_sect2_1}
\end{figure}

The three FPI system optical designs described in this and associated reports are based on an input F-ratio of 50 with a telecentric, or nearly telecentric, reimaging system. This is the so-called POP \citep[Pier Optical Path;][]{2022A&A...666A..21Q}, which is the transfer system that translates the focal plane (F2) of the primary optical system and its multi conjugate adaptive (MCAO) system to the optical tables of the science instrumentation (F3). We are using the presently (as of 23 April 2024) accepted version of the POP optical design by Álvaro Pérez García (private communication from the EST Project Office),  as defined in Table \ref{table_sect1_2}. There are particular aspects of that design, such as a small but significant variation of the image scale with wavelength, which are reflected in the outputs from the FPI systems. These might change somewhat if the POP design is revised, but our expectation is that POP will retain its output F-ratio of 50 with (nearly) telecentric re-imaging at F3, such that any revision of the optical designs of the FPI systems induced by a redesign of POP will be minor. Our design work in Zemax, uses a simplified representation of the EST telescope and POP, as described in Fig. \ref{fig:Fig_sect2_1} and Table \ref{table_sect1_2}.

\begin{table}[h]
  \centering
  \small
  \setlength{\tabcolsep}{4pt}
  \begin{tabular}{|l|c|c|}
        \hline
    \mathstrut
&Real system & Simplified Zemax model \\
\hline
Entrance pupil & 4200~mm & 2~mm \\
Focal length & 210,000~mm & 100~mm \\
F-number & 50 & 50 \\
Full circular FOV for FPI & 60 arcsec & $\pm$17 degrees \\
Image height (field radius) & 30.5~mm &30.5~mm \\
Telecentric output & yes & yes \\
     \hline
  \end{tabular}
  \vspace{1mm}
  \caption{Parameters of the EST telescope and POP model used.}
 \label{table_sect1_2}  
\end{table}

\subsection{FPI systems}

In setting up the FPI system, certain decisions were taken that define the final design of the system.
Foremost of these is the decision to base the re-imaging system on lenses, not mirrors, and to make the
system as compact as possible in order to eliminate the need for folding mirrors. These are concepts
that have been successfully implemented with CRISP and CHROMIS at the Swedish 1-m Solar
Telescope, and that have proven to deliver science data of unprecedented quality. The second decision
concerned the overall length, which is constrained by the available space in the optics lab to be less
than 6 m. Very early on, we developed a conceptual design of an FPI system for EST with an overall
length of only 3.5 m, to demonstrate the feasibility of a very compact system. In our present design we
have made a compromise and set the overall length to be about 4.5-4.7 m, depending on the pixel
size of the detector. The high Strehl obtained for the present designs suggests that it may be possible to
reduce the overall length further, if necessary.

A third and more complex decision was to use air-spaced doublets for the two largest lenses (that have
clear apertures in excess of 180 mm), even though discussions with a potential provider of such lenses
indicated the possibility of cementing such large doublets, provided that the coefficients of thermal
expansion of the glasses are well matched and that an elastic cement is used to bond the glasses. A
final decision on whether the two large lenses should be cemented or air-spaced requires an
investigation of thermal effects from possible combinations of optical glasses. Exploring this
possibility can be delayed until the final design phase, once decisions on funding have been taken.

A fourth decision was to compensate for the focus variation with wavelength of POP and (if present)
also of the FPI system itself by moving the camera lens of the FPI system. The rationale for this
decision is discussed by Scharmer et al. (2025, Sect. 5.2.1) and at depth in Sect. \ref{focusing_challenge} of this document.

The design described in this report is constrained by the chosen length of the FPI system, level of
image quality (Strehl), final selection of glasses, and final transmission values. We have verified that
the glasses (such as N-FK5 or FK5HTI and PBM2Y, diameter 170 mm, for EST-B) can be acquired in the needed sizes. We also have received feedback on the safe choice of edge thicknesses of the lenses
from one potential provider of the lenses, and on the thicknesses of the FPI etalons, given their
expected diameters and clear apertures. Thus, the optical designs presented are realistic in these
respects.

However, some simplifications were also made in order to speed up the work and avoid overworking a
design that anyway is likely to undergo changes during the final design. The system is set up as a
symmetrical system (around its optical axis). The FP plates were modeled as two glass blocks without
any wedges, tilts or air slits, and the optimisation was made with the symmetrical model. Our
experience with earlier designs (CRISP, CHROMIS and CRISP2) is that the introduction of realistic
FPI plates with wedges and one tilted etalon does not degrade the image quality.
What remains to be done in the final design is primarily to include the details of the FPI plates, to
refine the tolerance analysis, and also to model ghost images and polarization fringes.

\section{EST-B FPI system} \label{EST-B_FPI_system}
\subsection{EST-B system parameters and configuration} \label{EST-B_config}
\begin{figure}[htbp]
\center
\includegraphics[angle=0, width=0.99\linewidth,clip]{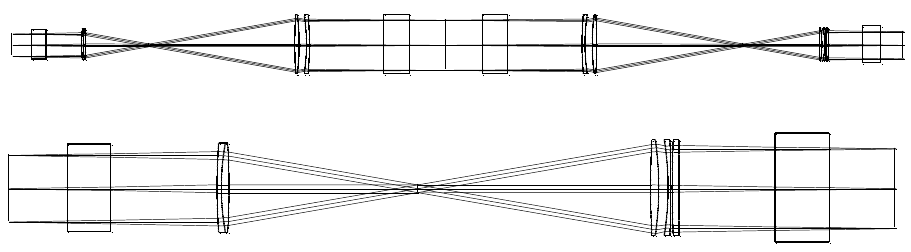}
 \caption{Shows the layout of the narrowband (top) and wideband (bottom) EST-B systems, designed for 12~$\mu$m pixel size. The figures are stretched 2x in the vertical direction for clarity. Both camera lenses are air spaced
triplet lenses. The rectangles represent the beam splitter cubes. The two figures are scaled separately.}
\label{fig:Fig_sect3_1}
\end{figure}

\begin{figure}[htbp]
\center
\includegraphics[angle=0, width=0.99\linewidth,clip]{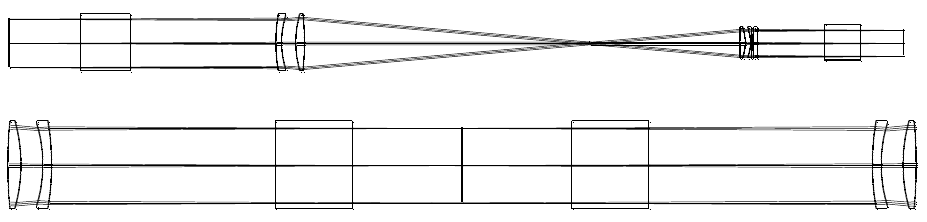}
 \caption{The rear (top) and central (bottom) of the 12~$\mu$m pixel narrowband EST-B system.}
\label{fig:Fig_sect3_2}
\end{figure}

Table  \ref{table_sect3_1} summarises the input parameters, which are set by requirements on minimum Strehl and spectral resolution (defining the focal ratio between the FPIs) and FOV diameter (1 arcmin) as decided by SAG, and the input focal ratio of POP. This Table also specifies the length of the EST-B, obtained during the optimisation of its optical design. Table  \ref{table_sect3_2} summarises data on the magnification and image scales needed with different choices of camera pixel sizes, assuming critical sampling as given by the top-level requirements summarised in Table \ref{table_sect1_1}. Figure \ref{fig:Fig_sect3_1} show overviews of the layout of the optical design of the EST-B narrowband and wideband systems, designed for 12~$\mu$m pixel size, and Fig. \ref{fig:Fig_sect3_2} shows details of the narrowband system.
\begin{table}[h]
  \centering
  \small
  \begin{tabular}{|l|c|c|c|c|cc}
        \hline
    \mathstrut
Wavelength range (nm)& 380-500\\
Input F-ratio (at F3)& 50*\\
F-number between FPIs& 110**\\
Output F-ratio & Table \ref{table_sect3_2}\\
Magnification (Science focus to between FPIs) & 2.2\\
Magnification (Science focus to detector)& Table \ref{table_sect3_2}\\
Field diameter on sky (arc.sec.)& 60\\
Field diameter at F3 (mm)& 61\\
Field diameter between FPIs &134.2\\
Field diameter at detector &Table \ref{table_sect3_2}\\
Length F3 to second pupil stop (mm) &4000\\
Length F3 to detector. 12~$\mu$m pixels (mm)& 4660\\
Length F3 to detector. 5~$\mu$m pixels (mm) &4377\\
 \hline
  \end{tabular}
  \vspace{1mm}
  \caption{Summary of input parameters for EST-B. * POP design by Álvaro Pérez García. ** Obtained from numerical simulations, see Scharmer et al. (2025)}
 \label{table_sect3_1}  
\end{table}

\begin{table}[h]
  \centering
  \small
  \begin{tabular}{|c|l|c|c|c|c|c|c}
        \hline
    \mathstrut
 Pixel size & Magn. & Image scale & Diameter & Output \\
 ($\mu$m) && at detector & at detector& focal \\
 & & ("/mm) & (mm) &ratio \\
\hline
5&0.5&2.0&30.5&25\\
8&0.8&1.25&48.8&40\\
12&1.2&0.833&73.2&60\\
14&1.4&0.714&85.4&70\\
 \hline
  \end{tabular}
  \vspace{1mm}
  \caption{Magnification, image scale and output focal ratios for different pixel sizes with EST-B. The input image scale is 1.0"/mm, the output pixel scale is 0.010"/pixel for all pixel sizes.}
 \label{table_sect3_2}  
\end{table}

\subsection{Optimisation of EST-B} \label{EST-B_optimisation}
 The model of the FPI system consists of a number of optical components in the following order
(starting on the input side): one 80 mm beam splitter cube (used to deflect a small fraction of the light
to the wideband re-imaging system), two doublet lenses, two FPI etalons, one doublet lens, one triplet
lens (the camera lens), one (polarizing) beam splitter cube, and a detector with a given pixel size (we
assumed two different pixel sizes: 5~$\mu$m  and 12~$\mu$m ) and designed the camera lens to match these pixel
sizes. The polarization modulator is not included in the design, since there is no available information
on that yet. We note that the polarization modulator can be placed either in front of the FPI system
(after the 80 mm wideband beam splitter), or close to the cameras (before the polarizing beam splitter),
and that the second option offers an advantage if the pixel size is small.

The final design presented here does not include the primary optical system of EST (M1-M6) but
includes a realistic representation of POP. The optimisation of the FPI system was made in steps, as
follows:

\begin{enumerate}

\item The FPI system was optimised with POP modeled as a single ideal (perfect) paraxial lens. The
camera lens was also modeled as a single ideal paraxial lens. This optimisation was made in order to
deliver the best image quality for all wavelengths in a single focal plane, independent of the POP. We
refer to this Zemax file as the “mother file”. The mother file is the same for all pixel sizes and camera
lenses, and the mother file contains all the expensive components.

\item The variables optimised previously were set to fixed values and the ideal paraxial camera lens
was replaced with triplet lenses. The file was set up with two configurations (narrowband and
wideband) in the same file, each with its own separate air spaced triplet as camera lens.
This made it possible to force the geometric distortion to be the same for the narrowband and
wideband systems. In the VIS system the narrowband and wideband systems were described in
separate files. The camera objectives were optimised for maximum image quality in one single image
plane for all wavelengths. One such two configuration file was created for cameras with 12~$\mu$m  pixels
and one with 5~$\mu$m pixels. For 12~$\mu$m  pixels, the performance was improved somewhat by optimising
the design for four configurations (narrowband and wideband at two different focus settings).
\par
\vspace{2mm}
The above-described files are still independent of POP. For the system with 5~$\mu$m pixels the lowest
Strehl value was 96\%. When POP is included in the design, the Strehl decreases to 0.91. To increase
that Strehl value, it would be possible to optimise the camera lens with POP included, thus taking
aberrations of POP into account. This would leave the mother file unchanged, only the camera lens
would be designed to compensate for aberrations in POP. This approach has not been attempted. For
the system with larger pixels, there appears to be no need to consider this possibility. If the pixel size
changes in the future, only this step has to be repeated, with or without POP.

\item The paraxial lens representing POP was replaced by the most recent design of POP (design as of
23 April 2024), but assuming an ideal telescope. The wideband and narrowband systems are described
in separate files. Since POP is not fully color corrected, the focal position of POP is different for
different wavelengths. However, we have chosen to not compensate the optical design of the FPI
system for the variation of the focal plane with wavelength that comes from POP. This decision is
made in order to avoid making the FPI design too dependent on details of the POP system that might
change during manufacture, installation or in operation (such as changes due to its thermal
environment). Relying on the optical design of the optical systems of the FPI systems to compensate
exactly the focus curve of POP, would have the downside of requiring a redesign of the FPI lenses in
case POP goes through even minor changes.
\end{enumerate}

Instead of compensating the focal curve of POP through a modification of the optical design of the FPI
system, the variation of focal position with wavelength for POP is taken care of by refocusing the
camera lens, or the camera itself, individually for each wavelength. So, during this stage of the FPI
system optimisation, the FPI system itself does not change, and the only change made is to allow the
camera lens or camera to be moved to compensate both for the focus curve of POP and any (small) focus curve
of the FPI system itself.

The optimisation of the FPI system, as described above, was made for two different pixel sizes, 5~$\mu$m 
and 12~$\mu$m , which should roughly cover the likely range of future pixel sizes for the EST FPI systems.
A small pixel size offers advantages, both in terms of cost and performance, relative to
cameras with a much larger pixel size. If anything, we suspect that the pixel size of EST-B could end up being smaller than 5~$\mu$m. Note that the full well of cameras with small pixels is not a problem for any of the 
proposed EST FPI systems, assuming (as must be the case) that short integration times are use to combat
seeing effects and boost image quality. This is in particular the case of EST-B, for which light levels will be very small.

It might be necessary at some stage later, when the tolerances are known, to set up a Zemax model of
the entire optical chain from the telescope to the detector.

\subsection{Performance and alignment of EST-B} \label{EST-B_performance}
\subsubsection{The focusing challenge} \label{focusing_challenge}
 
 As mentioned previously, the focus curve of POP leads to the need for a mechanism to refocus each
FPI system individually and at each wavelength that is within the operational limits of that FPI system.
This mechanism needs to be fast, such that the duty cycle and thus the overall efficiency of the FPI
system is not impacted. On the other hand, any FPI system that is tuned to a new wavelength will also
require a change of prefilter, so that the time needed to change prefilter constitutes an unavoidable
pause in the recording of science data. Therefore, as long as the time needed for refocusing does not
exceed the time needed to change prefilter, refocusing will not impact the duty cycle (overall
efficiency) of the FPI system.

Simulations show (see Sect. \ref{EST-B_image_scale_variations}) that moving the camera in principle leads to smaller image scale variations than when movement of the camera lens is used for refocusing, but we nevertheless advice against that solution for two reasons: the first is that a narrow- and wide-band combined FPI system will contain four cameras and thus needs four motorised translation stages per FPI system. The other objection is that moving cameras with connected cables for data transfer, and perhaps cooling, appears more cumbersome than moving a small lens.
 
Our preferred choice of compensating the focus curve of POP is therefore to translate the
last lens (the camera lens) of the FPI system. We are convinced that this can be done quickly enough,
in particular if the pixel size of the camera is small. This is because the camera lens is relatively small
and thus constitutes a small mass to move: for EST-B with 5~$\mu$m pixels, the lens design has a diameter
of 46 mm, and the total thickness of the glasses constituting the triplet camera lens is about 29 mm
(see Table \ref{table_EST-B_5my_camera_lens}). On the other hand, the pre-filters have diameters of 70 mm, and there will be the need for a filter wheel or translation stage to hold a minimum of 4 filters. This corresponds to a
filter wheel of at least 150 mm diameter, or a translation stage that is at least 280 mm long. The
needed movement of the filter wheel or translation stage obviously corresponds to a minimum of
about 80 mm (if shifting to a neighbouring filter). Within the wavelength range of EST-B, POPs focus
curve covers a range of only 6.4 mm. The range of refocusing EST-B scales as the POP focus change
multiplied with the square of the total magnification (counted from the input to the output sides of the
FPI system). For the 5~$\mu$m pixel size system, the magnification equals 0.5, and for the 8~$\mu$m pixel size
it is 0.8, such that required focusing range is smaller or much smaller than that of POP. However, for a
system with 12~$\mu$m pixel size, the magnification is 1.2 such that the refocusing range actually is
increased over that of POP. Moreover, the camera lens diameter is increased to 88 mm, and the total lens thickness to about 32 mm. If cameras with 12~$\mu$m pixel size is chosen, a detailed analysis of the
time required to refocus the FPI system should be made in order to confirm that refocusing can be
made within the time needed to change prefilter.

Because of the need to refocus the FPI systems in order to compensate for the focus curve of POP, we have investigated whether such refocusing leads to significant changes in the image scale. We emphasise that a variation of image scale with wavelength is not a problem by itself, since the three FPI systems will operate with very different image scales anyway, and mature techniques exist for combining and interpreting science data with widely different image scales. 

A relevant question therefore is how large input errors we can tolerate on the input side of the FPI system, and how to secure this level of focus error?  We argue, that the focus error on the input side should be very small, about 5 mm, which is comparable to the variation of the focus curve of POP within the wavelength range of one of the FPI systems. There are two reasons for requiring very small focus errors on the input sides of the FPI systems. The first is simply to minimise the range of movement of the FPI focusing mechanism, which will involve image scale variations and possibly even image quality degradation for large amounts of defocusing. The other reason is to ensure that the focal plane between the FPIs stays away from the surfaces of the two FPIs, such that the FPIs always appear defocused in the focal planes of the science cameras. The problem is that the movement of the focal plane scales as the square of the focal ratio, such that a 5 mm movement of the F/50 focal plane on the input side of the FPI system leads to movement of 43 mm in the F/147 beam between the etalons. Adding to that a 5 mm focus curve, means an overall movement of the focal plane between the FPIs of nearly 100 mm. This will be acceptable if the FPIs are located at least 200 mm from the nominal focal plane, midways between the two etalons, but obviously very large input focus errors will cause problems.

If indeed refocusing of the camera lens of the FPI system will be needed also to compensate any
drift of POPs focus curve, it could lead to the appearance of localised flatfield artefacts in science
data. A particular problem would be if the image scale changes significantly, which would complicate
co-alignment of data recorded at different wavelengths. To minimise the negative impact of such
drifts, we have made alternative designs of the camera lens, paying particular attention to minimising
image scale changes when refocusing this lens. Such designs may be preferred in order to make the
system as robust as possible. We discuss image scale changes when refocusing the FPI system for
three different optical designs in Sect. \ref{EST-B_image_scale_variations} and the image quality of these systems in Sect. \ref{EST-B_image_quality}.

To overcome the challenge of ensuring a very small focus error on the input side of the FPI systems, and 
to ensure a small range of movement of the camera lens, we strongly suggest that the mechanical structure supporting each FPI system, including its wideband system, filter wheel, polarimeter and cameras, must have a slow focus mechanism with adequate stroke. 

We here also point out the potential benefit of fine tuning the focus curve of POP, which in its
present version is constrained to providing two particular wavelengths at the same focus: 517 nm and
617 nm (this requirement, which was imposed before the solution was found to the problem of
refocusing the FPI systems, now has been removed). We anticipate a revision of the design of POP
at a later stage, and that revision could include the aim of producing a focus curve that is better optimised to combat any
shortcomings in the focusing of the FPIs and/or other instrumentation. The priority could for example
be to optimise POP to provide a better margin for ensuring that refocusing can be made within the
time needed to change prefilter, taking into account all 3 FPI systems, and for each of these with all
possible wavelength combinations.

Whereas we in no way exclude the possibility that the focus position and focus curve of POP can be stabilised with appropriate thermal control, our recommendation is the following, until we can fully rely on a strategy for such control:

\begin{itemize}
\item[$\bullet$] Back and forth focusing of the FPI system to compensate the focus curve of POP, by means of moving the camera lens, should be 100\% repetitive during a single observing day to avoid introducing artefacts into the data.
\item[$\bullet$] The stability of POP should ideally be such that any focus drift during one observing day can be compensated for with the AO system of EST without exhausting its stroke.
\item[$\bullet$] Any drifts of the input focal plane, beyond those that can be managed by the AO system, need to be compensated for by moving the entire mechanical structure holding the FPI system and its wideband system.
\item[$\bullet$] Whereas the focal plane of POP can be allowed to drift during the day, it is crucial that the thermal stability of POP is sufficiently small that the shape of the focus curve remains the same, such that the focus positions at different wavelengths drift by the same amount. 
\end{itemize}
 
 \begin{table*}[!h]
  \centering
  \small
  \begin{tabular}{|c|c|c|c|c|c|c|c|}
        \hline
    \mathstrut
Input Focus & Camera lens & Image height & Image height & Narrowb./Diff \\
shift (mm) & movement (mm) & change ($\mu$m) & change ($\mu$m)  & image scale (\%) \\
&&Narrowband & Differential & change (\%)\\[1ex]
\hline
$\pm$5 & $\pm$7 & 26 & 11 & 0.071/0.030 \\
$\pm$10 & $\pm$14 & 73 & 21 & 0.20/0.057 \\
\hline &&&&\\
& Camera body &&&\\
& movement (mm) &&&\\[1ex]
\hline
$\pm$5 & $\pm$7 & 10 & 5 & 0.027/0.014 \\
$\pm$10 & $\pm$14 & 20 & 10 & 0.055/0.027 \\
 \hline
  \end{tabular}
  \vspace{1mm}
  \caption{Performance of EST-B optimised for image quality. The Table shows the effects of focusing on image scale variations with 12~$\mu$m pixel size. The output field of view diameter is 73.3 mm. “Differential” refers to the difference between the narrowband and wideband FPIs.}
 \label{table_sect3_3}  
\end{table*}

\begin{table*}[!h]
  \centering
  \small
  \begin{tabular}{|c|c|c|c|c|c|c|c|}
        \hline
    \mathstrut
Input Focus & Camera lens & Image height & Image height & Narrowb./Diff \\
shift (mm) & movement (mm) & change ($\mu$m) & change ($\mu$m)  & image scale  \\
&&Narrowband & Differential & change (\%)\\[1ex]
\hline
$\pm$5 & $\pm$7 & 19 & 8 & 0.052/0.022 \\
$\pm$10 & $\pm$14 & 58 & 17 & 0.16/0.046 \\
\hline &&&&\\
& Camera body &&&\\
& movement (mm) &&& \\[1ex]
\hline
$\pm$5 & $\pm$7 & 2 & 4 & 0.005/0.011 \\
$\pm$10 & $\pm$14 & 4 & 6 & 0.011/0.016 \\
 \hline
  \end{tabular}
  \vspace{1mm}
  \caption{Performance of a EST-B optimised for minimum image scale variations when refocusing. The Table
shows the effects of focusing on image scale variations with 12~$\mu$m pixel size. The output field of view diameter is 73.3 mm. “Differential” refers to the difference between the narrowband and wideband FPIs.}
 \label{table_sect3_4}  
\end{table*}

\subsubsection{Focusing and image scale variations} \label{EST-B_image_scale_variations}

To investigate the effect of refocusing, we have assumed a focus shift of $\pm$5 and $\pm$10 mm on the
input side of the FPI system, and then calculated the needed compensation (by moving the camera or the camera lens) on the output side, and the corresponding maximum change in the image height. Note that the calculated focus curve
of POP, according to its present design, is 6.4 mm within the wavelength range of EST-B (380-580
nm), so that our calculations cover an adequate focus range. Note also, that the change in image height
involves both image scale variations and geometric distortion, and that the maximum change in image
height may not occur at the edges but rather in the interior part of the field of view. Nevertheless, we have converted the change in image height to an equivalent change in image scale, to give an impression of the image scale change, by assuming that the maximum image height occurs at the edge of the 1 arc min field of view.
Results are presented below for the system with 12~$\mu$m  pixel size, which is more challenging than for
the system with 5~$\mu$m pixel size. This is because the focus shift on the input side needs a compensation
on the output side that scales as the square of the magnification, such the compensation needed for 5~$\mu$m pixel size is nearly 6 times smaller than with 12~$\mu$m  pixel size. In Table  \ref{table_sect3_3}, we give the result of
these calculations when either the camera lens or the camera body is moved to compensate the focus
shift on the input side.
Table  \ref{table_sect3_3} shows that the difference in image scale, which is the most critical quantity, changes by
0.057\% when movement of the camera lens is used to compensate 10 mm focus shift on the input side.
If image reconstruction is based on the segmentation of the image into 100x100 pixel subimages,
which is similar to what is used for SST data, then this corresponds to an image displacement of less
than 0.06 pixel at the edge of the subfield which would be negligible, meaning that a simple shift of
the entire sub-field of either the wideband or narrowband image could be used for compensation.

We have also investigated whether the camera lens can be further optimised to reduce any image scale variations, and the result is presxented in Table \ref{table_sect3_4}. As can be seen in that Table, the image scale variations are reduced, but not by a very large amount. However, this reduction of image scale variations comes at the prize of a significantly reduced image quality. As shown in Fig. \ref{fig:EST-B_spot_narrowband_FPI+POP_imqual_12mu_pix_A2.9} and Table \ref{table_EST-B_spot_narrowband_FPI+POP_imqual_12mu_pix_A2.9}, the original design of the camera lens delivers a minimum Strehl of 95\% over the wavelength range and FOV of EST-B. The design that minimises image scale variations delivers a minimum Strehl of only 0.87, as shown in Fig. \ref{fig:EST-B_spot_narrowband_FPI+POP_imscale_12mu_pix_A3.3} and Table \ref{table_EST-B_spot_narrowband_FPI+POP_imscale_12mu_pix_A3.3}, which we consider unacceptable. Our conclusion is that the original design of the camera lens is strongly preferable, and that small image scale variations when refocusing must be expected and dealt with. In this context, it should be noted that by requiring that camera lens refocusing is forced to be 100\% repetitive during an individual observing day (Sect. \ref{focusing_challenge}), we strongly reduce the risk of introducing flat-field artefacts from any thermal drifts of POP. 

Tables \ref{table_sect3_3} and \ref{table_sect3_4} also show that moving the camera body, rather than the camera lens, gives rise to even smaller relative image scale changes. However, for reasons explained earlier, our preference remains to be that of moving the camera lens rather than the camera body to provide compensation of the focus
curve of POP. The main argument is that the wideband and narrowband systems involve two camera lenses but a
total of four cameras (if a phase diversity camera is used with the wideband system), such that moving
the lenses requires fewer moving units than if the camera bodies are used for refocusing.

To put the change in image scale from refocusing into perspective, we note that the image scale of
POP itself varies by almost 1\% over the wavelength range of EST-B (see Figure. A2.1).

We also have made calculations for a design that optimises image quality with 5~$\mu$m pixel size. For
that system, both the absolute and differential image heights vary by less than 2~$\mu$m, when the input
focal plane is shifted by 5 mm. This small variation of the image height is considered satisfactory, and we 
have made no efforts to design a system with even smaller variations in image
height. That the required movement of the camera body or lens is nearly 6 times smaller for the 5~$\mu$m
pixel system than for the 12~$\mu$m pixel size system, constitute arguments for preferring the use of
cameras with small rather than large pixel size.

The noticeable image scale variations associated with refocusing for the 12~$\mu$m pixel size system
motivated us to investigate an alternate design that targets a minimisation of image scale
variations as a consequence of refocusing the camera lens (or body). In the next section, we discuss 
the image quality of the different camera lens designs.

\subsubsection{EST-B image quality} \label{EST-B_image_quality}

Appendices 2 and 3 provides detailed information about the image quality for the three narrowband
and wideband EST-B systems, with and without POP.

\begin{table}[h]
  \centering
  \small
  \begin{tabular}{|c|c|c|c|}
        \hline
    \mathstrut
 & 5~$\mu$m & 12~$\mu$m & 12~$\mu$m\\
$\lambda$ (nm) && Image qual. & Image scale \\[1ex]
 \hline 
 380 & 91\% & 95\% & 87\% \\
 400 & 94\% & 95\% & 90\% \\
 420 & 95\% & 95\% & 91\% \\
 440 & 96\% & 95\% & 92\% \\
 460 & 96\% & 95\% & 93\% \\
 480 & 96\% & 96\% & 93\% \\
 500 & 96\% & 96\% & 94\% \\
 \hline
  \end{tabular}
  \vspace{1mm}
  \caption{Minimum Strehl values for EST-B Narrowband, for 5~$\mu$m pixel size and for 12~$\mu$m pixel size when the optical design is optimised for image quality and with more emphasis on reducing image scale variation when refocusing.}
 \label{table_sect3_5}  
\end{table}

\begin{table}[h]
  \centering
  \small
  \begin{tabular}{|c|c|c|c|}
        \hline
    \mathstrut
 & 5~$\mu$m & 12~$\mu$m & 12~$\mu$m\\
$\lambda$ (nm) && Image qual. & Image scale \\[1ex]
 \hline 
 380 & 95\% & 95\% & 90\%\\
 400 & 96\% & 95\% & 91\% \\
 420 & 96\% & 95\% & 91\% \\
 440 & 96\% & 95\% & 92\% \\
 460 & 97\% & 96\% & 92\% \\
 480 & 97\% & 96\% & 92\% \\
 500 & 97\% & 96\% & 92\% \\
 \hline
  \end{tabular}
  \vspace{1mm}
  \caption{Minimum Strehl values for EST-B Wideband, for 5~$\mu$m pixel size and for 12~$\mu$m pixel size when the optical design is optimised for image quality and with more emphasis on reducing image scale variation when refocusing.}
 \label{table_sect3_6}  
\end{table}

Table \ref{table_sect3_5} summarises the minimum Strehl values for the three narrowband EST-B
systems considered here: the 5~$\mu$m pixel size system and the two designs for the 12~$\mu$m pixel size
system. Except for the outermost field point at 380 nm wavelength, the 5~$\mu$m and the 12~$\mu$m system
optimised for image quality perform equally well. Instead optimising the 12~$\mu$m system for
reduced image scale variation when refocusing, carries a noticeable penalty in terms of image quality,
with a lowest Strehl of 87\%. Table \ref{table_sect3_6} summarises the
performance of the corresponding wideband systems, which to a large extent mirrors the performance
of the narrowband systems.

\subsubsection{EST-B Telecentricity between the FPI plates}
An F-ratio of 110 corresponds to a marginal ray angle of 0.26\textdegree. The telecentricity of POP is
important here. If POP would not be telecentric, this number would be affected. In reality, POP is not
exactly telecentric at all wavelengths – the maximum departure from telecentricity is about 0.02\textdegree
for a field angle of 0.5 arc minute. This small departure from telecentricity is demagnified 2.2 times by
the collimator (doublet A plus doublet B) at the location of the FPI plates, so the contribution to the
telecentricity error from POP is only 0.009\textdegree. Additionally, EST-B itself contributes to this error
with 0.006\textdegree, giving a total of 0.015\textdegree which is negligible (6\%) when comparing to the marginal
ray angle of the F/110 beam.

\subsubsection{EST-B Transmission}
Tables \ref{table_sect3_7} and \ref{table_sect3_8}   show the transmission values of the instrument for each glass and wavelength, and
the total for each wavelength. The actual thickness of each glass is taken into account, but the lenses
are modelled as parallel plates with a thickness corresponding to that of each lens at its centre.
Absorption data for silica was missing and a value of 1.0 is assumed. The reflection losses are not
included. The numbers are valid for the narrow and wideband systems with 12~$\mu$m pixels. The
numbers for the 5~$\mu$m systems are not shown, but are very similar.

\begin{table}[h]
  \centering
  \small
   \setlength{\tabcolsep}{4.8pt}
  \begin{tabular}{|c|c|c|c|c|c|c|c|}
        \hline
    \mathstrut
$\lambda$ (nm)&380&400&420&440&460&480&500\\
Glass&&&&&&&\\
        \hline
SILICA&1.000&1.000&1.000&1.000&1.000&1.000&1.000 \\
PBL1Y&0.998&0.999&0.999&0.999&0.999&0.999&0.999 \\
N-FK51A&0.996&0.997&0.997&0.997&0.997&0.998&0.999 \\
N-FK5&0.994&0.996&0.995&0.995&0.993&0.993&0.993 \\
PBM2Y&0.991&0.995&0.996&0.996&0.996&0.996&0.998 \\
SILICA&1.000&1.000&1.000&1.000&1.000&1.000&1.000 \\
SILICA&1.000&1.000&1.000&1.000&1.000&1.000&1.000 \\
PBM2Y&0.991&0.995&0.996&0.996&0.996&0.996&0.998 \\
N-FK5&0.994&0.996&0.995&0.995&0.993&0.993&0.993 \\
N-PK51&0.991&0.995&0.994&0.995&0.996&0.997&0.99 \\
PBL1Y&0.998&0.999&0.999&0.999&0.999&0.999&0.999 \\
N-PK51&0.990&0.995&0.994&0.995&0.996&0.997&0.998 \\
SILICA&1.000&1.000&1.000&1.000&1.000&1.000&1.000 \\
\hline
TOTAL&0.944&0.967&0.968&0.967&0.967&0.969&0.974 \\
 \hline
  \end{tabular}
  \vspace{1mm}
  \caption{Transmission of optical glasses used for the beam splitters, FPIs and lenses in EST-B Narrowband
 system. Reflective losses depend on chosen anti-reflection coatings and are not included.}
 \label{table_sect3_7}  
\end{table}

\begin{table}[h]
  \centering
  \small
   \setlength{\tabcolsep}{4.8pt}
  \begin{tabular}{|c|c|c|c|c|c|c|c|}
        \hline
    \mathstrut
$\lambda$ (nm)&380&400&420&440&460&480&500\\
Glass&&&&&&&\\
        \hline
SILICA&1.000&1.000&1.000&1.000&1.000&1.000&1.000\\
PBL1Y&0.998&0.999&0.999&0.999&0.999&0.999&0.999\\
N-FK51A&0.996&0.997&0.997&0.997&0.997&0.998&0.999\\
N-PK51&0.990&0.995&0.994&0.994&0.995&0.996&0.997\\
PBL1Y&0.999&0.999&0.999&0.999&0.999&0.999&0.999\\~$\mu$
N-PK51&0.990&0.995&0.994&0.994&0.996&0.997&0.997\\
SILICA&1.000&1.000&1.000&1.000&1.000&1.000&1.000\\
\hline
TOTAL&0.972&0.985&0.983&0.984&0.987&0.989&0.992\\
 \hline
  \end{tabular}
  \vspace{1mm}
  \caption{Transmission of  beam splitters and lenses in the Wideband system. Reflective losses depend on chosen
anti-reflection coatings and are not included.}
 \label{table_sect3_8}  
\end{table}

\clearpage
\section{Design and performance of EST-V} \label{EST-V_FPI_system}

\begin{figure}[!h]
\center
\includegraphics[angle=0, width=0.99\linewidth,clip]{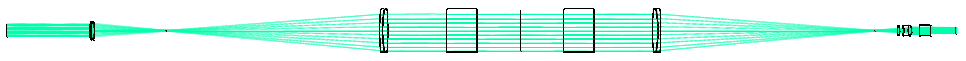}
\includegraphics[angle=0, width=0.99\linewidth,clip]{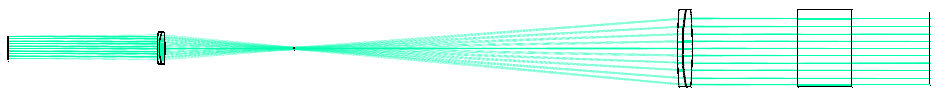}
\includegraphics[angle=0, width=0.99\linewidth,clip]{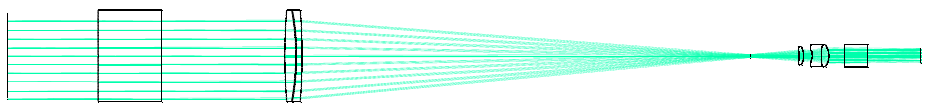}
 \caption{The layout of the entire narrowband  EST-V system (top), designed for 12~$\mu$m pixel size. The middle and bottom panels show the first and rear parts of the same system in more detail. The figures are stretched 2x in the vertical direction for clarity. F1-F3 are focal planes, L1-L4 are lenses, P1-P2 pupil planes, and PBS is the polarising beam splitter. The camera lens (L4) is an air spaced triplet lens, the other lenses are doublets.}
\label{fig:Fig_sect4_1}
\end{figure}

\begin{figure}[!h]
\center
\includegraphics[angle=0, width=0.99\linewidth,clip]{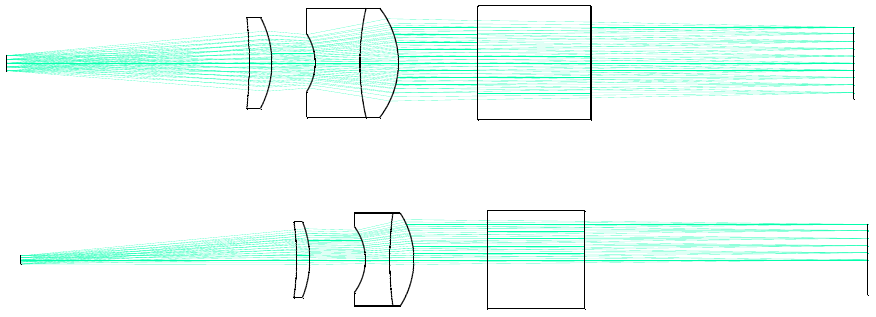}
 \caption{The layout of the camera lenses designed for the 6.5$\mu$m pixel size (top) and for the 12~$\mu$m pixel size system (bottom) for EST-V. Shown is also the polarising beam splitter. The scaling is different for the two panels.}
\label{fig:Fig_sect4_2}
\end{figure}

The approach to designing EST-V is similar to that of EST-B, and in order not to overburden the reader with information, we move most of the information about the design and performance of the narrowband and wideband systems of EST-V to Appendix \ref{EST-V_design2}. Here, we give the basic input parameters of the system in Tables \ref{table_sect4_1} and \ref{table_sect4_2}. We also show the optical layout of the EST-V system in Fig. \ref{fig:Fig_sect4_1} and that of the two the camera lens in Fig. \ref{fig:Fig_sect4_2}. Both designs are shown for a pixel size of both 6.5~$\mu$m and 12~$\mu$m.

Section \ref{EST-V_wideband_performance} in the Appendix contains Strehl tables and spot diagrams for EST-V with both 6.5~$\mu$m and 12~$\mu$m pixel size, with and without POP. The Strehl values are excellent, 97\% or more at all field points and wavelengths, with the designs of these camera lenses. The focus curve of POP is very flat in this wavelength range, so very little refocusing is needed for all wavelengths. 
\subsection{EST-V design}
\begin{table}[!h]
  \centering
  \small
  \begin{tabular}{|l|c|c|c|c|cc}
        \hline
    \mathstrut
Wavelength range (nm)& 500-680\\
Input F-ratio (at F3)& 50*\\
F-number between FPIs& 147**\\
Output F-ratio & Table \ref{table_sect4_2}\\
Magnification (Science focus to between FPIs) & 2.94\\
Magnification (Science focus to detector)& Table \ref{table_sect4_2}\\
Field diameter on sky (arc.sec.)& 60\\
Field diameter at F3 (mm)& 61\\
Field diameter between FPIs &179.3\\
Field diameter at detector &Table \ref{table_sect4_2}\\
Length F3 to second pupil stop (mm) &4000\\
Length F3 to detector. 12~$\mu$m pixels (mm)& 4694\\
Length F3 to detector. 5~$\mu$m pixels (mm) &4479\\
 \hline
  \end{tabular}
  \vspace{1mm}
  \caption{Summary of input parameters for EST-V. * POP design by Álvaro Pérez García. ** Obtained from numerical simulations, see Scharmer et al. (2025)}
 \label{table_sect4_1}  
\end{table}

\begin{table}[!h]
  \centering
  \small
  \begin{tabular}{|c|l|c|c|c|c|c|c}
        \hline
    \mathstrut
 Pixel size & Magn. & Image scale & Diameter & Output \\
 ($\mu$m) && at detector & at detector& focal \\
 & & ("/mm) & (mm) &ratio \\
\hline
4.0&0.307&3.250&18.7&15.35\\
6.5&0.500&2.000&30.5&25.00\\
8.0&0.615&1.625&37.5&30.70\\
12.0&0.923&1.083&56.3&46.15\\
 \hline
  \end{tabular}
  \vspace{1mm}
  \caption{Magnification, image scale and output focal ratios for different pixel sizes with EST-V. The input image scale is 1.0"/mm, the output pixel scale is 0.013"/pixel for all pixel sizes.}
 \label{table_sect4_2}  
\end{table}

In Appendix \ref{EST-V_design2}, we show spot diagrams and Strehl tables for EST-V with both pixel sizes. This is done both for EST-V as a stand alone system and when connected to POP.

\subsection{EST-V performance}
\begin{figure}[htbp]
\center
\includegraphics[angle=0, width=0.99\linewidth,clip]{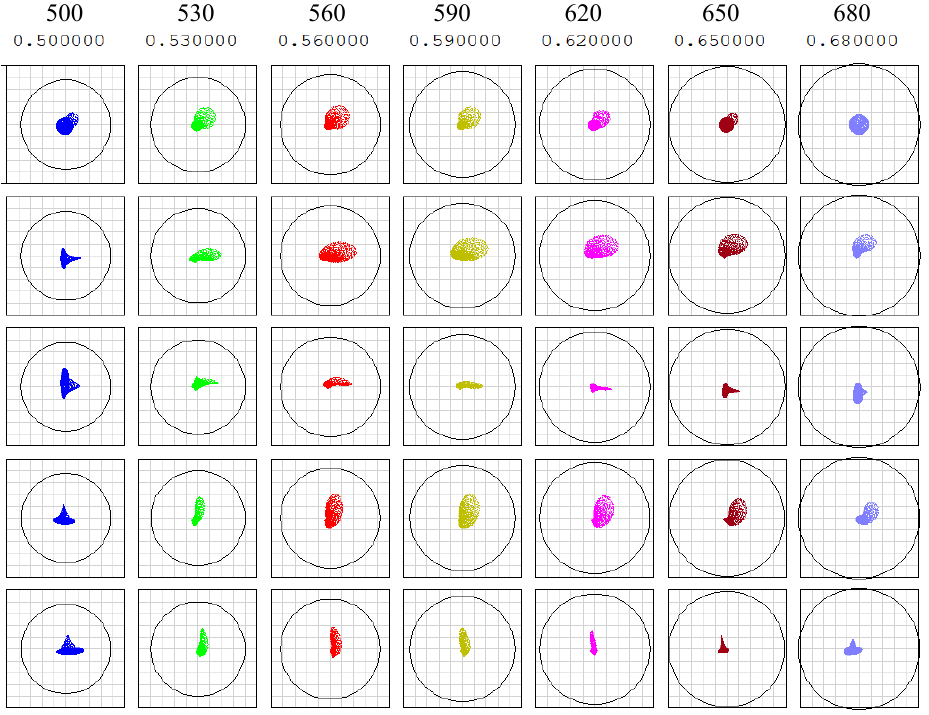}
 \caption{Monochromatic spot diagrams for EST-V stand alone (without POP), based on 100 simulations of random errors taken from Tables \ref{table_EST-V_tolerances1}-\ref{table_EST-V_tolerances3}, and choosing the sample with the lowest Strehl. The Strehl values are given i Table \ref{table_Strehl_EST-V_simulations}.}
\label{fig:Spot_diagram_EST-V_simulations}
\end{figure}

We made 100 simulations of the image quality with all tolerances given previously in Tables \ref{table_EST-V_tolerances1}-\ref{table_EST-V_tolerances3},
including those for the (difficult) triplet camera lens. The tolerances were drawn randomly from these
tolerance tables, with equal probability for all values within the given interval. In Fig. \ref{fig:Spot_diagram_EST-V_simulations}, the spot
diagrams corresponding to the poorest image quality from these 100 simulations are shown, and in
Table \ref{table_Strehl_EST-V_simulations} we give the corresponding Strehl values. The telecentricity at the location of the FPIs was unaffected by these tolerances, but the whole beam was tilted by an angle of 0.027 degrees relative to the optical axis.
\begin{table}[!h]
  \centering
  \small
  \begin{tabular}{|c|l|c|c|c|c|c|c}
        \hline
    \mathstrut
$\lambda$~(nm)&500&530&560&590&620&650&680\\
FOV&&&&&&&\\
\hline
Center of FOV&0.99&0.99&1.00&1.00&1.00&0.99&0.99\\
Edge North&0.98&0.99&0.99&0.99&0.99&0.99&1.00\\
Edge South&0.97&0.99&0.99&0.99&0.99&0.99&0.99\\
Edge East&0.98&0.99&0.99&0.99&0.99&0.99&1.00\\
Edge West&0.97&0.99&0.99&0.99&1.00&1.00&0.99\\
\hline
 \end{tabular}
  \vspace{1mm}
  \caption{Strehl values for EST-V alone (without POP), based on 100 simulations of random errors taken from Tables \ref{table_EST-V_tolerances1}-\ref{table_EST-V_tolerances3}, and choosing the sample with the lowest Strehl. Spot diagrams are shown in Fig. \ref{fig:Spot_diagram_EST-V_simulations}.}
 \label{table_Strehl_EST-V_simulations}  
\end{table}
\subsubsection{Telecentricity between the FPI plates}
An F-ratio of 147 corresponds to a marginal ray angle of 0.194\textdegree. The telecentricity of POP is
important here. If POP would not be telecentric, this number would be affected. In reality, POP is not
exactly telecentric at all wavelengths – the maximum departure from telecentricity is about 0.007\textdegree
for a field angle of 0.5 arc minute. This small departures from telecentricity is demagnified 2.94 times
by the collimator (doublet L1 plus doublet L2) at the location of the FPI plates, so the contribution to the
telecentricity error from POP is only 0.001\textdegree. Additionally, EST-V itself contributes to this error
with 0.005\textdegree, giving a total of 0.006\textdegree, which is negligible when comparing to the marginal ray angle of the F/147 beam.

The maximum deviation from telecentricity in detector space is about 0.035 degrees. With POP this is
slightly larger. This is of minor importance.

\subsubsection{Refocusing and image scale variations}
For the EST-V system with 6.5~$\mu$m pixels, the position of the input focal plane (corresponding to the
output focal plane of POP - F3) can be refocused at least $\pm$ 20~mm without any significant
deterioration of image quality. The camera lens needs to move $\pm$4 mm to compensate for that
refocus. The telecentricity error (in detector space), however, grows from 0.015\textdegree to about 0.18\textdegree for a 20~mm focus change. For the 12~$\mu$m system, the corresponding movement of the camera lens is about $\pm$18~mm, and the telecentricity error (in detector space) increases to 0.3\textdegree. The Strehl value is then reduced by 0.01.

The focus curve of EST-V with POP varies by only $\pm$0.15~mm for the 6.5~$\mu$m pixel system and $\pm$0.46~mm for the 12~$\mu$m pixel system. However, in spite of these small variations of the focal position, the image scale varies by a total of $\pm$ 0.5\% when EST-V is connected to POP. These image scale
variations are entirely coming from POP, and not EST-V. This is an intrinsic property of POP that may
be possible to reduce in a forthcoming revision of its design. In our opinion, however, this image scale
variation is not a concern. Remember, that the three FPI systems are intended to allow joint
diagnostics of the photosphere and chromosphere by allowing the simultaneous recording of
spectropolarimetric data at three or more wavelengths. These three different systems operate at
completely different image scales, ranging from 0.010" to 0.017" per pixel. The most probable
observing mode is that each FPI system will observe at a single wavelength, such that the variation of
image scale within each FPI system is comparatively small compared to the variation of image scale
for the entire data set, recorded with the three FPIs.

\subsubsection{Transmission}
Table \ref{table_EST-V_transmission} show the transmission values of EST-V with the 6.5~$\mu$m pixel size camera lens, for each glass individually and the total of the entire system. The
actual thickness of each glass is taken into account, but the lenses are modelled as parallel plates with
a thickness corresponding to that of each lens at its centre. Absorption data for silica was missing and
a value of 1.0 is assumed. The reflection losses are not included.

\begin{table}[!h]
  \centering
  \small
   \setlength{\tabcolsep}{4pt}
  \begin{tabular}{|c|l|c|c|c|c|c|c|}
        \hline
    \mathstrut
$\lambda  (nm)$&500&530&560&590&620&650&680\\
Glass&&&&&&&\\
\hline
SILICA&1.000&1.000&1.000&1.000&1.000&1.000&1.000\\
S-NBH8&0.994&0.997&0.998&0.998&0.998&0.998&0.999\\
N-PSK53A&0.995&0.996&0.997&0.996&0.995&0.996&0.996\\
SF6&0.996&0.997&0.998&0.998&0.998&0.998&0.998\\
S-LAL18&0.995&0.997&0.997&0.996&0.996&0.998&0.998\\
SILICA&1.000&1.000&1.000&1.000&1.000&1.000&1.000\\
SILICA&1.000&1.000&1.000&1.000&1.000&1.000&1.000\\
S-LAL18&0.995&0.997&0.997&0.996&0.996&0.998&0.998\\
SF6&0.996&0.997&0.998&0.998&0.998&0.998&0.998\\
S-BSM18&0.997&0.998&0.998&0.998&0.998&0.998&0.998\\
KZFSN5&0.994&0.995&0.996&0.997&0.997&0.997&0.997\\
N-SK15&0.993&0.995&0.995&0.995&0.995&0.995&0.995\\
SILICA&1.000&1.000&1.000&1.000&1.000&1.000&1.000\\
\hline
TOTAL&0.956&0.969&0.975&0.972&0.971&0.975&0.977\\
\hline
 \end{tabular}
  \vspace{1mm}
  \caption{Transmission of the EST-V FPI system designed for 6.5~$\mu$m pixels. Reflection losses are not included since these depend on choices of anti-reflection coatings. The transmission of the 12~$\mu$m pixel system is almost identical to that of the 6.5~$\mu$m pixel system, and is not shown.}
 \label{table_EST-V_transmission}  
\end{table}

\subsubsection{CTE variations}
For the 6.5~$\mu$m pixel system, the difference in CTE between the glasses in the first cemented doublet is 1.46
and 2.2 for doublet of the camera lens (L4). For the 12~$\mu$m pixel system, the difference for the doublet of the
L4 is 1.4.

\subsection{Concluding comments relating to EST-V}
We have presented a preliminary but quite detailed conceptual design of a telecentric FPI system for
EST, intended for use in the 500-680 nm wavelength range. The design goal is to minimise the clear
apertures of the FPIs in order to reduce their costs and the challenges of their manufacture, and to
make the optical path length as short as possible, such that there is no need for using mirrors to fold
the beam. At the same time, no compromise is made as regards image quality. The overall Strehl,
which is limited by a combination of apodisation effects at the etalons and the design, manufacture
and alignment of the optics, can be kept in excess of 0.9. Assuming a pixel size in the range 
6.5-12~$\mu$m, the overall length from the input focal plane (F1)  to the camera focal plane is in the range
4.5-4.7 m, which is well within the constraints of the available optics lab at EST.

The FPI system is modular in the sense of allowing an adoption to a wide range of pixel scales by only replacing the last
lens (the camera lens) and the tube connecting that lens to the nearby pupil stop.

Based on contacts with suppliers of optical glass, lenses and FPIs, the tolerance analysis presented
here, and previous experience from the design and production of two similar but smaller FPI systems
for the Swedish 1-m Solar Telescope (CRISP and CHROMIS), we firmly believe that the system can
be produced along the directions given by our design.

The weakest part of the design is the camera lens, which is a triplet lens with very demanding
tolerances, but that has the potential of delivering a Strehl close to 1 if these tolerances can be
respected. This lens design needs further discussions with potential suppliers of this lens. It is possible
that the outcome of these discussions will be to adopt a safer solution instead, at the prize of a small
reduction of the Strehl, by either redesigning the triplet lens or replacing this lens with an air-spaced or
even cemented doublet lens. If so, the tolerance requirements are relaxed to those of CRISP and
CHROMIS, both of which are highly successful installations.

\clearpage
\section{Design and performance of EST-R} \label{EST-R_design3}

\begin{figure}[!h]
\center
\includegraphics[angle=0, width=0.99\linewidth,clip]{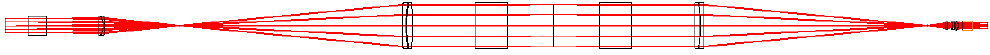}
\includegraphics[angle=0, width=0.99\linewidth,clip]{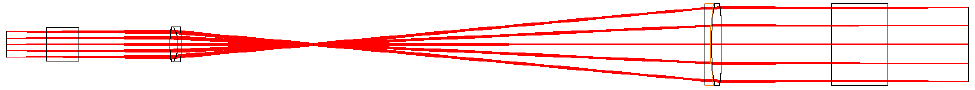}
\includegraphics[angle=0, width=0.99\linewidth,clip]{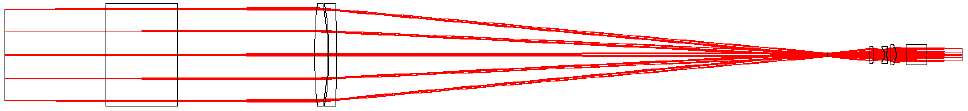}
 \caption{The layout of the entire narrowband  EST-V system (top), designed for 12~$\mu$m pixel size. The middle and bottom panels show the first and rear parts of the same system in more detail. The figures are stretched 2x in the vertical direction for clarity. F1-F3 are focal planes, L1-L4 are lenses, P1-P2 pupil planes, and PBS is the polarising beam splitter. The camera lens (L4) is an air spaced triplet lens, the other lenses are doublets.}
\label{fig:Fig_sect5_1}
\end{figure}

\begin{figure}[!h]
\center
\includegraphics[angle=0, width=0.99\linewidth,clip]{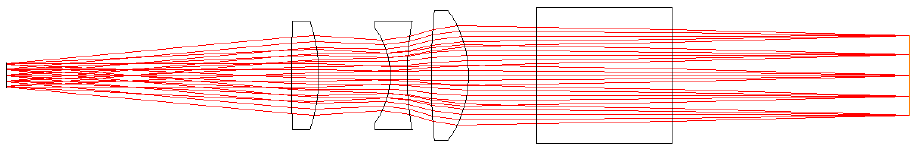}
\includegraphics[angle=0, width=0.99\linewidth,clip]{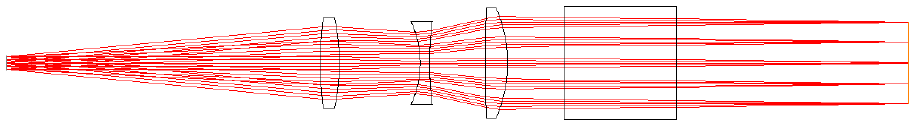}
 \caption{The layout of the camera lenses designed for the 6.5$\mu$m pixel size (top) and for the 12~$\mu$m pixel size system (bottom) for EST-V. Shown is also the polarising beam splitter. The scaling is different for the two panels.}
\label{fig:Fig_sect5_2}
\end{figure}

The approach to designing EST-R is similar to that of EST-B and EST-V, and in order not to overburden the reader with information, we move most of the information about the design and performance of the narrowband and wideband systems of EST-R to Appendix \ref{ESTR2}. Here, we give the basic input parameters of the system in Tables \ref{table_sect5_1} and \ref{table_sect5_2}. We also show the optical layout of the EST-R system in Fig. \ref{fig:Fig_sect5_1} and that of the two the camera lens in Fig. \ref{fig:Fig_sect5_2}. Both designs are shown for a pixel size of both 6.5~$\mu$m and 12~$\mu$m.

\subsection{EST-R design}
\begin{table}[!h]
  \centering
  \small
  \begin{tabular}{|l|c|c|c|c|cc}
        \hline
    \mathstrut
Wavelength range (nm)& 680-1000\\
Input F-ratio (at F3)& 50*\\
F-number between FPIs& 147**\\
Output F-ratio & Table \ref{table_sect5_2}\\
Magnification (Science focus to between FPIs) & 2.94\\
Magnification (Science focus to detector)& Table \ref{table_sect5_2}\\
Field diameter on sky (arc.sec.)& 60\\
Field diameter at F3 (mm)& 61\\
Field diameter between FPIs &179.3\\
Field diameter at detector &Table \ref{table_sect5_2}\\
Length F3 to second pupil stop (mm) &4000\\
Length F3 to detector. 12~$\mu$m pixels (mm)& 4266\\
Length F3 to detector. 5~$\mu$m pixels (mm) &4480\\
 \hline
  \end{tabular}
  \vspace{1mm}
  \caption{Summary of input parameters for EST-R. * POP design by Álvaro Pérez García. ** Obtained from numerical simulations, see Scharmer et al. (2025)}
 \label{table_sect5_1}  
\end{table}

\begin{table}[!h]
  \centering
  \small
  \begin{tabular}{|c|l|c|c|c|c|c|c}
        \hline
    \mathstrut
 Pixel size & Magn. & Image scale & Diameter & Output \\
 ($\mu$m) && at detector & at detector& focal \\
 & & ("/mm) & (mm) &ratio \\
\hline
4.0&0.235&4.25&14.33&11.75\\
6.5&0.382&2.62&23.30&19.10\\
8.0&0.472&2.12&28.79&23.60\\
12.0&0.704&1.42&42.94&35.20\\
 \hline
  \end{tabular}
  \vspace{1mm}
  \caption{Magnification, image scale and output focal ratios for different pixel sizes with EST-R. The input image scale is 1.0"/mm, the output pixel scale is 0.017"/pixel for all pixel sizes.}
 \label{table_sect5_2}  
\end{table}

\subsection{EST-R performance}
Sections \ref{EST-R_narrowband_performance} and \ref{EST-R_wideband_performance} in Appendix \ref{ESTR2} contains Strehl tables and spot diagrams for the narrowband and wideband systems of EST-R with both 6.5~$\mu$m and 12~$\mu$m pixel size, with and without POP. The Strehl values are excellent, 97\% or more at all field points and wavelengths, with the present designs of the camera lenses. The focus curve of POP is very flat in this wavelength range, so very little refocusing is needed for all wavelengths. A tolerance analysis for EST-R has not yet been carried out, but we expect more relaxed tolerances than for EST-B and EST-V.

\clearpage
\section{Further comments on image scale variations} \label{im_scale_variations}
In this Section we summarise calculated variations of image scales for all three narrowband and wideband FPI systems proposed for EST. We also calculate the more important differences  of these variations between the narrowband and wideband systems at each wavelength. Tables \ref{table_im_scale_all_5-6.5mu} and \ref{table_im_scale_all_12mu} summarises the results of these calculations.

\begin{table}[!h]
  \centering
  \small
  \begin{tabular}{|l|c|c|c|c|c|c|c}
        \hline
    \mathstrut
    System & EST-B & EST-V & EST-R \\
    Pixel size ($\mu$m) & 5.0 & 6.5 & 6.5 \\
    \hline
    Without POP &&&\\
    \hline
    Intern. variation narrowband ($\mu$m) & 87 & 1.4 & 1.7\\
    Internal variation wideband ($\mu$m) & 81 & 0.6 & 1.9 \\
    Difference at each wavelength ($\mu$m) & 8  & small & small \\
    Difference at each wavelength (\%) & 0.05\% & - & -\\
    \hline
    With POP &&&\\
    \hline
    POP itself ($\mu$m) & 293 & 148 & 123 \\
    Intern. variation narrowband ($\mu$m) & 52 & 72 & 51 \\
    Intern. variation wideband ($\mu$m) & 67 & 67 & 37 \\
    Difference at each wavelength ($\mu$m) & 20  & 1 & 14 \\
    Difference at each wavelength (\%)& 0.13\% & 0.006\%& 0.12\%\\
 \hline
  \end{tabular}
  \vspace{1mm}
  \caption{Chromatic variation of image scale over the wavelength ranges of the three FPI systems developed for EST, for a pixel size of 5.0 or 6.5~$\mu$m.}
 \label{table_im_scale_all_5-6.5mu}  
\end{table}

\begin{table}[!h]
  \centering
  \small
  \begin{tabular}{|l|c|c|c|c|c|c|c}
        \hline
    \mathstrut
    System & EST-B & EST-V & EST-R \\
    Pixel size ($\mu$m) & 12.0 & 12.0 & 12.0 \\
    \hline
    Without POP &&&\\
    \hline
    Intern. variation narrowband ($\mu$m) & 3 & 7.3 & 3.5 \\
    Internal variation wideband ($\mu$m) & 4 & 3.2 & 5.9 \\
    Difference at each wavelength ($\mu$m) & small  & small & small \\
    Difference at each wavelength (\%)& - & - & -\\
    \hline
    With POP &&&\\
    \hline
    POP itself ($\mu$m) & 293 & 148 & 123 \\
    Intern. variation narrowband ($\mu$m) & 269 & 134 & 132 \\
    Intern. variation wideband ($\mu$m) & 458 & 138 & 48 \\
    Difference at each wavelength ($\mu$m) & 212  & 11 & 110 \\
    Difference at each wavelength (\%) & 0.58\% & 0.04\%& 0.5\%\\
 \hline
  \end{tabular}
  \vspace{1mm}
  \caption{Chromatic variation of image scale over the wavelength ranges of the three FPI systems developed for EST, for a pixel size of 12~$\mu$m.}
 \label{table_im_scale_all_12mu}  
\end{table}

We can conclude from these Tables, that the difference between the intrinsic narrowband and wideband image scale variations of the FPI systems are small. However, when connected to POP, these differential variations are much larger. The main reason for this is the large image scale variation of POP itself. In this regard, the EST-V system behaves better than the other two FPI systems. The reason for that is that the camera lenses for the narrowband and wideband systems of  EST-V were optimised separately, with this improvement in mind. In general, the co-variation of image scale of the narrowband and wideband systems when the EST FPI systems are refocused have not been studied in sufficient depth so far. This should be done, but is probably not meaningful until the optical design of POP is in a more final state. As discussed earlier, it is also not yet clear just how accurate the co-focusing of the narrowband and wideband systems needs to be, and that must be clarified as well.

\section{Comments relating to the manufacture, installation and alignment of the FPI system} \label{comments_manufacture}
Based on experience with CRISP and CHROMIS, we make the following comments and suggestions,
of relevance to the manufacture, transportation, installation and adaptability of the narrowband FPI
systems.

Each FPI system contains the following points of particular interest: the input focal plane F1, lens L1,
pupil plane P1, L2, FPI1, FPI2, L3, P2, L4 and the output focal plane F2. We will refer to the part of
the system that begins with pupil plane P1 and that ends with pupil plane P2 as the “core system” and the part that includes L1 plus the core system as the “mother system”. This division of the full FPI
system into two subsystems makes sense from the point of its flexibility to modifications:

The core system, which for the EST-B system has a total length of 3257 mm, is defined by an input
pupil position P1, L2, FPI1, FPI2, L3 and its output pupil position P2. This allows some flexibility on
both the input side and the output side. In principle, we can use a lens L1 of any focal length. As long
as this lens re-images the pupil precisely on P1, the pupil will be re-imaged at infinity by the fixed lens
L2 at FPI1 and FPI2, and onto P2 by L3. This gives flexibility in modifying the f-ratio of the beam at
the location of the etalons, thus impacting the diameter of the useful field of view and (because of
pupil apodization effects) the spectral resolution. Although we expect the system to be well designed
initially, moderate changes to the optical system may later turn up to be desirable. Another aspect is
testing of the system before installation at EST: by modifying L1 and the diameter of the pupil stop,
there should be good possibilities to temporarily adopting the FPI system to a different telescope, or
other test setup, for initial tests and optimisation of its control software and camera system.

The mother system is defined by L1 plus the core system. By proper choice of design and location of
L4, the mother system allows adoption of its final image scale to the cameras chosen, while making
the re-imaging system telecentric (if P2 is re-imaged to infinity by L4). In our design, we have chosen
to optimize the overall design for two different pixel sizes that we believe cover the range of future
cameras.

For transport and installation, it is convenient to split the instrument into five pre-fabricated parts: 1)
the lens tube that holds lens L1 (doublet A); 2) the part that holds the first pupil stop and the tube
connecting to the container part; 3) the container (middle) part that holds L2 (the first doublet B), the
14
two FPIs, and L3 (the second doublet B); 4) the part that includes the tube connecting to the container
part and also holding the second pupil stop P2; and 5) the part that holds the remainder of that tube
and that ends with the camera lens L4. All parts must be manufactured to tolerances set by the optical
design and the previous tolerance analysis. Assembly of the five parts must be possible to within these
tolerances without recourse to additional alignment procedures but possibly includes help tools to
ensure proper mounting. Of particular importance is to design and manufacture the core system to be a
robust and self supporting mechanical structure, which is sufficiently stiff and that allows dismounting
and easy reassembly after transportation without losing its critical alignment tolerances. Note also that
the second pupil stop should be easily removable to simplify initial alignment of the system with the
optical axis of the telescope.

The container part is sealed as a vacuum system with L2 and L3 acting as vacuum windows through
mounting with O-rings on the sides of these lenses facing the FPI etalons. Vacuum proof connectors
are used for the cabling needed to tune and control the etalons, and for controlling the tip-tilt angles of
the etalons. However, the container operates at a pressure close to the adjacent air pressure. but is
flushed and filled with dry clean nitrogen gas.

We consider it mandatory to keep the etalon surfaces, and their external surfaces, well away from the
focal plane, which nominally is located at the center of the FPI container (midways between L2 and
L3). This is in order to ensure that any defects, including small-scale cavity errors, appear strongly
defocused at the focal plane of the cameras. Our recommendation is to have the surface of the etalons
that is closest to that focal plane at a distance for which a point object would be smeared to about 1
mm. If the f-ratio at the etalons is 110, that corresponds to a distance of about 110 mm. However, to
achieve that in practice, we have to take into account a small focus error on the input side (POP) of up
to 5 mm plus the focus curve of POP which corresponds to 6.4 mm for EST-B, which gives a total of
12 mm (rounded up). This input focus error is multiplied by the square of the f-ratio at the etalons
(110 for EST-B and 150 for EST-V and EST-R) divided by the input f-ratio (50). This multiplies the
input focus error by almost a factor 5 for EST-B and 9 for EST-V and EST-R. The FPI container
therefore must be long enough to allow the etalons to be mounted at least 200-250~mm from the midpoint
of the container. As Table \ref{table_EST-B} demonstrates, the distance between the innermost surfaces of the two
lenses L2 and L3 (doublets B) is 1.5 m, such that there is ample space to mount the FPIs away from
the FPI focal plane.

There are two pupil locations in the system, and it is crucial to use pupil stops to ensure the
elimination of ghost images and unwanted interference fringes. The first pupil stop could have a
diameter that is about 20\% larger than the actual pupil image, and the second about 10\% larger. Both
pupil stops must be mounted on x/y translation stages with micrometer screws to allow a minor
tweaking of the pupil stop location, once the overall system is well aligned. The initial centering of the
pupil stops must be very accurate, and it must be possible to return the pupil stop location to that
center position with high accuracy, in case a complete realignment of the system is needed.

Once all parts are assembled, alignment of the entire system is made by centering the front part on the
center of field-of-view of POP, and centering the rear part by ensuring that the pupil image ends up
well centered on the two pupil stops. To aid this alignment, it is convenient if the first pupil stop can
be viewed through a sort of retractable periscope that on the input side contains a small 50\% beam
splitter cube that allows the pupil image to pass through the beam splitter to the pupil stop, while the
same pupil stop can be viewed from the outside via the periscope. Once the pupil image is roughly
centered on the pupil stop, it is more convenient to view the pupil stop from the other side of the same,
and to ensure that the pupil image is very well centered on its pupil stop.

Once the FPI system is adequately aligned with respect to the first pupil stop, a similar procedure is
used to fine tune the alignment by ensuring that the pupil image is well centered on the second pupil
stop. At that point, a minor tweak of the pupil stop position with the micrometer screws may be
necessary. The system is now well aligned as regards the reimaging system.

As discussed above, the input focal plane of EST-B should coincide with the F3 focal plane (the
output focal plane of POP) with an error of less than only 5 mm. Even if the location with F3 can be
measured accurately, it seems likely (as far as we know now), that the F3 focus will show e.g.
seasonal or day-to-day drifts that cannot be compensated with the SCAO system. We have argued, that
either POP must have some kind of focusing mechanism (which is preferable but challenging, since
the lenses of POP are used as vacuum windows), or the optical table holding the narrowband FPI system 
and its wideband counter part must be mounted on a supporting structure that allows translation along its optical axis.

To align the FPIs inside their container needs motors that control tip and tilt of the FPIs, arranged such
that their alignment does not change and remains stable when the power to the tip-tilt motors is switched off. The high
resolution FPI must first be accurately aligned such that its reflecting surfaces are exactly
perpendicular to the input optical axis, which is easily accomplished if this FPI is the first etalon in the
beam. By checking that the reflected light from the etalon surfaces goes precisely back in the direction
of the incoming beam, it is easy to ensure its accurate tip-tilt alignment.

Assuming that the high-res FPI is accurately aligned with the optical axis, it is imperative that the
second low resolution etalon is tilted slightly to eliminate ghost images and associated unwanted
(unstable) interference fringes. This is done most easily by observing the inter etalon reflections
between the two etalons, using the previously described periscope first on the side of the low-
resolution etalon facing the incoming beam. These reflections produce multiple ghosts of the pupil
image close to the second pupil stop. When these ghost pupil images are aligned on top of each other,
the optical axis of the low-resolution etalon is aligned with the optical axis. Tilting that etalon slightly
gives a row of pupil images with decreasing intensity away from the primary pupil image. When the
first ghost image barely touches the primary ghost image, the tilt angle is about 1/(2 F\#) radians,
where F\# is the focal ratio (=110 for EST-B, and 147 for EST-V and EST-R) at the location of the etalons. This is the minimum (and
optimum!) tilt angle needed to eliminate the ghost images. By switching the periscope to the output
side of the low resolution etalon, and possibly tweaking the x,y position of the pupil stop slightly, we
can now assure proper alignment of the FPI system.

The previous procedure for mounting and aligning an FPI system has, but without the use of the
periscopes described above, been applied to CHROMIS and took less than one day to complete
when the etalons were pre-installed in the container, and the lenses pre-installed in their tubes or the
FPI container.

The camera lens L4 (singlet C plus doublet C) must be mounted in a single lens tube that can be
translated along the optical axis by a sufficient amount to compensate for the focus curves of
POP+EST-V plus any error in the co-alignment of the focus positions of POP (F3) and the input focal
plane of EST-V. This requires a fast and highly repeatable translation stage with a travel range of at
least $\pm$ 15 mm for the camera lens tube.

We also comment on the space available on the input and output sides of EST-V. On the input side,
there needs to be space for the 80 mm wideband beam splitter cube, the filter wheel, and the
polarisation modulator (in the order mentioned, but it likely is a better idea to put this modulator on
the output side between the camera lens and the polarising beam splitter if the camera pixel size is 8~$\mu$m or smaller), close to but not exactly at the input focal plane (corresponding to the output of POP –
F3). For example, in Table \ref{table_EST-B}, which describes the mother system of EST-B, the distance from the focal plane to the first surface of doublet A is given as 195 mm,
so there is enough room to place the modulator between that focal plane and doublet A. The 80 mm
wideband beam splitter cube and the filter wheel are then placed in front of, but close to, F3. At the
input focus position of EST-B (exactly!!) there needs to be a mechanism that can fold in a pinhole
array, a field stop, and likely other targets for calibrating image scales and alignments of data sets
taken at different wavelengths. Right after the 80 mm wideband beam splitter cube, there probably
needs to be a second beam splitter cube for sending light to the wavefront sensor for the EST-V
system, but not for EST-B or EST-R, so this is not an issue here.

On the output side after the camera lens, there needs to be space for the polarisation modulator (this is
likely the best location for that if the pixel size is 8~$\mu$m or smaller), a 50 mm polarising beam splitter
cube with the 5~$\mu$m pixel size system and for a 100 mm cube with the 12~$\mu$m system, plus two cameras
at 90 deg angle to each other. The latter requires a minimum distance of W/2, where W is the width of
the camera, from the center of the cube to the focal plane of the camera. For the 5~$\mu$m system of EST-B (see
Table \ref{table_EST-B_5my_camera_lens}) the distance from the last surface of doublet C to the focal plane is 181 mm such that the
distance from the center of the 50 mm cube to the focal plane could be up to a maximum of 156 mm.
For the 12~$\mu$m system, the distance from the last surface of doublet C to the focal plane is (see Table \ref{table_EST-B_12my_camera_lens1})
232~mm, such that the distance from the center of the 90 mm cube to the focal plane could be a
maximum of 140 mm. In both cases, there will be no problem with space for the cameras. For the
wideband system, the space available on the input and output sides is somewhat less than that for
EST-V.

The 80 mm wideband beam splitter splits the input beam into two beams, of which the wideband beam
is rotated by 90 deg relative to the input beam. The wideband beam obviously needs sufficient space
in the direction perpendicular to the input optical axis. From Tables \ref{table_EST-B_5my_camera_lens_wideband}--\ref{table_EST-B_12my_camera_lens_wideband_imscale}, we can add up the
“thicknesses” given to calculate the total length from the input focal plane (F1) to the camera focal
plane (F3). To that must be added approximately 235~mm, corresponding to the distance from the center of the wideband beam splitter to the first surface of L1. For the 5~$\mu$m system, this length is about 590~mm but for the 12~$\mu$m system optimised for reduced image scale variations, this distance is about 1110~mm. The mechanical structure holding the narrowband and the wideband systems should be wide enough to allow the overall all length of the wideband system with cameras to expand even further than that required with a 12~$\mu$m pixel size. Possibly, there could be need for folding the beam of the wideband system.

\section{Tolerances} \label{tolerances}
A detailed tolerance analysis is needed for the final system chosen for manufacture. In Appendix \ref{tolerances}, we
carry out a preliminary discussion of this for EST-B. Here, we summarise the most important results.
It is assumed that the design of the lenses will be iterated on the basis of melt data from the producer
of the optical glasses, such that there is no need to set tolerances on the refractive index. Because the
lenses are very large compared to the footprint of the pupil projected on the lens surfaces, the
requirement on polishing errors is very relaxed: about ½-1 wave peak to valley is sufficient, and that is
easy to achieve. As regards the lenses, the tolerances on curvatures, thicknesses, wedge angles and
mounting are tight but achievable, as demonstrated by CRISP and CHROMIS.
A more demanding challenge is the manufacture and mounting of the camera lenses. For this, it is
necessary to contact the anticipated manufacturer and establish that the tolerances are indeed
achievable. We outline a possible approach to that in Sect. A4.4. Possibly, this dialogue with the
manufacturer could lead to the need of redesigning this lens in order to relax the tolerances.

\section{Discussion and conclusions} \label{discussion}
We have presented a preliminary but quite detailed conceptual design of three telecentric FPI system for
EST, covering the entire visible and NIR wavelength range, from 380~nm to 1000~nm. The design goals
involve minimising the clear apertures of the FPIs in order to reduce their costs and the challenges of
their manufacture, and to make the optical path length as short as possible, such that there is no need
for using mirrors to fold the beam. The choice made is also to use lenses rather than mirrors for all the
re-imaging optics in order to maximise the overall transmission and to make the system as
robust mechanically and optically as possible. At the same time, almost no compromise is made as regards image
quality.

The FPI systems are modular in the sense of allowing an adoption to a wide range of pixel scales by
only replacing the last lens (the camera lens) and the tube connecting that lens to the nearby pupil stop.
In addition, the modularity even allows the first lens to be replaced by one of a different focal length, thus
modifying the diameter of the field-of-view and the f-ratio at the location of the etalons. This
possibility is not explored any further at present, however.

Of these three systems, the most challenging is EST-B, which is intended for use in the 380-500 nm wavelength range. 
The most demanding part of the design of EST-B is the camera lens, which is a triplet lens with tight
tolerances, but that has the potential of delivering a Strehl close to 100\% if these tolerances can be
respected. This lens design needs further discussions with potential suppliers. It is possible that the outcome of
these discussions will be to adopt a safer solution instead, at the prize of a small reduction of the
Strehl, by either redesigning the triplet lens or replacing this lens with an air-spaced or even cemented
doublet lens. If so, the tolerance requirements are relaxed to those of CRISP and CHROMIS, both of
which are highly performing FPI systems, operating since many years (Scharmer et al. 2025).

A particular challenge in designing an FPI system for EST is its secondary re-imaging system, called
POP (Pier Optical Path). In its present preliminary design, POP consists of a large triplet lens on the input side, a beam splitter to divide
the input beam in two, plus a field lens and a doublet lens for each of the two beams. The challenge
consists of two parts: one is the focus curve of POP, which can be considered predictable and is likely
to show only small variations over the working wavelength range of each FPI system. This focus
curve requires refocusing for every spectral line used by the FPI systems. The second challenge is to 
ensure that the focus error on the input side of the FPI systems can be constrained to very small values - we have suggested 5~mm as an upper limit in order to achieve two goals. The first goal is to limit the movement of the FPI focusing mechanism (see below) to small values, and the second is to stabilise the movement of the focal plane midways between the FPIs.
We address these two challenges in different ways, as described below.

The focus curve of POP requires refocusing of any EST instrument that is tuned in wavelength. There
is thus the need for a rapid refocusing method that works on each FPI system individually, whenever
science data are obtained at high cadence from two or more spectral lines with the same FPI system.
We propose to use the camera lens (the last lens) of the FPI system for such refocusing. This is
because a) it is telecentric and thus should not change the image scale when refocused (but see below);
 b) this lens is relatively small and thus involves the movement of only a small mass; c) the required
translation along the optical axis is small; and d) highly repeatable translation stages are commodity
opto-mechanical hardware that is available from several manufacturers. We also investigate the
possibility of moving the cameras instead of the camera lenses but advice against that option. This is
primarily because each combination of a narrowband and wideband FPI system will have at least three
cameras and will have four cameras when phase diversity is used with the wideband system. Moving
cameras to refocus thus requires twice as many mechanisms to control when compared to moving the
two camera lenses. The other arguments against moving the cameras rather than the camera lenses are
their larger mass and complications from connected cables, and possibly (this needs confirmation) a
greater risk of producing flat-field artefacts when moving optics at or close to the focal plane.

In our design study, we pay particular attention to the image quality of the FPI system when used at
different wavelengths, which requires refocusing to compensate for the focus curve of POP. We do
this for optical designs with pixel sizes of both 5 or 6.5~$\mu$m and 12~$\mu$m, which covers the range of likely pixel
sizes, where the only difference is the design of the camera lens. Both designs deliver minimum
Strehls close to 95\% at all wavelengths and field angles, demonstrating a desirable flexibility to future
changes of cameras and pixel sizes. However, the design with smaller pixel size offers several
advantages over a larger pixel size. The first is that the camera lens is smaller and thus easier to move
fast. The second advantage is that the translation of the lens needed to compensate POPs focus curve
scales as the square of the magnification (counted from the input side of the FPI system to the final
focal plane). Specifically, the range needed to compensate POPs 6.3 mm focus curve within the
working wavelength range of EST-B is only 1.5 mm with the design for 5~$\mu$m pixel size and 9.3 mm
for the 12~$\mu$m pixel size system. This should be compared to the minimum distance needed to translate
a linear stage with 70 mm pre-filters ($\pm$80 mm), or rotating a filter wheel with such filters, which
involves a large movement. We are confident that refocusing can be made well within the time needed
to change pre-filter, which means that refocusing does not carry any penalty in terms of reduced duty
cycle or overall efficiency of the FPI system, beyond that induced by a necessary pre-filter change.

In addition to the arguments given above, the use of cameras with a small rather than large pixel size
offers advantages in terms of costs and availability of commercially available cameras, and the
possibility of using a polarization modulator with relatively small aperture close to the focal plane,
instead of a much larger modulator in front of the entire FPI system (in the F/50 output beam of  POP).

Slow drifts (from one day to the next, or seasonal variations) in the focus curve of POP are assumed to
be compensated for by translation of the entire mechanical structure holding the FPI system with its
polarimeter and filter wheel, wideband counterpart, and all cameras, along the optical axis. We
rather arbitrarily propose allowance for a focus range of $\pm$100 mm for that translation. Our
expectation is that efforts will be spent on stabilising focus drifts of POP to much smaller
values than that.

Faster drifts in the focus curve, occurring within a single observing day, should preferably be handled
by the M2 AO system (SCAO) of EST without exhausting more than 10\% of its available stroke. This
is in order to change the optical setup as little as possible, such that (polarimetric) flatfield artefacts
arising from any such changes can be avoided. Should there be a need be to compensate for focus
drifts beyond the capability of the AO system during the day, there are two choices: either to focus
the last (doublet) lens of POP, or to translate the entire construction holding the narrowband and wideband
systems with their cameras, filter wheels and beam splitters. We strongly advice against using the camera 
lens to compensate for thermal drifts.

A relevant question is whether the image scale changes significantly when the camera lens is
refocused. Note that the three FPI systems will by design be optimized for their individual operating
wavelengths, and thus operate at (widely) different image scales. Co-aligning and registering such
diversely sampled data is routinely made with data from e.g., SST. Image scale changes resulting from
wavelength tuning within each FPI system should therefore not be a concern. However, if during an
observing day the focal plane of POP drifts by an amount that is so large that it cannot be handled by
the AO system, it "must" be handled by a slow focusing mechanisms as discussed previously. 
By thus forcing the FPI focusing to only deal with the focus curve of POP, and not its thermal drifts, 
we make the FPI focusing 100\% repetitive, such that we minimise (but not completely eliminate) the risk of flatfield artefacts caused
by FPI focusing not being consistent when recording science data and calibration data (flats). This 
recommendation is in part a consequence of our only moderately successful attempt to redesign the
camera lens of the 12~$\mu$m pixel size system to reduce the image scale variations when refocusing,
and the penalty in Strehl this gives rise to.

As a response to the above challenge, we have developed preliminary, but quite detailed, conceptual designs of three
telecentric FPI systems for EST-B. These three systems differ only in their design of
the camera lens. The first such system is optimised for a pixel size of 5~$\mu$m and the two remaining
systems are for a pixel size of 12~$\mu$m. The first of the two 12~$\mu$m systems is optimised with respect to
image quality, and the second system is optimised with attention paid also to minimising image scale
variations when the camera lens is used to compensate for the focus curve of POP. Prioritising this
type of constraint carries a noticeable drop in Strehl from about 95\% to 90\%. Considering the  
limited reduction of the image scale variations, and the prize it carries in terms of reduced Strehl, 
we do not recommend this type of design. 

A major effort has gone into designing the wideband re-imaging system such that its
image scale changes by the same amount as that of the narrowband system when it is refocused. This
is because the wideband system serves as an anchor channel for the processing of all narrowband data (Scharmer et al. 2025)
By recording images with narrowband and wideband cameras that are synchronized,
we can ensure that seeing induced aberrations, including image shifts and geometric distortions, are identical for the
narrowband and wideband images. This is crucial in processing all data obtained with FPI systems.
We have found that to achieve this goal, the camera lenses of the narrowband and wideband systems
need to be slightly different. Although that somewhat disturbs the elegance of previous SST designs
(CRISP, CRISP2 and CHROMIS), which uses duplicates of L1 and L4 of the narrowband system to construct the
wideband system, it should increase the overall cost of the FPI system only marginally.

However, the relevance of the above discussion about the importance of minimising image scale differences 
between the narrowband and wideband systems fully depends on whether or not this is important for 
processing the narrowband and wideband images together. The present MOMFBD code used to process all CRISP and 
CHROMIS data at SST requires the image scales to be the same. However, this is merely a feature 
of this code, and there are no obvious reasons for why a code allowing differences in image scale should not work
as well (Mats Löfdahl, private communication). There are several questions that should be answered before we force the narrowband FPI 
system and its wideband companion to deliver the same image scale: i) how big differences in image 
scale between the narrowband and wideband systems can the present MOMFBD code tolerate without
delivering noticeably poorer results? ii) Is it perhaps possible to process first the wideband images, and then
the narrowband images using the results of the previous processing? If so, differences in image scale
are of no consequence. iii) If narrowband and wideband images need to be processed together, what is 
the cost and effort of rewriting the MOMFBD code to handle different image scales? These 
questions should be answered before we decide whether we should force the image scales of the 
narrowband and wideband systems to be identical, if such a constraint induces significant costs or 
challenges in the design or manufacture of FPI systems for EST.

To conclude, we have presented designs and an analysis of designs of FPI systems for EST that clarify the nature
of the main challenges, but also suggest their solution. In spite of some concerns expressed above, we conclude that 
it is fully feasible to construct highly performing,
compact and robust FPI systems for EST, which can easily be adopted to future pixel sizes.
The overall Strehl of EST-B, which is limited by a combination of apodization effects at the etalons
and the design, manufacture and alignment of the optics, can be kept in excess of 90\%. 
The EST-V and EST-R FPI systems are less challenging than EST-B, and can more easily reach their target Strehls.
The overall lengths of these systems, from the input focal plane (F1) to the camera focal plane (F3) is in the range 4.4--4,7~m for all systems. 

Continued developments of designs for these FPI systems need a more detailed investigation of tolerances 
for the manufacture and assembly of the lenses, in particular the camera lenses for EST-B. Further design work for the camera lenses of EST-V and EST-R may (or may not!) be desirable in order to reduce their image scale variations when refocusing, which was not considered a concern at the time when these lenses were designed.

\begin{acknowledgements}
The Swedish 1-m Solar Telescope is operated on the island of La Palma by the Institute for Solar Physics of
Stockholm University in the Spanish Observatorio del Roque de los Muchachos of the Instituto de Astrofísica
de Canarias. The Institute for Solar Physics is supported by a grant for research infrastructures of national importance from the Swedish Research Council (registration number 2021-00169).  This work also has received funding from the European Union’s Horizon 2020 research and innovation programme under grant agreement
No 824135.

The European Solar Telescope project is supported by a grant for research infrastructures from the Swedish Research Council (registration number 2023-00169).  

CRISP and CHROMIS were funded by the Wallenberg Foundations, registration numbers 2003.0037 and 2012.1005. 

G. Scharmer acknowledges valuable information related to EST, POP and requirements for the FPI systems by the EST Science Advisory Group (SAG) and the EST Project Office (in particular Claudia Ruiz de Galarreta) and discussions with and feedback from TIS consortium members Javier Bailén, Luis Bellot Rubio, Luca Giovanelli, Matteo Munari and Javier Sanchez. Álvaro Pérez García is thanked for providing details of the design of POP. Mats Löfdahl is thanked for comments on the use of the MOMFBD code.
\end{acknowledgements}


\begin{thebibliography}{57}
\expandafter\ifx\csname natexlab\endcsname\relax\def\natexlab#1{#1}\fi
\bibitem[{{Quintero Noda} {et~al.}(2022){Quintero Noda}, {Schlichenmaier},
  {Bellot Rubio}, {L{\"o}fdahl}, {Khomenko}, {Jur{\v{c}}{\'a}k}, {Leenaarts},
  {Kuckein}, {Gonz{\'a}lez Manrique}, {Gun{\'a}r}, {Nelson}, {de la Cruz
  Rodr{\'\i}guez}, {Tziotziou}, {Tsiropoula}, {Aulanier}, {Aboudarham},
  {Allegri}, {Alsina Ballester}, {Amans}, {Asensio Ramos}, {Bail{\'e}n},
  {Balaguer}, {Baldini}, {Balthasar}, {Barata}, {Barczynski}, {Barreto
  Cabrera}, {Baur}, {B{\'e}chet}, {Beck}, {Bel{\'\i}o-As{\'\i}n},
  {Bello-Gonz{\'a}lez}, {Belluzzi}, {Bentley}, {Berdyugina}, {Berghmans},
  {Berlicki}, {Berrilli}, {Berkefeld}, {Bettonvil}, {Bianda}, {Bienes
  P{\'e}rez}, {Bonaque-Gonz{\'a}lez}, {Braj{\v{s}}a}, {Bommier}, {Bourdin},
  {Burgos Mart{\'\i}n}, {Calchetti}, {Calcines}, {Calvo Tovar}, {Campbell},
  {Carballo-Mart{\'\i}n}, {Carbone}, {Carlin}, {Carlsson}, {Castro L{\'o}pez},
  {Cavaller}, {Cavallini}, {Cauzzi}, {Cecconi}, {Chulani}, {Cirami},
  {Consolini}, {Coretti}, {Cosentino}, {C{\'o}zar-Castellano}, {Dalmasse},
  {Danilovic}, {De Juan Ovelar}, {Del Moro}, {del Pino Alem{\'a}n}, {del Toro
  Iniesta}, {Denker}, {Dhara}, {Di Marcantonio}, {D{\'\i}az Baso}, {Diercke},
  {Dineva}, {D{\'\i}az-Garc{\'\i}a}, {Doerr}, {Doyle}, {Erdelyi}, {Ermolli},
  {Escobar Rodr{\'\i}guez}, {Esteban Pozuelo}, {Faurobert}, {Felipe}, {Feller},
  {Feijoo Amoedo}, {Femen{\'\i}a Castell{\'a}}, {Fernandes}, {Ferro
  Rodr{\'\i}guez}, {Figueroa}, {Fletcher}, {Franco Ordovas}, {Gafeira},
  {Gardenghi}, {Gelly}, {Giorgi}, {Gisler}, {Giovannelli}, {Gonz{\'a}lez},
  {Gonz{\'a}lez}, {Gonz{\'a}lez-Cava}, {Gonz{\'a}lez Garc{\'\i}a},
  {G{\"o}m{\"o}ry}, {Gracia}, {Grauf}, {Greco}, {Grivel}, {Guerreiro},
  {Guglielmino}, {Hammerschlag}, {Hanslmeier}, {Hansteen}, {Heinzel},
  {Hern{\'a}ndez-Delgado}, {Hern{\'a}ndez Su{\'a}rez}, {Hidalgo}, {Hill},
  {Hizberger}, {Hofmeister}, {J{\"a}gers}, {Janett}, {Jarolim}, {Jess},
  {Jim{\'e}nez Mej{\'\i}as}, {Jolissaint}, {Kamlah}, {Kapit{\'a}n},
  {Ka{\v{s}}parov{\'a}}, {Keller}, {Kentischer}, {Kiselman}, {Kleint},
  {Klvana}, {Kontogiannis}, {Krishnappa}, {Ku{\v{c}}era}, {Labrosse}, {Lagg},
  {Landi Degl'Innocenti}, {Langlois}, {Lafon}, {Laforgue}, {Le Men}, {Lepori},
  {Lepreti}, {Lindberg}, {Lilje}, {L{\'o}pez Ariste}, {L{\'o}pez
  Fern{\'a}ndez}, {L{\'o}pez Jim{\'e}nez}, {L{\'o}pez L{\'o}pez}, {Manso
  Sainz}, {Marassi}, {Marco de la Rosa}, {Marino}, {Marrero}, {Mart{\'\i}n},
  {Mart{\'\i}n G{\'a}lvez}, {Mart{\'\i}n Hernando}, {Masciadri}, {Mart{\'\i}nez
  Gonz{\'a}lez}, {Matta-G{\'o}mez}, {Mato}, {Mathioudakis}, {Matthews}, {Mein},
  {Merlos Garc{\'\i}a}, {Moity}, {Montilla}, {Molinaro}, {Molodij}, {Montoya},
  {Munari}, {Murabito}, {N{\'u}{\~n}ez Cagigal}, {Oliviero}, {Orozco
  Su{\'a}rez}, {Ortiz}, {Padilla-Hern{\'a}ndez}, {Pa{\'e}z Ma{\~n}{\'a}},
  {Paletou}, {Pancorbo}, {Pastor Ca{\~n}edo}, {Pastor Yabar}, {Peat},
  {Pedichini}, {Peixinho}, {Pe{\~n}ate}, {P{\'e}rez de Taoro}, {Peter},
  {Petrovay}, {Piazzesi}, {Pietropaolo}, {Pleier}, {Poedts}, {P{\"o}tzi},
  {Podladchikova}, {Prieto}, {Quintero Nehrkorn}, {Ramelli}, {Ramos Sapena},
  {Rasilla}, {Reardon}, {Rebolo}, {Regalado Olivares}, {Reyes
  Garc{\'\i}a-Talavera}, {Riethm{\"u}ller}, {Rimmele}, {Rodr{\'\i}guez
 Delgado}, {Rodr{\'\i}guez Gonz{\'a}lez}, {Rodr{\'\i}guez-Losada},
  {Rodr{\'\i}guez Ramos}, {Romano}, {Roth}, {Rouppe van der Voort}, {Rudawy},
  {Ruiz de Galarreta}, {Ryb{\'a}k}, {Salvade}, {S{\'a}nchez-Capuchino},
  {S{\'a}nchez Rodr{\'\i}guez}, {Sangiorgi}, {Say{\`e}de}, {Scharmer},
  {Scheiffelen}, {Schmidt}, {Schmieder}, {Scir{\`e}}, {Scuderi}, {Siegel},
  {Sigwarth}, {Sim{\~o}es}, {Snik}, {Sliepen}, {Sobotka}, {Socas-Navarro},
  {Sola La Serna}, {Solanki}, {Soler Trujillo}, {Soltau}, {Sordini}, {Sosa
  M{\'e}ndez}, {Stangalini}, {Steiner}, {Stenflo}, {{\v{S}}t{\v{e}}p{\'a}n},
  {Strassmeier}, {Sudar}, {Suematsu}, {S{\"u}tterlin}, {Tallon}, {Temmer},
  {Tenegi}, {Tritschler}, {Trujillo Bueno}, {Turchi}, {Utz}, {van Harten}, {van
  Noort}, {van Werkhoven}, {Vansintjan}, {Vaz Cedillo}, {Vega Reyes}, {Verma},
  {Veronig}, {Viavattene}, {Vitas}, {V{\"o}gler}, {von der L{\"u}he},
  {Volkmer}, {Waldmann}, {Walton}, {Wisniewska}, {Zeman}, {Zeuner}, {Zhang},
  {Zuccarello}, \& {Collados}}]{2022A&A...666A..21Q}
{Quintero Noda}, C., {Schlichenmaier}, R., {Bellot Rubio}, L.~R., {et~al.}
  2022, \aap, 666, A21

\bibitem[{{Scharmer}(2017)}]{2017psio.confE..85S}
{Scharmer}, G. 2017, in SOLARNET IV: The Physics of the Sun from the Interior
  to the Outer Atmosphere, 85

\bibitem[{{Scharmer} {et~al.}(2003){Scharmer}, {Bjelksj{\"o}}, {Korhonen},
  {Lindberg}, \& {Petterson}}]{2003SPIE.4853..341S}
{Scharmer}, G.~B., {Bjelksj{\"o}}, K., {Korhonen}, T.~K., {Lindberg}, B., \&
  {Petterson}, B. 2003, in Society of Photo-Optical Instrumentation Engineers
  (SPIE) Conference Series, Vol. 4853, Society of Photo-Optical Instrumentation
  Engineers (SPIE) Conference Series, ed. {S.~L.~Keil \& S.~V.~Avakyan},
  341--350

\bibitem[{{Scharmer} {et~al.}(2008){Scharmer}, {Narayan}, {Hillberg}, {de la
  Cruz Rodr{\'{\i}}guez}, {L{\"o}fdahl}, {Kiselman}, {S{\"u}tterlin}, {van
  Noort}, \& {Lagg}}]{2008ApJ...689L..69S}
{Scharmer}, G.~B., {Narayan}, G., {Hillberg}, T., {et~al.} 2008, \apjl, 689,
  L69
\bibitem[{{Scharmer} {et~al.}(2025){Scharmer},{de la
  Cruz Rodr{\'{\i}}guez}, {Leenaarts}, {Lindberg}, {S{\"u}tterlin}, {Hillberg},
  {Pietraszewski}, {de Wijn}, {Foster}, \& {Storey}}]{In prep.}
{Scharmer}, G.~B., {de la
  Cruz Rodr{\'{\i}}guez}, J., {Leenaarts}, J., {et~al.} 2025, in prep.

\end{thebibliography}

\appendix
\section{EST-B optical design and performance} \label{EST-B_design2}
\subsection{EST-B narrowband design}
\begin{figure}[h]
\center
\includegraphics[angle=0, width=0.99\linewidth,clip]{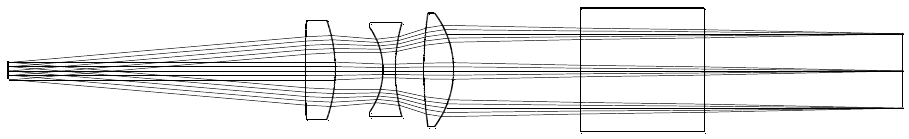}
\includegraphics[angle=0, width=0.99\linewidth,clip]{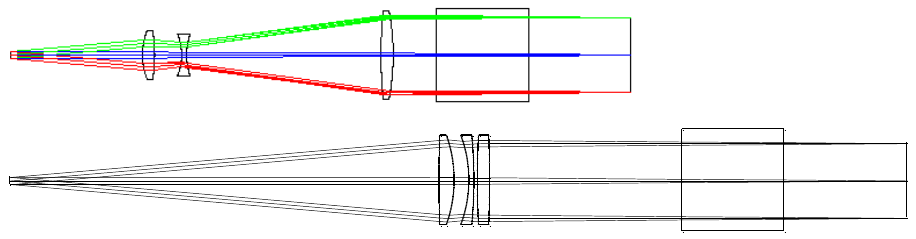}
 \caption{
Layout of three camera lenses designed for EST-B narrowband FPI system. Top: camera lens designed for 5~$\mu$m pixel size, middle: 12~$\mu$m pixel size camera lens optimised for image quality, bottom: 12~$\mu$m pixel size camera lens optimised for reduced image scale changes when refocusing. The 5~$\mu$m and 12~$\mu$m systems are drawn at different scales, but the two 12~$\mu$m systems are drawn at the same scale. For further details about the systems, see Tables \ref{table_EST-B_5my_camera_lens} and \ref{table_EST-B_12my_camera_lens1}.}
\label{fig:EST-B_camera_lenses}
\end{figure}

Table \ref{table_EST-B} shows the optical design of the fixed part of the optical system for EST-B, from L1 to P2, together with an idealised model of the telescope and POP. In Tables \ref{table_EST-B_5my_camera_lens}--\ref{table_EST-B_12my_camera_lens2}, we show the prescriptions of the optical design of three different camera lenses for EST-B, the layouts of which are shown in Fig. \ref{fig:EST-B_camera_lenses}. The first of these designs is of a camera lens for 5~$\mu$m pixel size, the remaining two are for cameras with 12~$\mu$m pixel size, one for which image quality is prioritised (Table \ref{table_EST-B_12my_camera_lens1}), the other prioritising reduced image scale variations when the camera lens is refocused (Table \ref{table_EST-B_12my_camera_lens2}).
\label{sec:EST-B}
\begin{table}[h]
  \centering
  \small
  \setlength{\tabcolsep}{2pt}
  \begin{tabular}{rrrrrcc}
    \hline
    \mathstrut
No. & Radius & Thickness & Lens dia & Beam dia & Glass & Label \\
& (mm) & (mm) & (mm) & (mm) & & \\
    \hline
 --&&&&&& \bf{Telescope} \\
 --&&&&&& \bf{+ POP} \\
0 && infinity&&&& \\
1 && 100 &&&& System stop \\ 
2 && 100 &&&& Ideal lens \\
-- &&&&&& \bf{FPI system} \\
3 && 100 &&&& F1 \\
4 && 80 && 64 & Silica & WBBS \\
5 && 195.33 && 64 \\
6 & 284.96 & 8 & 82 & 70 & PBL1Y & L1 \\
7 & 117.84 & 14 & 82 & 70 & N-FK51A & L1 \\
8 & -340.93 & 345.96 & 82 & 70 \\
9 && 800.82 && 7.9 && P1 \\
10 & 790.98 & 24 & 160 & 147 & N-FK5 ** & L2 \\
11 & -292.40 & 39.89 & 160 & 147 \\
12 & -253.37 & 16 & 150 & 142 & PBM2Y & L2 \\
13 & -463.09 & 407.93 & 160 & 142 \\
14 && 140 & 160 &139 & Silica & FPI1 \\
15 && 200 & 160 &139 \\
16 && 200 && 136 && F2 \\
17 && 140 & 160 & 139 & Silica &FPI2 \\
18 && 407.93 & 160 & 139 \\
19 & 463.09 & 16 & 160 & 143 & PBM2Y & L3 \\
20 & 253.37 & 39.89 & 150 & 143 \\
21 & 292.40 & 24 & 160 & 147 & N-FK5 ** & L3 \\
22 & -790.98 &  800.25 & 160 & 147 \\
23 &&&&&& P2 \\
 \hline
  \end{tabular}
    \vspace{1mm}
  \caption{Prescription of the EST-B FPI "mother" system (without camera lens). F1-F3 are focal planes, L1-L4 are cemented doublet lenses, P1-P2 are pupil planes, and WBBS is the wideband beam splitter. The total length from F1 to F3 is 4400 mm.}
  \label{table_EST-B}
\end{table}

\begin{table}[h]
  \centering
  \small
  \setlength{\tabcolsep}{3pt}
  \begin{tabular}{rrrrrcc}
    \hline
    \mathstrut
No. & Radius & Thickness & Lens dia & Beam dia & Glass & Label \\
& (mm) & (mm) & (mm) & (mm) & & \\
    \hline
23&&120.15&&8&&P2\\
24&850.56&12&40&29&S-BSM16&L4\\
25&-60.34&19.28&40&29&&L4\\
26&-29.73&5&34&28&PBL6Y&L4\\
27&69.08&11.50&38&28&&L4\\
28&158.60&12&46&37&N-FK51&L4\\
29&-38.15&51.22&46&37&&L4\\
30&&50&&35&SILICA&PBS\\
31&&80&&35&&F3\\
 \hline
  \end{tabular}
    \vspace{1mm}
  \caption{Prescription of the camera lens, designed for 5~$\mu$m pixel size for the EST-B narrowband FPI system.  P2 is the second pupil stop, F3 the final focal planes, L4 is an air spaced triplet lens, and PBS is the polarising beam splitter.}
  \label{table_EST-B_5my_camera_lens}
\end{table}

\begin{table}[h]
  \centering
  \small
  \setlength{\tabcolsep}{3pt}
  \begin{tabular}{rrrrrcc}
    \hline
    \mathstrut
No. & Radius & Thickness & Lens dia & Beam dia & Glass & Label \\
& (mm) & (mm) & (mm) & (mm) & & \\
    \hline
23&&128.881&&8&&P2\\
24&62.760&12&48&30&N-PK51&L4\\
25&-141.772&26.368&48&30&&L4\\
26&-55.048&5&40&24&PBL1Y&L4\\
27&63.972&190.494&40&24&&L4\\
28&412.253&12&86&78&N-PK51&L4\\
29&-223.835&42.417&86&78&&L4\\
30&&90&&77&SILICA&PBS\\
31&&100&&77&&F3\\
 \hline
  \end{tabular}
    \vspace{1mm}
  \caption{Prescription of the camera lens, designed for 12~$\mu$m pixel size, optimised for image quality, for the EST-B narrowband FPI system.  P2 is the second pupil stop, F3 the final focal planes, L4 is an air spaced triplet lens,, and PBS is the polarising beam splitter.}
  \label{table_EST-B_12my_camera_lens1}
\end{table}

\begin{table}[h]
  \centering
  \small
  \setlength{\tabcolsep}{3pt}
  \begin{tabular}{rrrrrcc}
    \hline
    \mathstrut
No. & Radius & Thickness & Lens dia & Beam dia & Glass & Label \\
& (mm) & (mm) & (mm) & (mm) & & \\
    \hline
23&&420.74&&&&P2\\
24&1248.34&15&88&&N-FK51A&L4\\
25&-130.41&14.48&88&&&L4\\
26&-124.35&5&88&&PBL1Y&L4\\
27&-419.51&3.15&88&&&L4\\
28&736.58&12&88&&N-SK5HTI&L4\\
29&-1328.11&188.27&88&&&L4\\
30&&100&&&SILICA&PDBS\\
31&&120&&&&F3\\
 \hline
  \end{tabular}
    \vspace{1mm}
  \caption{Prescription of the camera lens, designed for 12~$\mu$m pixel size, optimised to reduce image scale variations, for the EST-B narrowband FPI system.  P2 is the second pupil stop, F3 the final focal planes, L4 is an air spaced triplet lens,, and PDBS is the phase diversity beam splitter.}
  \label{table_EST-B_12my_camera_lens2}
\end{table}

\clearpage
\subsection{EST-B wideband design}
In this Appendix, we show the designs of the three wideband camera lenses designed for EST-B. The first lens of the wideband systems is identical to L1 in the narrowband system (the "mother" system) of Table \ref{table_EST-B}. The first thickness given in Tables \ref{table_EST-B_5my_camera_lens_wideband}--\ref{table_EST-B_12my_camera_lens_wideband_imscale} refers to the distance from the last surface of L1 to the first surface of L4.

\begin{figure}[h]
\center
\includegraphics[angle=0, width=0.99\linewidth,clip]{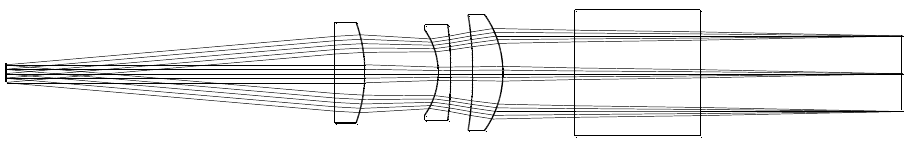}
\includegraphics[angle=0, width=0.75\linewidth,clip]{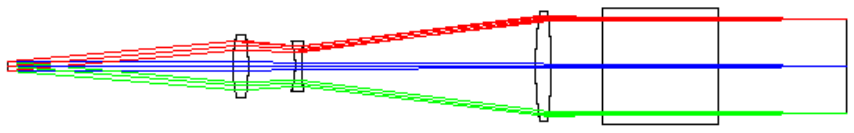}
\includegraphics[angle=0, width=0.99\linewidth,clip]{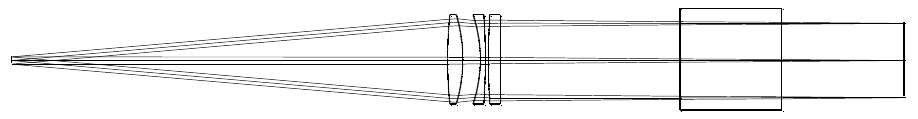}
 \caption{
Layout of three camera lenses designed for EST-B wideband FPI system. Top: camera lens designed for 5~$\mu$m pixel size, middle: 12~$\mu$m pixel size camera lens optimised for image quality, bottom: 12~$\mu$m pixel size camera lens optimised for reduced image scale changes when refocusing. The 5~$\mu$m and 12~$\mu$m systems are drawn at different scales, but the two 12~$\mu$m systems are drawn at the same scale. For further details about the systems, see Tables \ref{table_EST-B_5my_camera_lens_wideband}--\ref{table_EST-B_12my_camera_lens_wideband_imscale}.}
\label{fig:EST-B_wideband_camera_lenses}
\end{figure}

\begin{table}[h]
  \centering
  \small
  \setlength{\tabcolsep}{3pt}
  \begin{tabular}{rrrrrcc}
    \hline
    \mathstrut
No. & Radius & Thickness & Lens dia & Beam dia & Glass & Label \\
& (mm) & (mm) & (mm) & (mm) & & \\
    \hline
23&&130.65&&8&&P2\\
24&-6428.81&12&40&31&N-PK51&L4\\
25&-59.57&29.19&40&31&&L4\\
26&-29.35&5&34&30&PBL1Y&L4\\
27&-147.75&8.73&38&30&&L4\\
28&-143.21&12&46&36&N-PK51&L4\\
29&-39.89&28.44&46&36&&L4\\
30&&50&&35&SILICA&PBS\\
31&&80&&35&&F3\\
 \hline
  \end{tabular}
    \vspace{1mm}
  \caption{Prescription of the camera lens, designed for 5~$\mu$m pixel size for the EST-B wideband FPI system.  P2 is the second pupil stop, F3 the final focal planes, L4 is an air spaced triplet lens, and PBS is the polarising beam splitter.}
  \label{table_EST-B_5my_camera_lens_wideband}
\end{table}

\begin{table}[h]
  \centering
  \small
  \setlength{\tabcolsep}{3pt}
  \begin{tabular}{rrrrrcc}
    \hline
    \mathstrut
No. & Radius & Thickness & Lens dia & Beam dia & Glass & Label \\
& (mm) & (mm) & (mm) & (mm) & & \\
    \hline
  23&&175.00&&8&&P2\\
24&100.46&12&48&38&N-PK51&L4\\
25&-146.22&36.51&48&38&&L4\\
26&-70.66&5&40&32&PBL1Y&L4\\
27&114.32&182.27&40&32&&L4\\
28&273.19&12&86&77&N-PK51&L4\\
29&-392.05&40&86&77&&L4\\
30&&90&&77&SILICA&PDBS\\
31&&100&&77&&F3\\
 \hline
  \end{tabular}
    \vspace{1mm}
  \caption{Prescription of the camera lens, designed for 12~$\mu$m pixel size for the EST-B wideband FPI system, optimised  for best image quality. P2 is the second pupil stop, F3 the final focal planes, L4 is an air spaced triplet lens, and PDBS is the phase diversity beam splitter.}
  \label{table_EST-B_12my_camera_lens_wideband_imquality}
\end{table}

\begin{table}[h]
  \centering
  \small
  \setlength{\tabcolsep}{3pt}
  \begin{tabular}{rrrrrcc}
    \hline
    \mathstrut
No. & Radius & Thickness & Lens dia & Beam dia & Glass & Label \\
& (mm) & (mm) & (mm) & (mm) & & \\
    \hline
23&&428.89&&&&P2\\
24&418.65&15&88&&N-PK51A&L4\\
25&-152.38&17.17&88&&&L4\\
26&-138.66&5&88&&LLF1HTI&L4\\
27&-630.37&3.00&88&&&L4\\
28&686.81&12&88&&N-SK5HTI&L4\\
29&-6647.06&176.60&88&&&L4\\
30&&100&&&SILICA&PDBS\\
31&&120&&&&F3\\
 \hline
  \end{tabular}
    \vspace{1mm}
  \caption{Prescription of the camera lens, designed for 12~$\mu$m pixel size for the EST-B wideband FPI system, optimised to minimise image scale variations from refocusing.  P2 is the second pupil stop, F3 the final focal planes, L4 is an air spaced triplet lens, and PDBS is the phase diversity beam splitter.}
  \label{table_EST-B_12my_camera_lens_wideband_imscale}
\end{table}

\clearpage
\subsection{EST-B narrowband performance}

Below, we a large number of spot diagrams and Strehl tables to demonstrate the performance of the narrowband and wideband systems of EST-B in two different configurations for each pixel size. However, to focus attention on narrowband performance, which is the more interesting aspect of the system, we have moved the spot diagrams and Strehl tables of the wideband system to a separate section, while referring to these figures and data from the present section. Note, that the 12~$\mu$m system has been optimised in two variants: one with respect to image quality and the other with an attempt to reduce image scale variations associated with refocusing. We present data for both variants.

\begin{figure}[h]
\center
\includegraphics[angle=0, width=0.99\linewidth,clip]{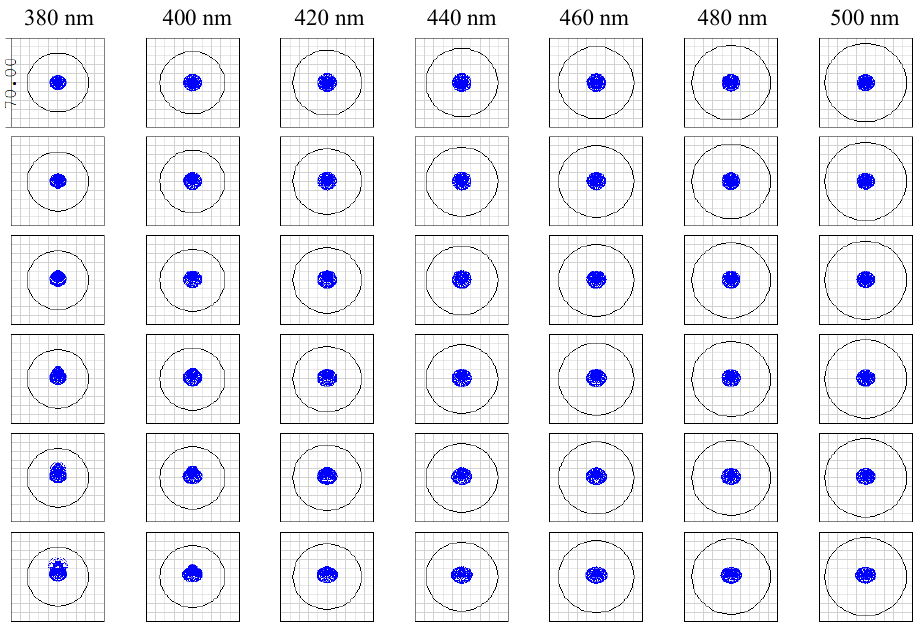}
 \caption{Calculated quantities, spot diagrams and Strehl values for POP alone, within the wavelength range of
EST-B (380-500 nm).}
\label{fig:EST-B_spot_POP_alone_A2.1}
\end{figure}

Figure \ref{fig:EST-B_spot_POP_alone_A2.1} and Table \ref{table_EST-B_12my_camera_lens2} show spot diagrams and Strehl values for POP alone, at several wavelengths in the range 380-500 nm. POP’s focus curve within that wavelength range is 6.4 mm and the image quality is
perfect (Strehl 0.98 or higher) after refocusing, but the image scale varies by about 1\% throughout this
wavelength range. This variation is a consequence of the field lens (a singlet lens) in the interior of
POP but since all observations made with EST will be narrowband, and since science data sets anyway
will normally consist of images recorded at widely different image scales, this should not constitute a
problem.

\begin{table}[h]
  \centering
  \small
  \begin{tabular}{cccccccc}
    \hline
    \mathstrut
 $\lambda~(nm)$&380&400&420&440&460&480&500 \\
FOV dist. \\
\hline
0.0"&0.99&0.99&0.99&0.99&0.99&0.99&0.99\\
6.0"&0.99&0.99&0.99&0.99&0.99&0.99&0.99\\
11.8"&0.99&0.99&0.99&0.99&0.99&0.99&0.99\\
18.0"&0.99&0.98&0.98&0.98&0.99&0.99&0.99\\
23.6"&0.98&0.98&0.98&0.98&0.99&0.99&0.99\\
30.0"&0.98&0.98&0.98&0.98&0.98&0.98&0.99\\  
 \hline
  \end{tabular}
    \vspace{1mm}
  \caption{Strehl values corresponding to POP alone, within the wavelength range of
EST-B (380-500 nm). Spot diagrams in Fig. \ref{fig:EST-B_spot_POP_alone_A2.1}.}
  \label{table_EST-B_12my_camera_lens2}
\end{table}

\begin{figure}[h]
\center
\includegraphics[angle=0, width=0.99\linewidth,clip]{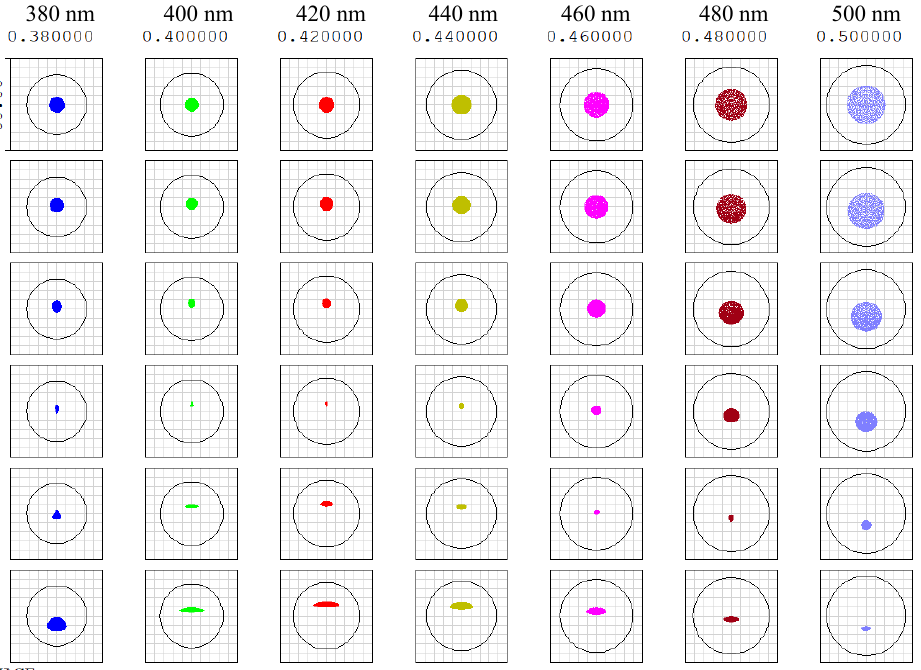}
 \caption{Calculated quantities, spot diagrams and Strehl values “mother system” alone, with an imaginary
perfect camera lens replacing an actual camera lens, within the wavelength range of EST-B (380-500 nm), at a
fixed focal plane. The reductions in Strehl near the center of the FOV at 460-500 nm is mostly from focus errors.}
\label{fig:EST-B_spot_mother_alone_A2.2}
\end{figure}

Figure \ref{fig:EST-B_spot_mother_alone_A2.2} and Table \ref{table_EST-B_spot_mother_alone_A2.2} show spot diagrams and Strehl values of the “mother” system, which contains all optics except the camera lens. The spot diagrams and Strehl values are calculated by replacing an actual
camera lens with an imaginary perfect thin lens that has a refractive index that is independent of
wavelength. This demonstrates the baseline performance of the system, which is limited by the design
of the three lenses preceding the camera lens. The main limitation of the mother system is its shallow
focus curve, which somewhat reduces the Strehl at the center of the FOV at 460-500 nm wavelengths.
With proper refocusing, the Strehl would be 97\% or higher at all wavelengths and field angles.

\begin{table}[h]
  \centering
  \small
  \begin{tabular}{cccccccc}
    \hline
    \mathstrut
 $\lambda~(nm)$&380&400&420&440&460&480&500 \\
FOV dist. \\
\hline
0.0"&0.98&0.99&0.98&0.98&0.96&0.95&0.93\\
6.0"&0.98&0.99&0.99&0.98&0.97&0.95&0.94\\
12.0"&0.99&1.00&0.99&0.99&0.98&0.97&0.96\\
18.0"&1.00&1.00&1.00&1.00&0.99&0.99&0.98\\
24.0"&0.99&0.99&0.99&1.00&1.00&1.00&1.00\\
30.0"&0.96&0.97&0.97&0.98&0.99&0.99&1.00\\
 \hline
  \end{tabular}
    \vspace{1mm}
  \caption{Strehl values corresponding to the "mother system" alone, with an imaginary
perfect camera lens replacing an actual camera lens, within the wavelength range of EST-B (380-500 nm), at a
fixed focal plane. The reductions in Strehl near the center of the FOV at 460-500 nm is mostly from focus errors. Spot diagrams in Fig. \ref{fig:EST-B_spot_mother_alone_A2.2}.}
  \label{table_EST-B_spot_mother_alone_A2.2}
\end{table}

\begin{figure}[h]
\center
\includegraphics[angle=0, width=0.99\linewidth,clip]{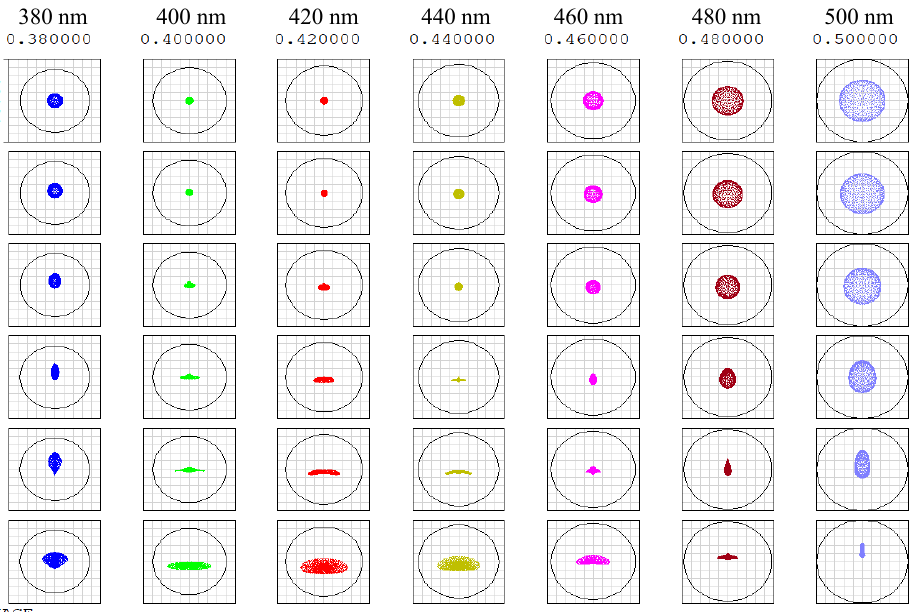}
 \caption{Calculated quantities, spot diagrams and Strehl values for EST-B Narrowband as a stand-alone
instrument (without POP) using 5~$\mu$m pixels. The optimization is made by imposing a common focus for all
wavelengths. The spot diagrams and Strehl values shown are obtained from that fixed focal plane. The image
quality improves by individual focusing for each wavelength.}
\label{fig:EST-B_spot_standalone_5mu_pix_A2.3}
\end{figure}

\begin{figure}[h]
\center
\includegraphics[angle=0, width=0.99\linewidth,clip]{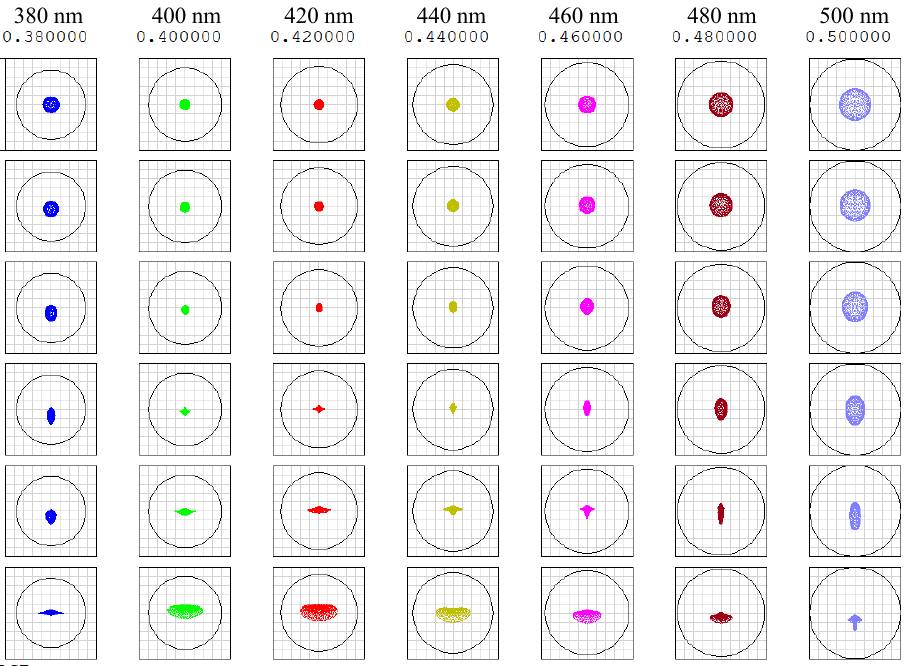}
 \caption{Calculated quantities, spot diagrams and Strehl values for EST-B Wideband as a stand-alone
instrument (without POP) using 5~$\mu$m pixels pixels. The optimization is made by imposing a common focus for all wavelengths. The spot diagrams and Strehl values shown are obtained from that fixed focal plane. The image quality improves by individual focusing for each wavelength.}
\label{fig:EST-B_spot_wideband_standalone_5mu_pix_A2.4}
\end{figure}

Figure  \ref{fig:EST-B_spot_standalone_5mu_pix_A2.3} and Table \ref{table_EST-B_spot_standalone_5mu_pix_A2.3} show spot diagrams and Strehl values for the EST-B narrowband system with 5~$\mu$m pixel size as a stand-alone instrument (without POP) and at a fixed focus. The lowest Strehl value is 0.89
and is limited by a small focus error. With proper refocusing, the Strehl would be 95\% or higher at all
wavelengths. The corresponding stand-alone wideband system (Fig. \ref{fig:EST-B_spot_wideband_standalone_5mu_pix_A2.4} and Table \ref{table_EST-B_spot_wideband_standalone_5mu_pix_A2.4}) performs even better.
Without any refocusing, the minimum Strehl is 0.94 and would be 97\% with proper focusing.

\begin{table}[h]
  \centering
  \small
  \begin{tabular}{cccccccc}
    \hline
    \mathstrut
 $\lambda~(nm)$&380&400&420&440&460&480&500 \\
FOV dist. \\
\hline
0.0"&0.97&0.99&1.00&0.99&0.97&0.93&0.89\\
6.0"&0.97&1.00&1.00&0.99&0.97&0.94&0.89\\
12.0"&0.98&1.00&1.00&1.00&0.98&0.95&0.92\\
18.0"&0.97&1.00&1.00&1.00&0.99&0.97&0.94\\
24.0"&0.96&0.98&0.98&0.99&1.00&0.99&0.97\\
30.0"&0.96&0.96&0.95&0.97&0.98&1.00&0.99\\
 \hline
  \end{tabular}
    \vspace{1mm}
  \caption{ Strehl values for EST-B Narrowband as a stand-alone
instrument (without POP) using 5~$\mu$m pixels. The optimization is made by imposing a common focus for all
wavelengths. The spot diagrams and Strehl values shown are obtained from that fixed focal plane. The image
quality improves by individual focusing for each wavelength. Spot diagrams in Fig. \ref{fig:EST-B_spot_standalone_5mu_pix_A2.3}.}
  \label{table_EST-B_spot_standalone_5mu_pix_A2.3}
\end{table}

\begin{figure}[h]
\center
\includegraphics[angle=0, width=0.99\linewidth,clip]{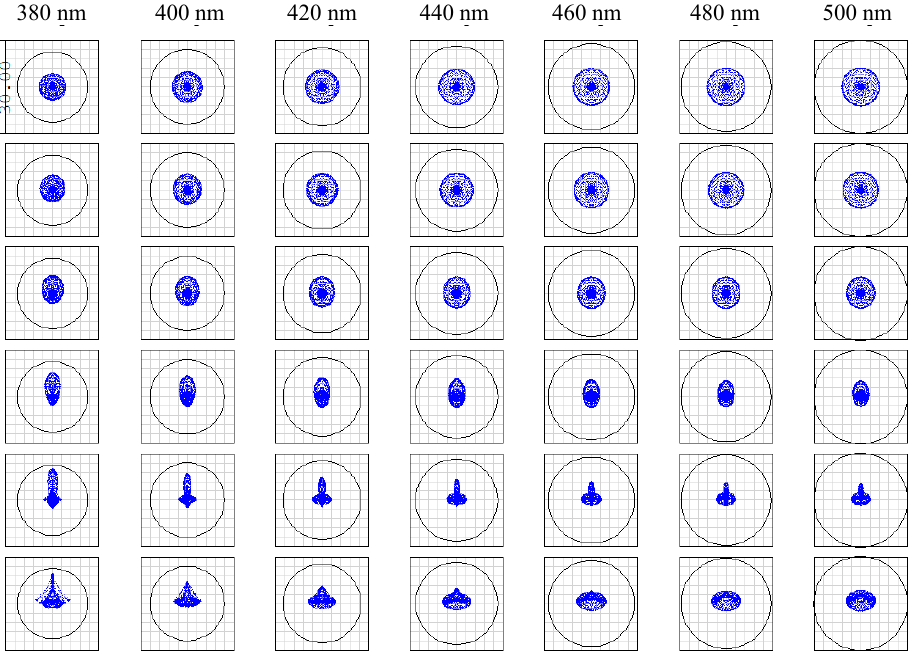}
 \caption{Calculated quantities, spot diagrams and Strehl values for EST-B Narrowband using 5~$\mu$m pixels,
when connected to POP with its 6.3 mm focus curve range. The spot diagrams and Strehl values are obtained by
individual focusing of the camera lens for each wavelength, which requires a total movement of the camera lens
by 1.5 mm.}
\label{fig:EST-B_spot_narrowband_FPI+POP_5mu_pix_A2.5}
\end{figure}

\begin{figure}[h]
\center
\includegraphics[angle=0, width=0.99\linewidth,clip]{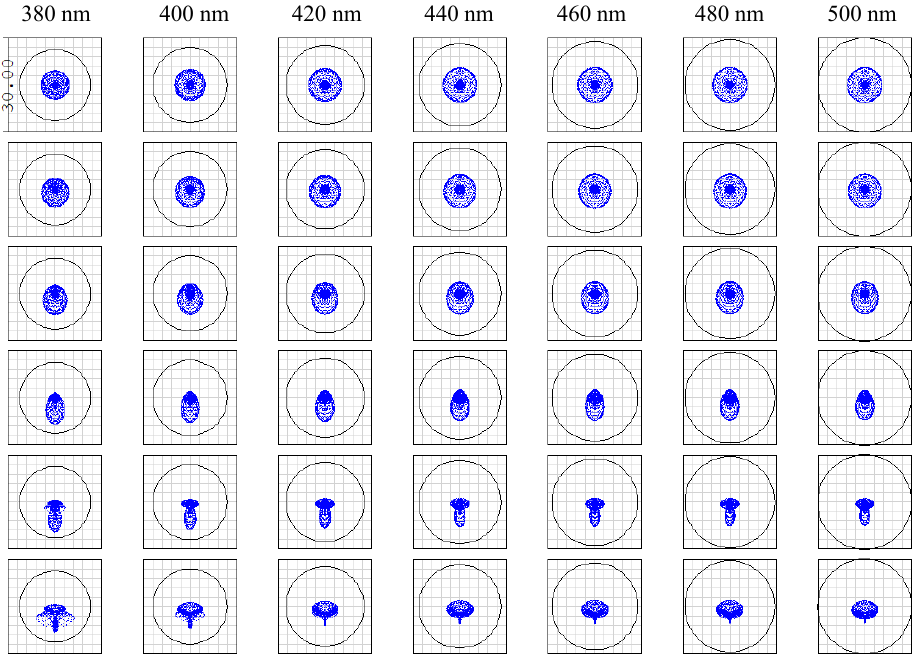}
 \caption{Calculated quantities, spot diagrams and Strehl values for EST-B Wideband using 5~$\mu$m pixels, when connected to POP with its 6.3 mm focus curve range. The spot diagrams and Strehl values are obtained by
individual focusing of the camera lens for each wavelength, which requires a total movement of the camera lens
by 1.5 mm.}
\label{fig:EST-B_spot_wideband_FPI+POP_5mu_pix_A2.6}
\end{figure}

Figure \ref{fig:EST-B_spot_narrowband_FPI+POP_5mu_pix_A2.5} and Table \ref{table_EST-B_spot_narrowband_FPI+POP_5mu_pix_A2.5} demonstrate the performance of EST-B narrowband with 5~$\mu$m pixel size when used with
POP to compensate its focus curve. Except for the outermost field point of the 1 arc min field-of-view
at 380 nm, the lowest Strehl is 0.94 or higher at all other field points and wavelengths. The
corresponding wideband system (Fig.  \ref{fig:EST-B_spot_wideband_FPI+POP_5mu_pix_A2.6} and Table \ref{table_EST-B_spot_wideband_FPI+POP_5mu_pix_A2.6}) delivers a minimum Strehl of 95\%. Considering that the 5~$\mu$m  pixel size also delivers very small image scale variations with refocusing, we consider the 5~$\mu$m
pixel size designs to need no further improvement.

\begin{table}[h]
  \centering
  \small
  \begin{tabular}{cccccccc}
    \hline
    \mathstrut
 $\lambda~(nm)$&380&400&420&440&460&480&500 \\
FOV dist. \\
\hline
0.0"&0.97&0.97&0.97&0.97&0.97&0.97&0.98\\
6.0"&0.97&0.98&0.98&0.98&0.98&0.98&0.98\\
11.8"&0.98&0.98&0.98&0.99&0.99&0.99&0.99\\
18.0"&0.97&0.98&0.99&0.99&0.99&0.99&0.99\\
23.6"&0.94&0.96&0.98&0.98&0.99&0.99&0.99\\
30.0"&0.91&0.94&0.95&0.96&0.96&0.96&0.96\\
 \hline
  \end{tabular}
    \vspace{1mm}
  \caption{Strehl values for EST-B Narrowband using 5~$\mu$m pixels,
when connected to POP with its 6.3 mm focus curve range. The spot diagrams and Strehl values are obtained by
individual focusing of the camera lens for each wavelength, which requires a total movement of the camera lens
by 1.5 mm. Spot diagrams in Fig. \ref{fig:EST-B_spot_narrowband_FPI+POP_5mu_pix_A2.5}.}
  \label{table_EST-B_spot_narrowband_FPI+POP_5mu_pix_A2.5}
\end{table}

\begin{figure}[h]
\center
\includegraphics[angle=0, width=0.99\linewidth,clip]{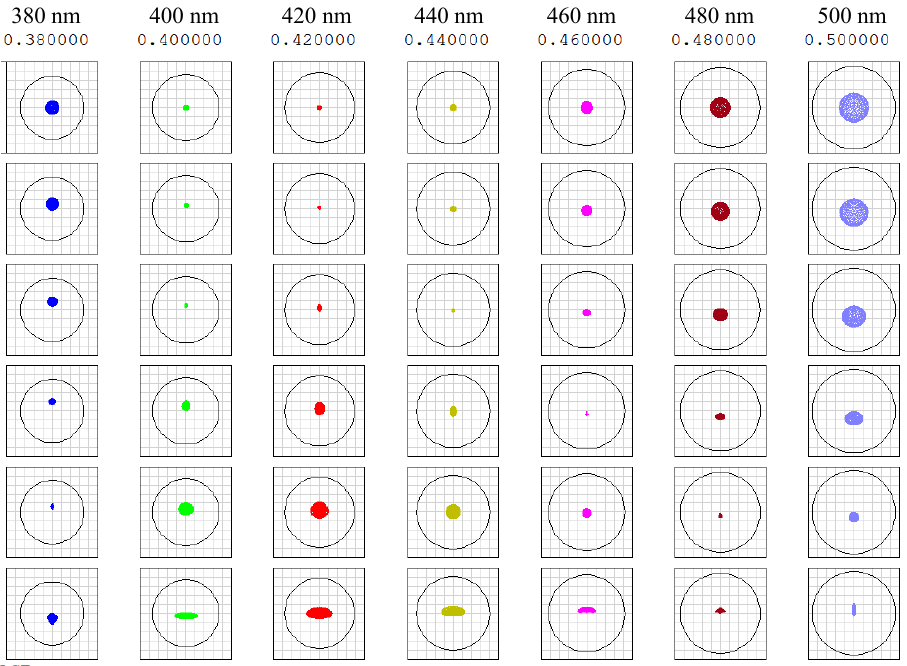}
 \caption{Calculated quantities, spot diagrams and Strehl values for EST-B Narrowband as a stand-alone
instrument (without POP) using 12~$\mu$m pixels. The optimization is for the best possible image quality and by
imposing a common focus for all wavelengths. The spot diagrams and Strehl values shown are obtained from
that fixed focal plane. The image quality improves by individual focusing for each wavelength.}
\label{fig:EST-B_spot_narrowband_standalone_12mu_pix_A2.7}
\end{figure}

\begin{figure}[h]
\center
\includegraphics[angle=0, width=0.99\linewidth,clip]{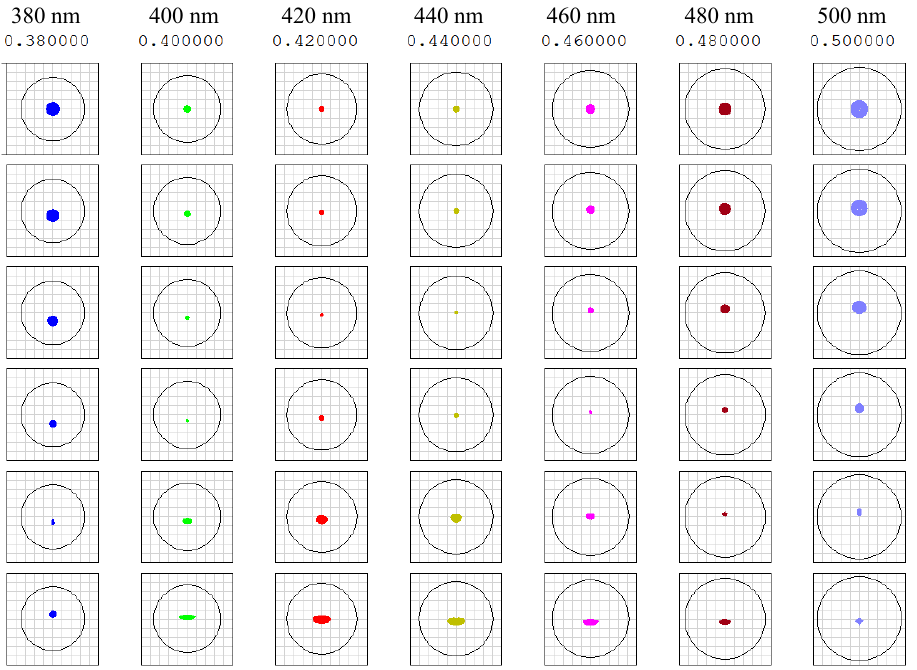}
 \caption{Calculated quantities, spot diagrams and Strehl values for EST-B Wideband as a stand-alone
instrument (without POP) using 12~$\mu$m pixels. The optimization is for the best possible image quality and by
imposing a common focus for all wavelengths. The spot diagrams and Strehl values shown are obtained from
that fixed focal plane. The image quality improves by individual focusing for each wavelength.}
\label{fig:EST-B_spot_wideband_standalone_12mu_pix_A2.8}
\end{figure}

Figure  \ref{fig:EST-B_spot_narrowband_standalone_12mu_pix_A2.7} and Table \ref{table_EST-B_spot_narrowband_standalone_12mu_pix_A2.7} show spot diagrams and Strehl values for the EST-B narrowband system with 12~$\mu$m
pixel size, when optimized for image quality and used as a stand-alone instrument (without POP) and
at a fixed focus. Even without refocusing, the lowest Strehl is 95\% and the spot diagram at 500 nm
shows that the dominant degradation is from a small focus error. The wideband system (Fig.  \ref{fig:EST-B_spot_wideband_standalone_12mu_pix_A2.8} and Table \ref{table_EST-B_spot_wideband_standalone_12mu_pix_A2.8})
performs even better – for that the lowest Strehl is 0.98 without refocusing.

\begin{table}[h]
  \centering
  \small
  \begin{tabular}{cccccccc}
    \hline
    \mathstrut
 $\lambda~(nm)$&380&400&420&440&460&480&500 \\
FOV dist. \\
\hline
0.0"&0.98&1.00&1.00&1.00&0.99&0.97&0.95\\
6.0"&0.98&1.00&1.00&1.00&0.99&0.98&0.96\\
12.0"&0.99&1.00&1.00&1.00&1.00&0.98&0.97\\
18.0"&0.99&1.00&1.00&1.00&1.00&0.99&0.98\\
24.0"&1.00&0.99&0.99&0.99&1.00&1.00&0.99\\
30.0"&0.98&0.99&0.98&0.99&0.99&1.00&0.99\\
 \hline
  \end{tabular}
    \vspace{1mm}
  \caption{Strehl values for EST-B Narrowband as a stand-alone
instrument (without POP) using 12~$\mu$m pixels. The optimization is for the best possible image quality and by
imposing a common focus for all wavelengths. The spot diagrams and Strehl values shown are obtained from
that fixed focal plane. The image quality improves by individual focusing for each wavelength. Spot diagrams in Fig. \ref{fig:EST-B_spot_narrowband_standalone_12mu_pix_A2.7}.}
  \label{table_EST-B_spot_narrowband_standalone_12mu_pix_A2.7}
\end{table}

\begin{figure}[h]
\center
\includegraphics[angle=0, width=0.99\linewidth,clip]{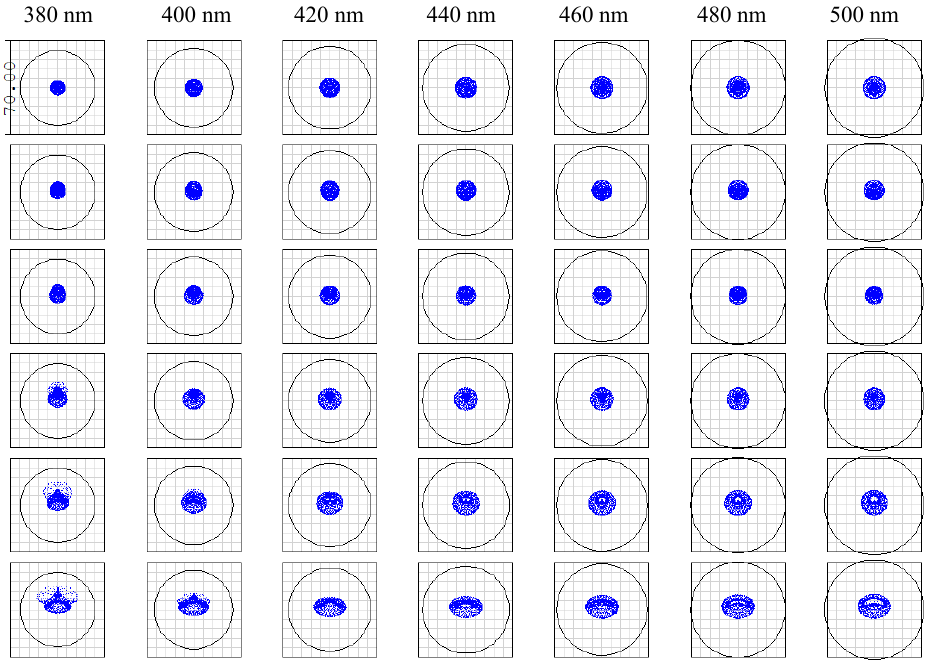}
 \caption{Calculated quantities, spot diagrams and Strehl values for EST-B Narrowband connected to POP,
optimized for image quality with 12~$\mu$m pixels, POPs 6.3 mm focus curve range, and optimized for image
quality. The spot diagrams and Strehl values are obtained by individual focusing of the camera lens for each
wavelength, which requires a total movement of the camera lens by 9.3 mm. The scale difference between the
narrow and broad system is 6 microns (0.016\%) at the worst point (the edge of the field for 380 nm), at all other
points it 0-2 microns (<0.005\%)}
\label{fig:EST-B_spot_narrowband_FPI+POP_imqual_12mu_pix_A2.9}
\end{figure}

\begin{figure}[h]
\center
\includegraphics[angle=0, width=0.99\linewidth,clip]{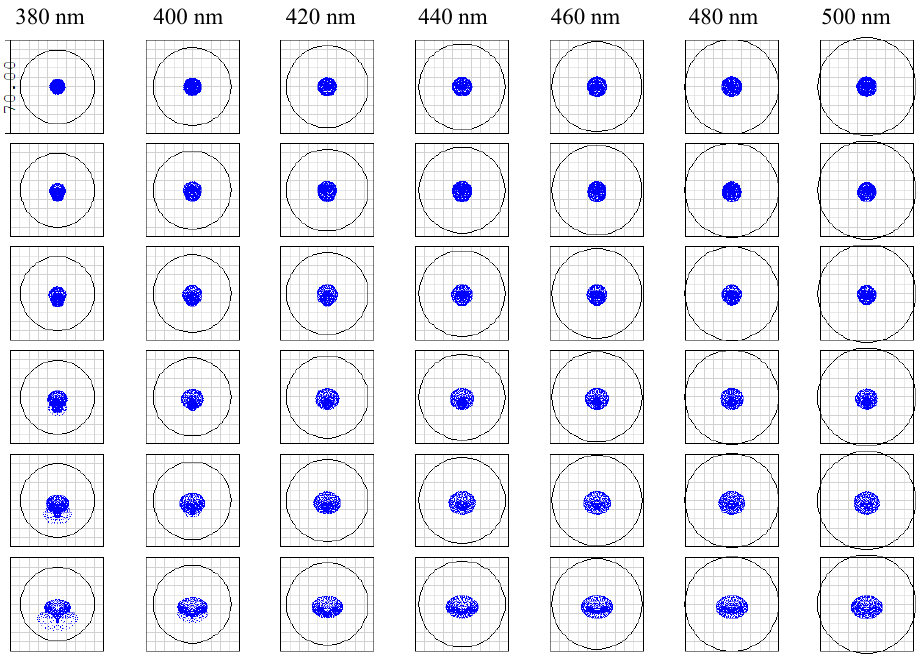}
 \caption{Calculated quantities, spot diagrams and Strehl values for EST-B Wideband+POP using 12~$\mu$m
pixels, with POPs 6.3 mm focus curve range. The design is optimized for best image quality. The spot diagrams
and Strehl values are obtained by individual focusing of the camera lens for each wavelength, which requires a
total movement of the camera lens by 9.3 mm.}
\label{fig:EST-B_spot_wideband_FPI+POP_imqual_12mu_pix_A2.10}
\end{figure}

Figure  \ref{fig:EST-B_spot_narrowband_FPI+POP_imqual_12mu_pix_A2.9} and Table \ref{table_EST-B_spot_narrowband_FPI+POP_imqual_12mu_pix_A2.9} demonstrate high performance of the above 12~$\mu$m system when connected to POP and
used to compensate for its focus curve. This requires a movement of the camera lens by 9.3 mm. The
image quality for this system is excellent – the lowest Strehl is 95\%. The performance of the wideband
system is equally good (Fig. \ref{fig:EST-B_spot_wideband_FPI+POP_imqual_12mu_pix_A2.10} and Table \ref{table_EST-B_spot_wideband_FPI+POP_imqual_12mu_pix_A2.10}).

\begin{table}[h]
  \centering
  \small
  \begin{tabular}{cccccccc}
    \hline
    \mathstrut
 $\lambda~(nm)$&380&400&420&440&460&480&500 \\
FOV dist. \\
\hline
0.0"&0.99&0.99&0.99&0.99&0.99&0.99&1.00\\
6.0"&0.99&0.99&0.99&0.99&0.99&0.99&1.00\\
11.8"&0.99&0.99&0.99&0.99&0.99&0.99&0.99\\
18.0"&0.99&0.98&0.98&0.98&0.98&0.99&0.99\\
23.6"&0.98&0.97&0.97&0.97&0.97&0.97&0.97\\
30.0"&0.95&0.95&0.95&0.95&0.95&0.96&0.96\\
 \hline
  \end{tabular}
    \vspace{1mm}
  \caption{Strehl values for EST-B Narrowband connected to POP,
optimized for image quality with 12~$\mu$m pixels, POPs 6.3 mm focus curve range, and optimized for image
quality. The spot diagrams and Strehl values are obtained by individual focusing of the camera lens for each
wavelength, which requires a total movement of the camera lens by 9.3 mm. The scale difference between the
narrow and broad system is 6 microns (0.016\%) at the worst point (the edge of the field for 380 nm), at all other
points it 0-2 microns (<0.005\%). Spot diagrams in Fig. \ref{fig:EST-B_spot_narrowband_FPI+POP_imqual_12mu_pix_A2.9}.}
  \label{table_EST-B_spot_narrowband_FPI+POP_imqual_12mu_pix_A2.9}
\end{table}

\begin{figure}[h]
\center
\includegraphics[angle=0, width=0.99\linewidth,clip]{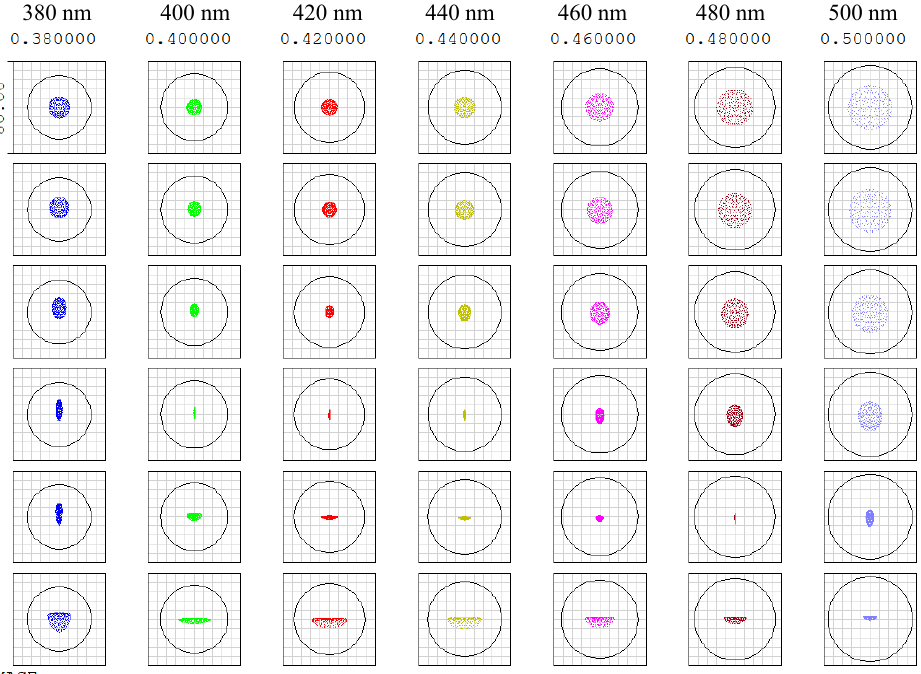}
 \caption{Calculated quantities, spot diagrams and Strehl values for EST-B Narrowband as a stand-alone
instrument (without POP) using 12~$\mu$m pixels. The optimization is made by minimizing image scale variations
from refocusing and by imposing a common focus for all wavelengths. The spot diagrams and Strehl values
shown are obtained from that fixed focal plane. The image quality improves by individual focusing for each
wavelength.}
\label{fig:EST-B_spot_narrowband_standalone_imscale_12mu_pix_A3.1}
\end{figure}

\begin{figure}[h]
\center
\includegraphics[angle=0, width=0.99\linewidth,clip]{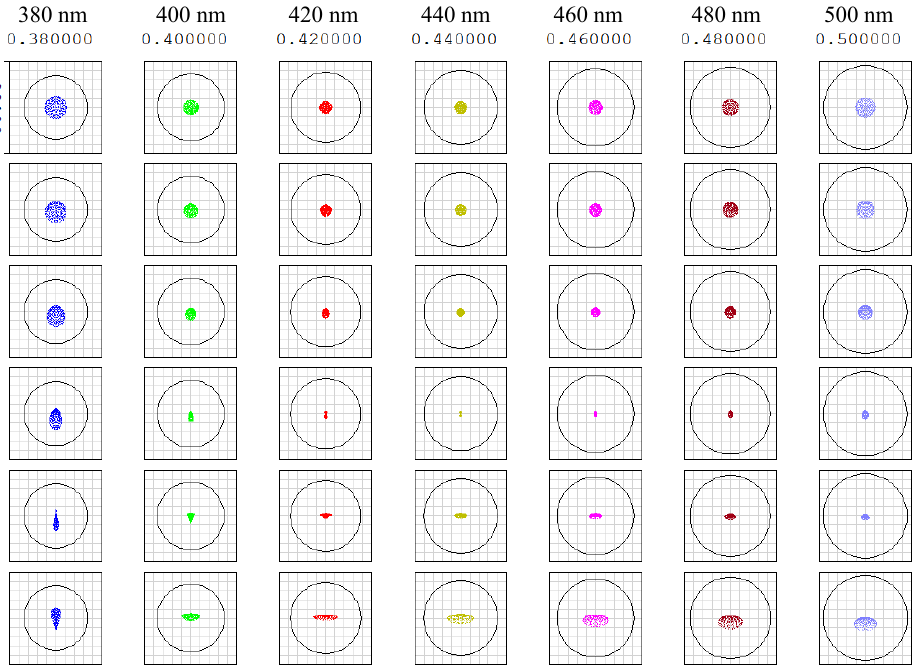}
 \caption{Calculated quantities, spot diagrams and Strehl values for EST-B Wideband as a stand-alone
instrument (without POP) using 12~$\mu$m pixels. The optimization is made by minimizing image scale variations
from refocusing and by imposing a common focus for all wavelengths. The spot diagrams and Strehl values
shown are obtained from that fixed focal plane. The image quality improves somewhat at 380 nm by individually
focusing at that wavelength.}
\label{fig:EST-B_spot_wideband_standalone_imscale_12mu_pix_A3.2}
\end{figure}

Figure  \ref{fig:EST-B_spot_narrowband_standalone_imscale_12mu_pix_A3.1} and Table \ref{table_EST-B_spot_narrowband_standalone_imscale_12mu_pix_A3.1} shows spot diagrams and Strehl values for the EST-B narrowband system used as a stand-
alone instrument (without POP) with 12~$\mu$m pixel size, when the design is optimized to reduce image
scale variations from refocusing. The spot diagrams and Strehl tables are obtained at a fixed focal
plane. For this design, performance is acceptable with a minimum Strehl of 0.91 even without any
refocusing. With refocusing to compensate for its intrinsic (shallow) focus curve, the Strehl would be
95\% or higher. The wideband system (Figure  \ref{fig:EST-B_spot_wideband_standalone_imscale_12mu_pix_A3.2} and Table \ref{table_EST-B_spot_wideband_standalone_imscale_12mu_pix_A3.2}) has an even flatter focus curve and a minimum Strehl of 96\% without refocusing.

\begin{table}[h]
  \centering
  \small
  \begin{tabular}{cccccccc}
    \hline
    \mathstrut
 $\lambda~(nm)$&380&400&420&440&460&480&500 \\
FOV dist. \\
\hline
0.0"&0.96&0.98&0.98&0.97&0.96&0.94&0.91\\
6.0"&0.97&0.98&0.98&0.98&0.96&0.94&0.92\\
12.0"&0.97&0.99&0.99&0.99&0.98&0.96&0.94\\
18.0"&0.97&0.99&0.99&0.99&0.99&0.98&0.97\\
24.0"&0.96&0.98&0.99&0.99&1.00&0.99&0.99\\
30.0"&0.94&0.95&0.95&0.96&0.97&0.99&0.99\\
 \hline
  \end{tabular}
    \vspace{1mm}
  \caption{Strehl values for EST-B Narrowband as a stand-alone instrument (without POP) using 12~$\mu$m pixels. The optimization is made by minimizing image scale variations from refocusing and by imposing a common focus for all wavelengths. The spot diagrams and Strehl values shown are obtained from that fixed focal plane. The image quality improves by individual focusing for each wavelength. Spot diagrams in Fig. \ref{fig:EST-B_spot_narrowband_standalone_imscale_12mu_pix_A3.1}.}
  \label{table_EST-B_spot_narrowband_standalone_imscale_12mu_pix_A3.1}
\end{table}

\begin{figure}[h]
\center
\includegraphics[angle=0, width=0.99\linewidth,clip]{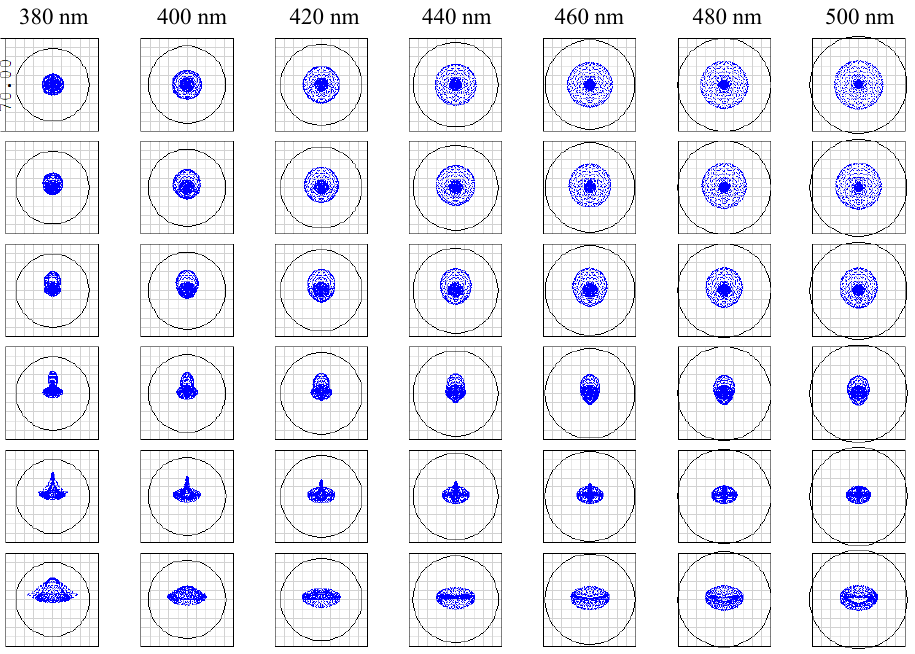}
 \caption{Calculated quantities, spot diagrams and Strehl values for EST-B Narrowband connected to POP
with its 6.3 mm focus curve range, and optimized for minimum image scale variations when refocusing with 12~$\mu$m pixels. The spot diagrams and Strehl values are obtained by individual focusing of the camera lens for each
wavelength, which requires a total movement of the camera lens by 9.3 mm.}
\label{fig:EST-B_spot_narrowband_FPI+POP_imscale_12mu_pix_A3.3}
\end{figure}

\begin{figure}[h]
\center
\includegraphics[angle=0, width=0.99\linewidth,clip]{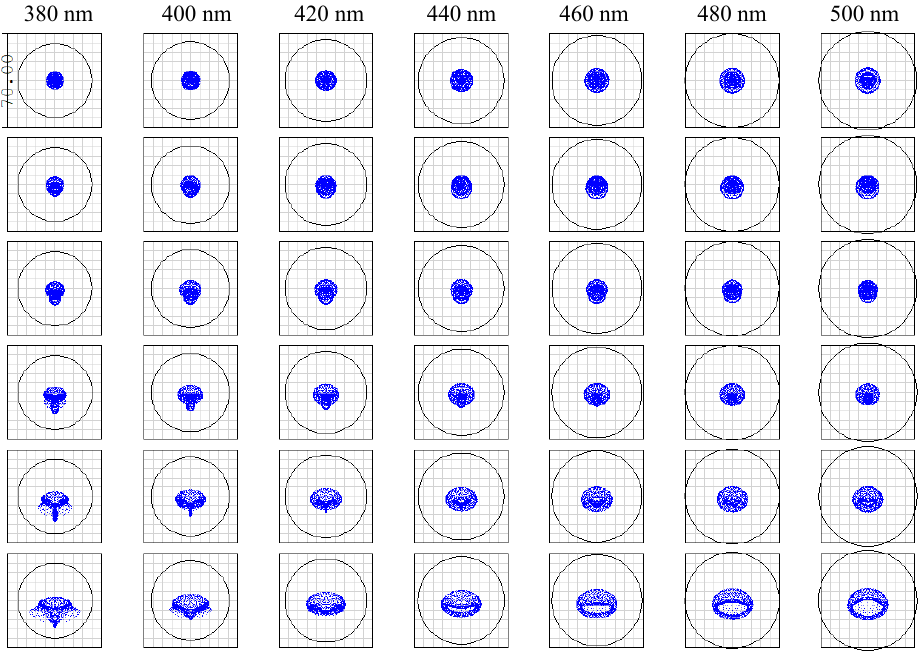}
 \caption{Calculated quantities, spot diagrams and Strehl values for EST-B Wideband optimized for minimum
image scale variations when refocusing with 12~$\mu$m pixels, when connected to POP with its 6.3 mm focus curve
range. The spot diagrams and Strehl values are obtained by individual focusing of the camera lens for each
wavelength, which requires a total movement of the camera lens by 9.3 mm.}
\label{fig:EST-B_spot_wideband_FPI+POP_imscale_12mu_pix_A3.4}
\end{figure}

Figure  \ref{fig:EST-B_spot_narrowband_FPI+POP_imscale_12mu_pix_A3.3} and Table \ref{table_EST-B_spot_narrowband_FPI+POP_imscale_12mu_pix_A3.3} show the performance of the 12~$\mu$m system connected to POP and optimised to reduce
image scale variations from refocusing. The minimum Strehl now is 0.87 at the edge of the field-of-view 
at 380 nm wavelength, but even disregarding this, the minimum Strehl is “only” 0.90. Since the
performance of both POP itself (shown in Fig. A2.1) and the present 12~$\mu$m system as a stand-alone instrument
is excellent, it is clear that the image quality degradation that occurs when EST-B is connected to POP
is a consequence of the refocusing. This image quality degradation is the prize that is paid for
enforcing a reduced image scale variation. The corresponding wideband system (Fig.  \ref{fig:EST-B_spot_wideband_FPI+POP_imscale_12mu_pix_A3.4} and Table \ref{table_EST-B_spot_wideband_FPI+POP_imscale_12mu_pix_A3.4}) show
similar performance and a minimum Strehl of 0.90.

\begin{table}[h]
  \centering
  \small
  \begin{tabular}{cccccccc}
    \hline
    \mathstrut
 $\lambda~(nm)$&380&400&420&440&460&480&500 \\
FOV dist. \\
\hline
0.0"&0.99&0.99&0.98&0.98&0.98&0.97&0.97\\
6.0"&0.99&0.99&0.98&0.98&0.98&0.98&0.98\\
11.8"&0.99&0.99&0.99&0.99&0.99&0.99&0.99\\
18.0"&0.97&0.98&0.98&0.99&0.99&0.99&0.99\\
23.6"&0.94&0.96&0.97&0.97&0.98&0.98&0.99\\
30.0"&0.87&0.90&0.91&0.92&0.93&0.93&0.94\\
 \hline
  \end{tabular}
    \vspace{1mm}
  \caption{ Strehl values for EST-B Narrowband connected to POP
with its 6.3 mm focus curve range, and optimized for minimum image scale variations when refocusing with 12~$\mu$m pixels. The spot diagrams and Strehl values are obtained by individual focusing of the camera lens for each wavelength, which requires a total movement of the camera lens by 9.3 mm. Spot diagrams in Fig. \ref{fig:EST-B_spot_narrowband_FPI+POP_imscale_12mu_pix_A3.3}.}
  \label{table_EST-B_spot_narrowband_FPI+POP_imscale_12mu_pix_A3.3}
\end{table}

\newpage
\subsection{EST-B wideband performance}

\begin{table}[h]
  \centering
  \small
  \begin{tabular}{cccccccc}
    \hline
    \mathstrut
 $\lambda~(nm)$&380&400&420&440&460&480&500 \\
FOV dist. \\
\hline
0.0"&0.97&0.99&0.99&0.99&0.97&0.96&0.94\\
6.0"&0.97&0.99&0.99&0.99&0.98&0.96&0.94\\
12.0"&0.97&0.99&0.99&0.99&0.98&0.97&0.95\\
18.0"&0.98&0.99&1.00&0.99&0.99&0.98&0.96\\
24.0"&0.98&0.99&0.99&0.99&0.99&0.99&0.98\\
30.0"&0.98&0.98&0.97&0.98&0.99&0.99&0.99\\
 \hline
  \end{tabular}
    \vspace{1mm}
  \caption{Strehl values for EST-B Wideband as a stand-alone
instrument (without POP) using 5~$\mu$m pixels pixels. The optimization is made by imposing a common focus for all wavelengths. The spot diagrams and Strehl values shown are obtained from that fixed focal plane. The image quality improves by individual focusing for each wavelength. Spot diagrams in Fig. \ref{fig:EST-B_spot_wideband_standalone_5mu_pix_A2.4}.}
  \label{table_EST-B_spot_wideband_standalone_5mu_pix_A2.4}
\end{table}

\begin{table}[h]
  \centering
  \small
  \begin{tabular}{cccccccc}
    \hline
    \mathstrut
 $\lambda~(nm)$&380&400&420&440&460&480&500 \\
FOV dist. \\
\hline
0.0"&0.97&0.97&0.98&0.98&0.98&0.98&0.98\\
6.0"&0.97&0.98&0.98&0.98&0.98&0.98&0.98\\
11.8"&0.97&0.98&0.98&0.98&0.99&0.99&0.99\\
18.0"&0.97&0.98&0.98&0.99&0.99&0.99&0.99\\
23.6"&0.96&0.97&0.98&0.98&0.98&0.99&0.99\\
30.0"&0.95&0.96&0.96&0.96&0.97&0.97&0.97\\
 \hline
  \end{tabular}
    \vspace{1mm}
  \caption{Strehl values for EST-B Wideband using 5~$\mu$m pixels, when connected to POP with its 6.3 mm focus curve range. The spot diagrams and Strehl values are obtained by individual focusing of the camera lens for each wavelength, which requires a total movement of the camera lens by 1.5 mm. Spot diagrams in Fig. \ref{fig:EST-B_spot_wideband_FPI+POP_5mu_pix_A2.6}.}
  \label{table_EST-B_spot_wideband_FPI+POP_5mu_pix_A2.6}
\end{table}

\begin{table}[h]
  \centering
  \small
  \begin{tabular}{cccccccc}
    \hline
    \mathstrut
 $\lambda~(nm)$&380&400&420&440&460&480&500 \\
FOV dist. \\
\hline
0.0"&0.98&0.99&1.00&1.00&0.99&0.99&0.98\\
6.0"&0.98&1.00&1.00&1.00&0.99&0.99&0.98\\
12.0"&0.99&1.00&1.00&1.00&1.00&0.99&0.99\\
18.0"&0.99&1.00&1.00&1.00&1.00&1.00&0.99\\
24.0"&1.00&1.00&1.00&1.00&1.00&1.00&1.00\\
30.0"&0.99&0.99&0.99&0.99&1.00&1.00&1.00\\
 \hline
  \end{tabular}
    \vspace{1mm}
  \caption{Strehl values for EST-B Wideband as a stand-alone
instrument (without POP) using 12~$\mu$m pixels. The optimization is for the best possible image quality and by
imposing a common focus for all wavelengths. The spot diagrams and Strehl values shown are obtained from
that fixed focal plane. The image quality improves by individual focusing for each wavelength. Spot diagrams in Fig. \ref{fig:EST-B_spot_wideband_standalone_12mu_pix_A2.8}.}
  \label{table_EST-B_spot_wideband_standalone_12mu_pix_A2.8}
\end{table}

\begin{table}[h]
  \centering
  \small
  \begin{tabular}{cccccccc}
    \hline
    \mathstrut
 $\lambda~(nm)$&380&400&420&440&460&480&500 \\
FOV dist. \\
\hline
0.0"&0.99&0.99&0.99&0.99&0.99&0.99&1.00\\
6.0"&0.99&0.99&0.99&0.99&0.99&0.99&1.00\\
11.8"&0.99&0.99&0.99&0.99&0.99&0.99&0.99\\
18.0"&0.99&0.98&0.98&0.98&0.98&0.99&0.99\\
23.6"&0.97&0.97&0.97&0.97&0.97&0.98&0.98\\
30.0"&0.95&0.95&0.95&0.95&0.96&0.96&0.96\\
 \hline
  \end{tabular}
    \vspace{1mm}
  \caption{Strehl values for EST-B Wideband+POP using 12~$\mu$m pixels, with POPs 6.3 mm focus curve range. The design is optimized for best image quality. The spot diagrams and Strehl values are obtained by individual focusing of the camera lens for each wavelength, which requires a total movement of the camera lens by 9.3 mm. Spot diagrams in Fig. \ref{fig:EST-B_spot_wideband_FPI+POP_imqual_12mu_pix_A2.10}.}
  \label{table_EST-B_spot_wideband_FPI+POP_imqual_12mu_pix_A2.10}
\end{table}

\begin{table}[h]
  \centering
  \small
  \begin{tabular}{cccccccc}
    \hline
    \mathstrut
 $\lambda~(nm)$&380&400&420&440&460&480&500 \\
FOV dist. \\
\hline
0.0"&0.96&0.98&0.99&0.99&0.99&0.99&0.98\\
6.0"&0.96&0.99&0.99&0.99&0.99&0.99&0.98\\
12.0"&0.97&0.99&1.00&1.00&1.00&0.99&0.99\\
18.0"&0.97&0.99&1.00&1.00&1.00&1.00&1.00\\
24.0"&0.97&0.99&0.99&1.00&1.00&1.00&1.00\\
30.0"&0.96&0.98&0.98&0.98&0.98&0.98&0.99\\
 \hline
  \end{tabular}
    \vspace{1mm}
  \caption{Strehl values for EST-B Wideband as a stand-alone
instrument (without POP) using 12~$\mu$m pixels. The optimization is made by minimizing image scale variations
from refocusing and by imposing a common focus for all wavelengths. The spot diagrams and Strehl values
shown are obtained from that fixed focal plane. The image quality improves somewhat at 380 nm by individually
focusing at that wavelength. Spot diagrams in Fig. \ref{fig:EST-B_spot_wideband_standalone_imscale_12mu_pix_A3.2}.}
  \label{table_EST-B_spot_wideband_standalone_imscale_12mu_pix_A3.2}
\end{table}

\begin{table}[h]
  \centering
  \small
  \begin{tabular}{cccccccc}
    \hline
    \mathstrut
 $\lambda~(nm)$&380&400&420&440&460&480&500 \\
FOV dist. \\
\hline
0.0"&0.99&0.99&0.99&0.99&0.99&0.99&0.99\\
6.0"&0.99&0.99&0.99&0.99&0.99&0.99&0.99\\
11.8"&0.99&0.99&0.99&0.99&0.99&0.99&0.99\\
18.0"&0.98&0.98&0.98&0.98&0.98&0.98&0.99\\
23.6"&0.95&0.96&0.96&0.96&0.96&0.96&0.97\\
30.0"&0.90&0.91&0.91&0.92&0.92&0.92&0.92\\
 \hline
  \end{tabular}
    \vspace{1mm}
  \caption{Strehl values for EST-B Wideband optimized for minimum image scale variations when refocusing with 12~$\mu$m pixels, when connected to POP with its 6.3 mm focus curve range. The spot diagrams and Strehl values are obtained by individual focusing of the camera lens for each
wavelength, which requires a total movement of the camera lens by 9.3 mm. Spot diagrams in Fig. \ref{fig:EST-B_spot_wideband_FPI+POP_imscale_12mu_pix_A3.4}.}
  \label{table_EST-B_spot_wideband_FPI+POP_imscale_12mu_pix_A3.4}
\end{table}

\clearpage
\section{EST-V optical design and performance}\label{EST-V_design2}
\subsection{EST-V narrowband design}

\begin{table}[h]
  \centering
  \small
  \setlength{\tabcolsep}{2pt}
  \begin{tabular}{rrrrrcc}
    \hline
    \mathstrut
No. & Radius & Thickness & Lens dia & Beam dia & Glass & Label \\
& (mm) & (mm) & (mm) & (mm) & & \\
    \hline
 --&&&&&& \bf{Telescope} \\
 --&&&&&& \bf{+ POP} \\
 0&&infinity\\
1&&100&&&&System stop\\
2&&20&&&&Ideal lens\\
-\\
3&&80&&&Silica\\
4&&25.19&&&&WBBS\\
5&&383.11&63*&&&F1\\
6&325.66&5&82&70.8&S-NBH8&L1\\
7&124.93&15.07&82&&N-PSK53A&L1\\
8&-392.07&331.35&82\\
9&&984.97&&&&P1\\
10&1241.01&12&200&&SF6&L2\\
11&528.73&2\\
12&572.40&23.94&200&&S-LAL18&L2\\
13&-1210.53&270&200&190.4\\
14&&140&200&188.6&Silica&FPI1\\
15&&200&200\\
16&&200&200&186.8&&F2\\
17&&140&200&&Silica&FPI2\\
18&&270&&188.6\\
19&1210.53&23.94&200&190.4&S-LAL18&L3\\
20&-572.40&2&200\\
21&-528.73&12&200&&SF6&L3\\
22&-1241.01&984.62\\
23 &&107.26&&&& P2 \\
 \hline
  \end{tabular}
    \vspace{1mm}
  \caption{Prescription of the EST-V FPI "mother" system (without camera lens). F1-F3 are focal planes, L1-L4 are cemented doublet lenses, P1-P2 are pupil planes, and WBBS is the wideband beam splitter.}
  \label{table_EST-V}
\end{table}

\begin{table}[h]
  \centering
  \small
  \setlength{\tabcolsep}{3pt}
  \begin{tabular}{rrrrrcc}
    \hline
    \mathstrut
No. & Radius & Thickness & Lens dia & Beam dia & Glass & Label \\
& (mm) & (mm) & (mm) & (mm) & & \\
    \hline
23&&107.26&&&&P2\\   
24&-186.56&9.98&40&27.2&S-BSM18&L4\\
25&-42.129&19.09&40&28.6&&L4\\
26&-23.27&19.76&&25.4&KZFSN5&L4\\
27&107.65&16.99&48&35.2&S-SK15&L4\\
28&-38.93&35&48&39.4&&L4\\
29&&50&50&&SILICA&PBS\\
30&&115.67&50&&&F3\\
31&0\\
 \hline
  \end{tabular}
    \vspace{1mm}
  \caption{Prescription of the camera lens, designed for 6.5~$\mu$m pixel size for the EST-V narrowband FPI system.  P2 is the second pupil stop, F3 the final focal planes, L4 is an air spaced triplet lens, and PBS is the polarising beam splitter.}
  \label{table_EST-V_6.5my_camera_lens}
\end{table}

\begin{table}[h]
  \centering
  \small
  \setlength{\tabcolsep}{3pt}
  \begin{tabular}{rrrrrcc}
    \hline
    \mathstrut
No. & Radius & Thickness & Lens dia & Beam dia & Glass & Label \\
& (mm) & (mm) & (mm) & (mm) & & \\
    \hline
23&&226.32&&&&P2\\
24&-211.59&10.72&62&49.4&S-BSM2&L4\\
25&-83.06&45.54&62&51&&L4\\
26&-44.90&20&54&48.8&KZFSN5&L4\\
27&283.07&19.78&76&62&S-LAL18&L4\\
28&-69.96&60&76&66&L4\\
29&&80&80&64.5&SILICA&PBS\\
30&&231.45&80&63.2\\
31&&0\\
 \hline
  \end{tabular}
    \vspace{1mm}
  \caption{Prescription of the camera lens, designed for 12~$\mu$m pixel size for the EST-V narrowband FPI system.  P2 is the second pupil stop, F3 the final focal planes, L4 is an air spaced triplet lens, and PBS is the polarising beam splitter.}
  \label{table_EST-V_12my_camera_lens}
\end{table}

\subsection{EST-V narrowband performance}
\begin{figure}[h]
\center
\includegraphics[angle=0, width=0.99\linewidth,clip]{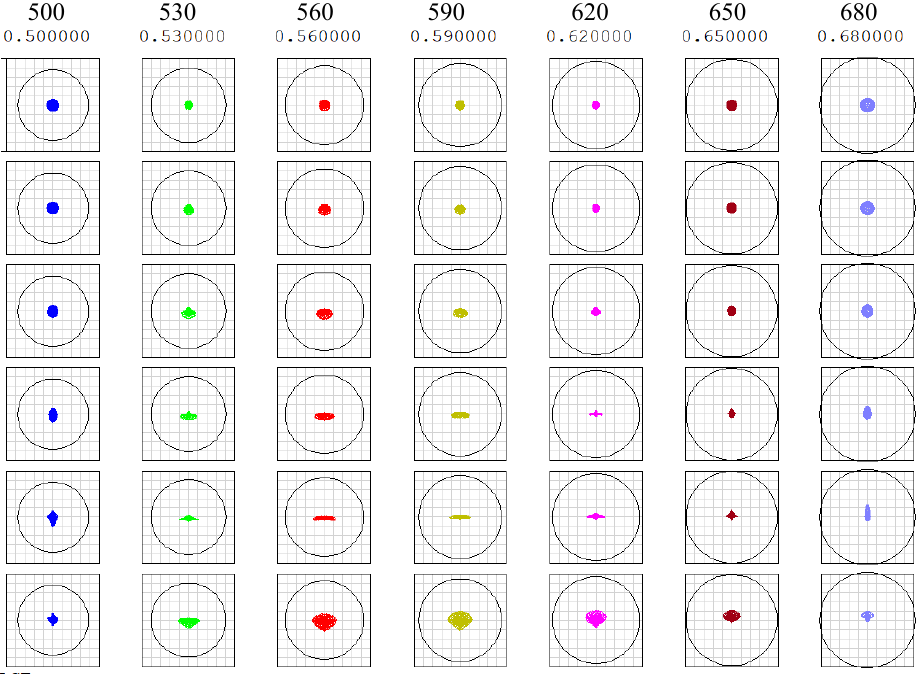}
 \caption{Calculated spot diagrams for EST-V alone, without POP, within the wavelength range of EST-V (500-680 nm), at a fixed focal plane.}
\label{fig:EST-V_spot_standalone_6.5mu_pix_T8}
\end{figure}
\begin{table}[h]
  \centering
  \small
  \begin{tabular}{cccccccc}
    \hline
    \mathstrut
 $\lambda~(nm)$&500&530&560&590&620&650&680 \\
FOV dist. \\
\hline
0.0"&0.98&1.00&1.00&1.00&1.00&0.99&0.99\\
6.2"&0.99&1.00&1.00&1.00&1.00&0.99&0.99\\
12.3"&0.99&1.00&1.00&1.00&1.00&1.00&0.99\\
18.5"&0.99&1.00&1.00&1.00&1.00&1.00&0.99\\
25.0"&0.98&0.99&0.99&1.00&1.00&1.00&0.99\\
30.0"&0.99&1.00&0.99&0.99&1.00&1.00&1.00\\
 \hline
  \end{tabular}
    \vspace{1mm}
  \caption{Strehl values for EST-V alone, without POP, within the wavelength range of EST-V (500-680 nm), at a fixed focal plane. Spot diagrams in Fig. \ref{fig:EST-V_spot_standalone_6.5mu_pix_T8}.}
  \label{table_EST-V_spot_mother_alone_A2.2}
\end{table}

\begin{figure}[h]
\center
\includegraphics[angle=0, width=0.99\linewidth,clip]{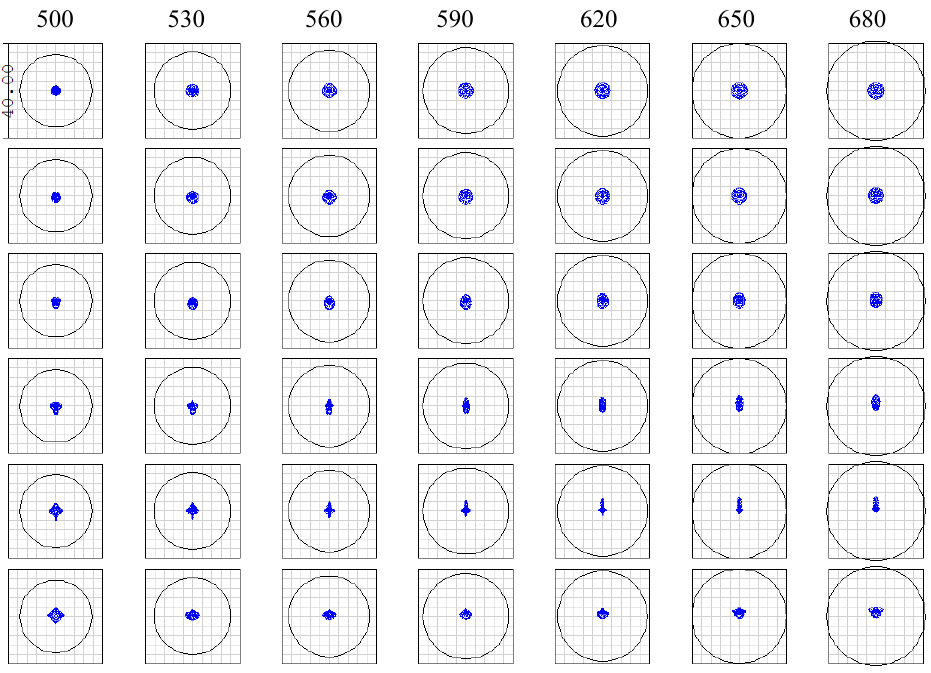}
 \caption{Spot diagrams for EST-V connected to POP using 6.5 µm pixels, within the 500-680~nm wavelength range of EST-V.}
\label{fig:EST-V_spot_narrowband_FPI+POP_imscale_6.5mu_pix_T9}
\end{figure}
\begin{table}[h]
  \centering
  \small
  \begin{tabular}{cccccccc}
    \hline
    \mathstrut
 $\lambda~(nm)$&500&530&560&590&620&650&680 \\
FOV dist. \\
\hline
0.0"&1.00&1.00&1.00&0.99&0.99&0.99&0.99\\
6.0"&1.00&1.00&1.00&0.99&0.99&0.99&0.99\\
11.8"&1.00&1.00&1.00&1.00&1.00&1.00&0.99\\
18.0"&0.99&1.00&1.00&1.00&1.00&1.00&1.00\\
23.6"&0.99&0.99&0.99&0.99&0.99&0.99&1.00\\
30.0"&0.98&0.99&0.99&0.99&1.00&1.00&1.00\\
 \hline
  \end{tabular}
    \vspace{1mm}
  \caption{Strehl values for EST-V Narrowband using with 6.5~$\mu$m pixels,
when connected to POP with its ?? mm focus curve range. The spot
diagrams and Strehl values are obtained by individual focusing of the
camera lens for each wavelength, which requires a total movement of
the camera lens by only 0.3~mm. Spot diagrams in \ref{EST-V_spot_narrowband_FPI+POP_imscale_6.5mu_pix_T9}.}
\end{table}

\begin{figure}[h]
\center
\includegraphics[angle=0, width=0.99\linewidth,clip]{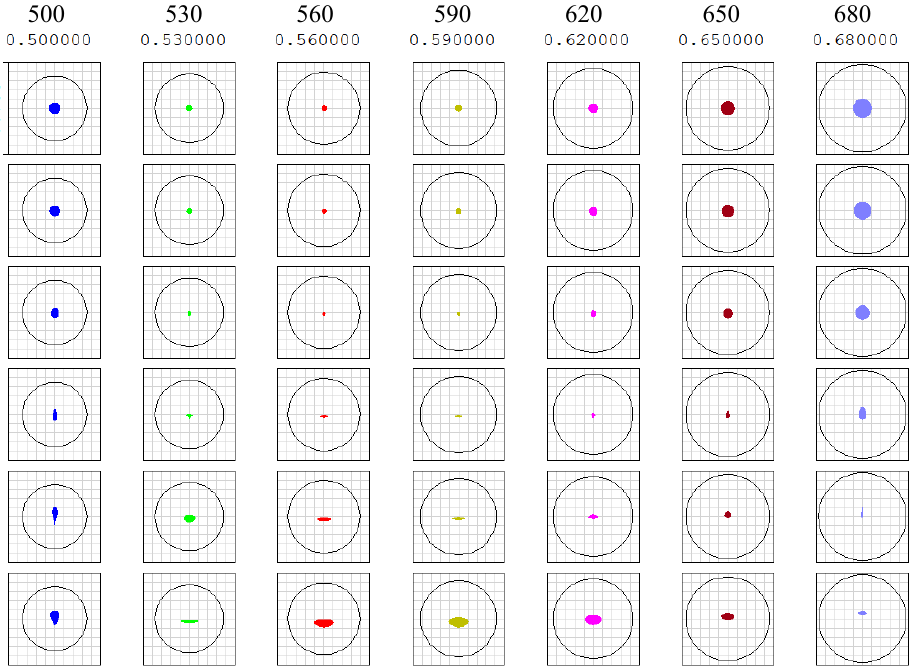}
 \caption{Calculated spot diagrams for EST-V alone with 12~$\mu$m pixels, without POP, within the wavelength range of EST-V (500-680 nm), at a fixed focal plane.}
\label{fig:EST-V_spot_standalone_12mu_pix_T10}
\end{figure}
\begin{table}[h]
  \centering
  \small
  \begin{tabular}{cccccccc}
    \hline
    \mathstrut
 $\lambda~(nm)$&500&530&560&590&620&650&680 \\
FOV dist. \\
\hline
0.0&0.99&1.00&1.00&1.00&0.99&0.99&0.98\\
6.2&0.99&1.00&1.00&1.00&1.00&0.99&0.99\\
12.3&0.99&1.00&1.00&1.00&1.00&1.00&0.99\\
18.5&0.99&1.00&1.00&1.00&1.00&1.00&1.00\\
25.0&0.98&0.99&0.99&1.00&1.00&1.00&1.00\\
30.0&0.98&0.99&0.99&0.99&0.99&1.00&1.00\\
 \hline
  \end{tabular}
    \vspace{1mm}
  \caption{Strehl values for EST-V alone, without POP, with 12~$\mu$m pixels within the wavelength range of EST-V (500-680 nm), at a fixed focal plane. Spot diagrams in Fig. \ref{fig:EST-V_spot_standalone_12mu_pix_T10}.}
  \label{table_EST-V_spot_FPI_alone_12mu_pix_T10}
\end{table}

\begin{figure}[h]
\center
\includegraphics[angle=0, width=0.99\linewidth,clip]{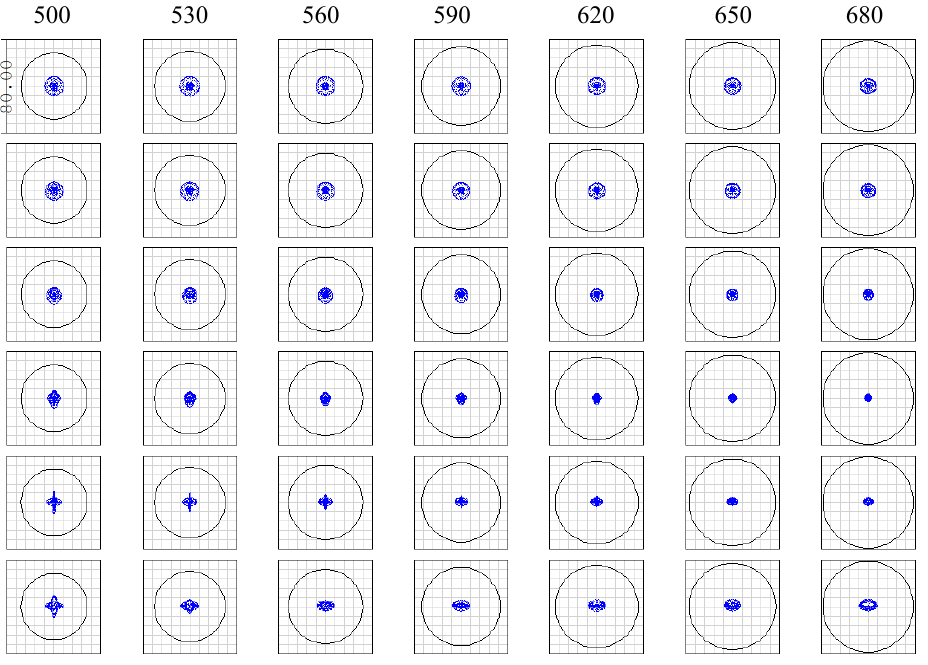}
 \caption{Spot diagrams for EST-V connected to POP using 12~$\mu$m pixels, within the 500-680~nm wavelength range of EST-V.}
\label{fig:EST-V_spot_narrowband_FPI+POP_imscale_12mu_pix_T11}
\end{figure}
\begin{table}[h]
  \centering
  \small
  \begin{tabular}{cccccccc}
    \hline
    \mathstrut
 $\lambda~(nm)$&500&530&560&590&620&650&680 \\
FOV dist. \\
\hline
0.0"&0.99&0.99&0.99&0.99&0.99&0.99&1.00\\
6.0"&0.99&0.99&1.00&1.00&1.00&1.00&1.00\\
11.8"&1.00&1.00&1.00&1.00&1.00&1.00&1.00\\
18.0"&0.99&0.99&1.00&1.00&1.00&1.00&1.00\\
23.6"&0.98&0.99&0.99&0.99&0.99&1.00&1.00\\
30.0"&0.97&0.98&0.98&0.98&0.98&0.99&0.99\\
 \hline
  \end{tabular}
    \vspace{1mm}
  \caption{Strehl values for EST-V Narrowband using with 12~$\mu$m pixels,
when connected to POP with its ?? mm focus curve range. The spot
diagrams and Strehl values are obtained by individual focusing of the
camera lens for each wavelength, which requires a total movement of
the camera lens by only 0.92~mm. Spot diagrams in \ref{fig:EST-V_spot_narrowband_FPI+POP_imscale_12mu_pix_T11}.}
\label{table_EST-V_spot_narrowband_FPI+POP_imscale_12mu_pix_T11}
\end{table}

\newpage
\subsection{EST-V wideband design}

\begin{figure}[h]
\center
\includegraphics[angle=0, width=0.8\linewidth,clip]{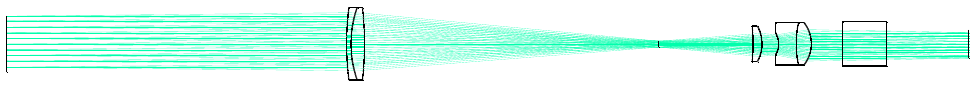}
\includegraphics[angle=0, width=0.99\linewidth,clip]{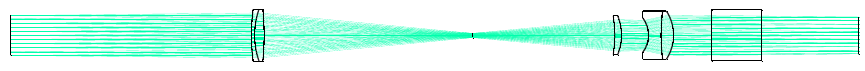}
\caption{Optical layout of EST-V wideband for 6.5~$\mu$m (top) and 12~$\mu$m (bottom) pixel size. The layouts are drawn to roughly the same scale. The first lenses in both layouts are the same and duplicates of L1 in the narrowband system. The second lenses are similar to L4 in the narrowband systems, but the first component of these triplet lenses have slightly different curvatures from those of the narrowband system. Also shown is the phase diversity beam splitter on the output side. The wideband beam splitter cube on the input side is not shown but is included in the layout of the narrowband system.}
\label{fig:EST-V_wideband_layout}
\end{figure}

\begin{table}[h]
  \centering
  \small
  \setlength{\tabcolsep}{2pt}
  \begin{tabular}{rrrrrcc}
    \hline
    \mathstrut
No. & Radius & Thickness & Lens dia & Beam dia & Glass & Label \\
& (mm) & (mm) & (mm) & (mm) & & \\
    \hline
 --&&&&&& \bf{Telescope} \\
 --&&&&&& \bf{+ POP} \\
 0&&infinity\\
1&&100&&&&System stop\\
2&&20&&&&Ideal lens\\
-\\
3&&80&&&Silica&WBBS\\
4&&25.19&&&&\\
5&&383.11&&63&&F1\\
6&325.66&5&82&&S-NBH8&L1\\
7&124.93&15.07&82&&N-PSK53A&L1\\
8&-392.07&331.35&82&&L1\\
9&&106.57&&&&P1/P2\\
10&-180.70&9.98&40&&S-BSM18&L4\\
11&-41.84&18.85&40&&&L4\\
12&-23.27&19.76&&&KZFSN5&L4\\
13&107.65&16.99&48&&S-SK15&L4\\
14&-38.93&35&48&&&L4\\
15&&50&50&&SILICA&PDBS\\
16&&92.34&50\\
17&&0&&&&F3\\
18&&&&31.4&\\
 \hline
  \end{tabular}
    \vspace{1mm}
  \caption{Prescription of the EST-V wideband system for 6.5~$\mu$m pixel size. F1,F3 are focal planes, L1 is a cemented doublet lens, L4 a triplet lens, P1/P2 is the pupil plane. WBBS is the wideband beam splitter and PDBS is the phase diversity beam splitter. The total length from F1 to F3 is ??.}
  \label{table_EST-V_wideband_6.5mu}
\end{table}

\begin{table}[h]
  \centering
  \small
  \setlength{\tabcolsep}{2pt}
  \begin{tabular}{rrrrrcc}
    \hline
    \mathstrut
No. & Radius & Thickness & Lens dia & Beam dia & Glass & Label \\
& (mm) & (mm) & (mm) & (mm) & & \\
    \hline
 9&&225.94&&&&P1/P2\\
10&-202.09&10.719&62&&S-BSM18&L4\\
11&-81.63&42.978&62\\
12&-44.90&20&54&&KZFSN5&L4\\
13&283.07&19.782&76&&S-SK15&L4\\
14&-69.96&60&76&L4\\
15&&80&80&&SILICA&PDBS\\
16&&153.84&80\\
17&&0\\
18&&&&58\\   
 \hline
  \end{tabular}
    \vspace{1mm}
  \caption{Prescription of the camera lens of the EST-V wideband system for 12~$\mu$m pixel size. The input side of the system is identical to that of the 6.5~$\mu$m wideband system in Table \ref{table_EST-V_wideband_6.5mu}. F3 is the focal plane, L4 is a triplet lens, P1/P2 is the pupil plane. WBBS is the wideband beam splitter and PDBS is the phase diversity beam splitter. The total length from F1 to F3 is ??.}
  \label{table_EST-V_wideband_12mu}
\end{table}
\newpage
\subsection{EST-V wideband performance}\label{EST-V_wideband_performance}

\begin{figure}[h]
\center
\includegraphics[angle=0, width=0.99\linewidth,clip]{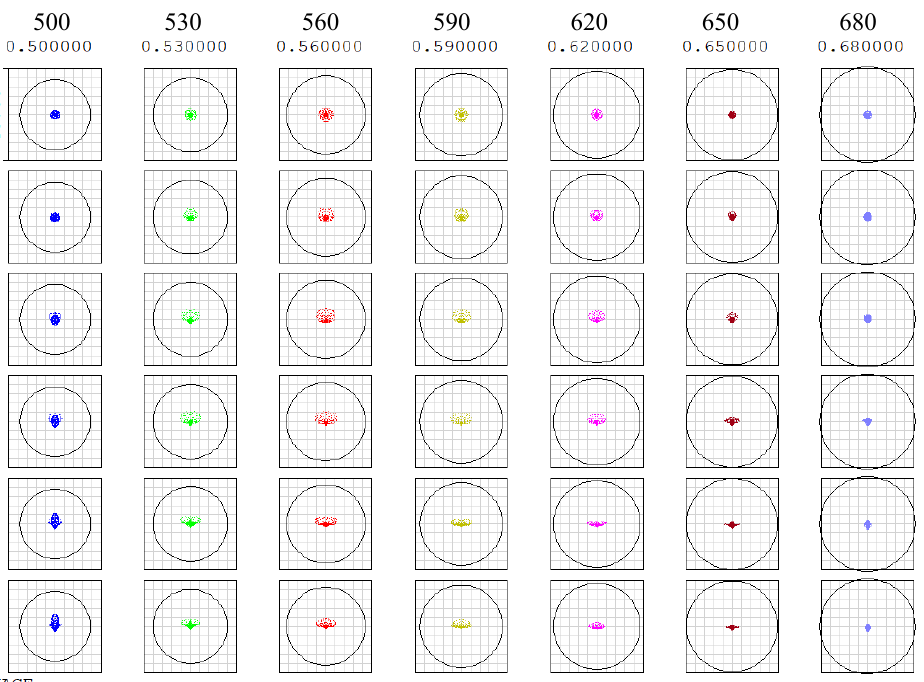}
 \caption{Calculated spot diagrams for EST-V wideband system for 6.5~$\mu$m pixel size, without POP, within the wavelength range of EST-V (500-680 nm), at a fixed focal plane.}
\label{fig:EST-V_wideband_spot_standalone_6.5mu_pix_T14}
\end{figure}
\begin{table}[h]
  \centering
  \small
  \begin{tabular}{cccccccc}
    \hline
    \mathstrut
 $\lambda~(nm)$&500&530&560&590&620&650&680 \\
FOV dist. \\
\hline
0.0"&0.99&1.00&1.00&1.00&1.00&1.00&1.00\\
6.2"&0.99&1.00&1.00&1.00&1.00&1.00&1.00\\
12.3"&0.99&1.00&1.00&1.00&1.00&1.00&1.00\\
18.5"&0.99&1.00&1.00&1.00&1.00&1.00&1.00\\
25.0"&0.99&1.00&1.00&1.00&1.00&1.00&1.00\\
30.0"&0.98&1.00&1.00&1.00&1.00&1.00&1.00\\
 \hline
  \end{tabular}
    \vspace{1mm}
  \caption{Strehl values for EST-V wideband system for 6.5~$\mu$m pixels, without POP, within the wavelength range of EST-V (500-680 nm), at a fixed focal plane. Spot diagrams in Fig. \ref{fig:EST-V_spot_standalone_6.5mu_pix_T14}.}
  \label{table_EST-V_wideband_spot_FPI_alone_6.5mu_pix_T14}
\end{table}

\begin{figure}[h]
\center
\includegraphics[angle=0, width=0.99\linewidth,clip]{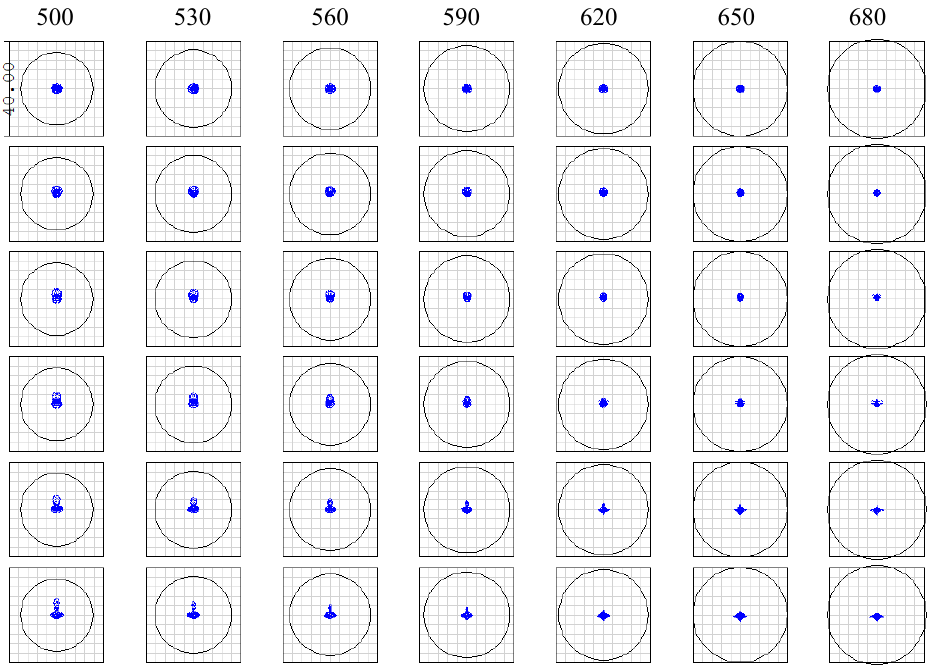}
 \caption{Spot diagrams for EST-V wideband connected to POP using 6.5~$\mu$m pixels, within the 500-680~nm wavelength range of EST-V.}
\label{fig:EST-V_spot_wideband_FPI+POP_imscale_6.5mu_pix_T15}
\end{figure}
\begin{table}[h]
  \centering
  \small
  \begin{tabular}{cccccccc}
    \hline
    \mathstrut
 $\lambda~(nm)$&500&530&560&590&620&650&680 \\
FOV dist. \\
\hline
0.0"&1.00&1.00&1.00&1.00&1.00&1.00&1.00\\
6.0"&1.00&1.00&1.00&1.00&1.00&1.00&1.00\\
11.8"&1.00&1.00&1.00&1.00&1.00&1.00&1.00\\
18.0"&1.00&1.00&1.00&1.00&1.00&1.00&1.00\\
23.6"&0.99&0.99&1.00&1.00&1.00&1.00&1.00\\
30.0"&0.98&0.99&0.99&0.99&1.00&1.00&1.00\\
 \hline
  \end{tabular}
    \vspace{1mm}
  \caption{Strehl values for EST-V Narrowband using with 6.5~$\mu$m pixels,
when connected to POP with its ?? mm focus curve range. The spot
diagrams and Strehl values are obtained by individual focusing of the
camera lens for each wavelength. Spot diagrams in \ref{fig:EST-V_spot_wideband_FPI+POP_imscale_6.5mu_pix_T15}.}
\label{table_EST-V_spot_wideband_FPI+POP_imscale_6.5mu_pix_T15}
\end{table}

\begin{figure}[h]
\center
\includegraphics[angle=0, width=0.99\linewidth,clip]{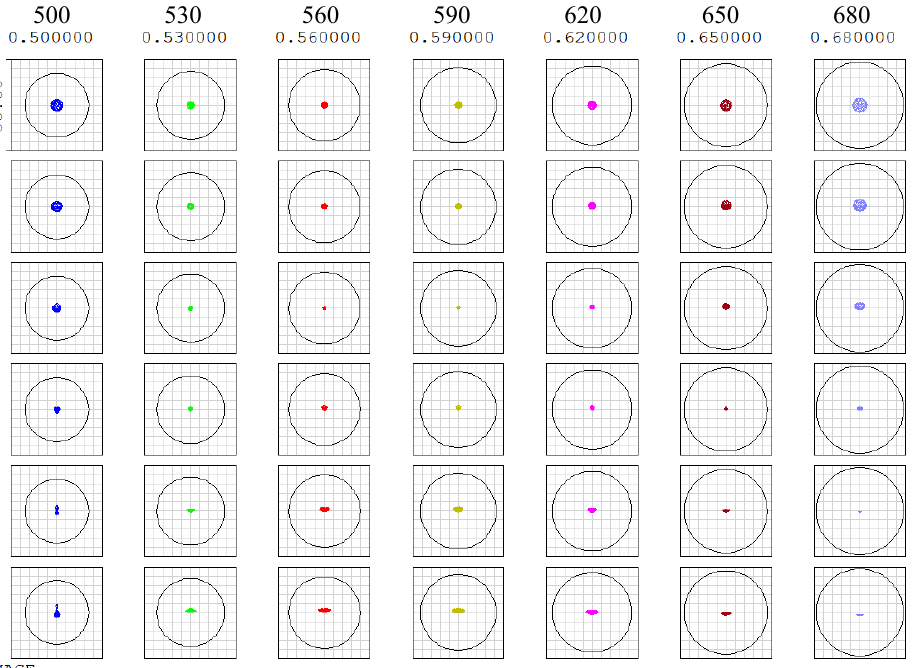}
 \caption{Calculated spot diagrams for EST-V wideband system for 12~$\mu$m pixel size, without POP, within the wavelength range of EST-V (500-680 nm), at a fixed focal plane.}
\label{fig:EST-V_wideband_spot_standalone_12mu_pix_T16}
\end{figure}
\begin{table}[h]
  \centering
  \small
  \begin{tabular}{cccccccc}
    \hline
    \mathstrut
 $\lambda~(nm)$&500&530&560&590&620&650&680 \\
FOV dist. \\
\hline
0.0"&0.99&0.99&1.00&1.00&0.99&0.99&0.99\\
6.2"&0.99&1.00&1.00&1.00&1.00&0.99&0.99\\
12.3"&0.99&1.00&1.00&1.00&1.00&1.00&1.00\\
18.5"&1.00&1.00&1.00&1.00&1.00&1.00&1.00\\
25.0"&1.00&1.00&1.00&1.00&1.00&1.00&1.00\\
30.0"&0.99&0.99&0.99&1.00&1.00&1.00&1.00\\
 \hline
  \end{tabular}
    \vspace{1mm}
  \caption{Strehl values for EST-V wideband system for 12~$\mu$m pixels, without POP, within the wavelength range of EST-V (500-680 nm), at a fixed focal plane. Spot diagrams in Fig. \ref{fig:EST-V_wideband_spot_standalone_12mu_pix_T16}.}
  \label{table_EST-V_wideband_spot_FPI_alone_12mu_pix_T16}
\end{table}

\begin{figure}[h]
\center
\includegraphics[angle=0, width=0.99\linewidth,clip]{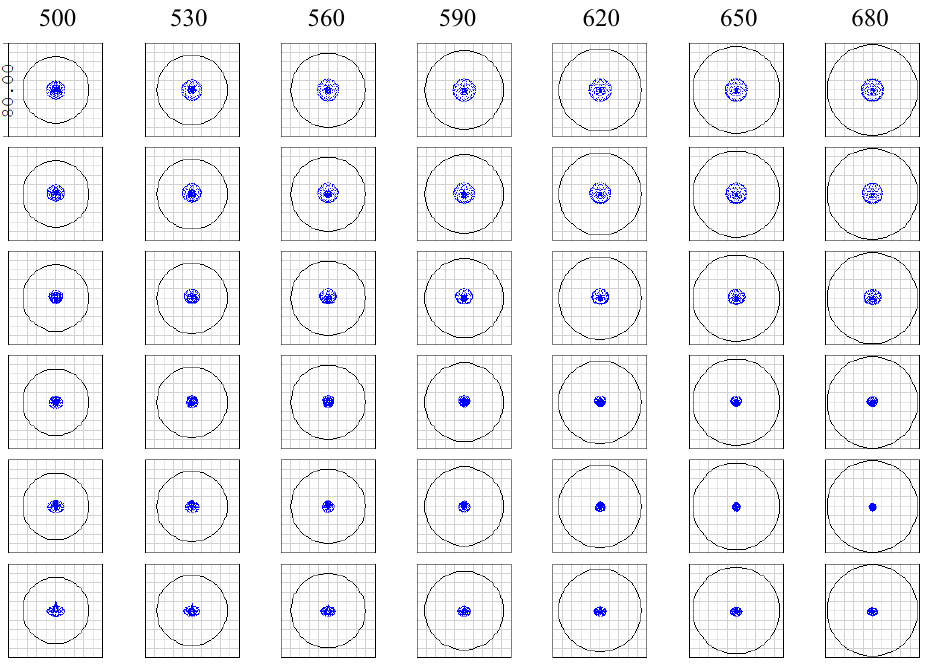}
 \caption{Spot diagrams for EST-V wideband connected to POP using 12~$\mu$m pixels, within the 500-680~nm wavelength range of EST-V.}
\label{fig:EST-V_spot_wideband_FPI+POP_imscale_12mu_pix_T17}
\end{figure}
\begin{table}[h]
  \centering
  \small
  \begin{tabular}{cccccccc}
    \hline
    \mathstrut
 $\lambda~(nm)$&500&530&560&590&620&650&680 \\
FOV dist. \\
\hline
0.0"&0.99&0.99&0.99&0.99&0.99&0.99&0.99\\
6.0"&1.00&0.99&0.99&0.99&0.99&0.99&0.99\\
11.8"&1.00&1.00&1.00&1.00&1.00&1.00&1.00\\
18.0"&0.99&1.00&1.00&1.00&1.00&1.00&1.00\\
23.6"&0.99&0.99&0.99&1.00&1.00&1.00&1.00\\
30.0"&0.98&0.98&0.99&0.99&0.99&0.99&1.00\\
 \hline
  \end{tabular}
    \vspace{1mm}
  \caption{Strehl values for EST-V wideband using with 12~$\mu$m pixels,
when connected to POP with its ?? mm focus curve range. The spot
diagrams and Strehl values are obtained by individual focusing of the
camera lens for each wavelength. Spot diagrams in \ref{fig:EST-V_spot_wideband_FPI+POP_imscale_12mu_pix_T17}.}
\label{table_EST-V_spot_wideband_FPI+POP_imscale_12mu_pix_T17}
\end{table}

\clearpage

\section{EST-R optical design and performance} \label{ESTR2}
\subsection{EST-R narrowband design}

\begin{table}[h]
  \centering
  \small
  \setlength{\tabcolsep}{2pt}
  \begin{tabular}{rrrrrcc}
    \hline
    \mathstrut
No. & Radius & Thickness & Lens dia & Beam dia & Glass & Label \\
& (mm) & (mm) & (mm) & (mm) & & \\
    \hline
 --&&&&&& \bf{Telescope} \\
 --&&&&&& \bf{+ POP} \\
 0&&infinity\\
1&&100&&&&System stop\\
2&&100&&&&Ideal lens\\
--\\
3&&100&&&&F1\\
4&&80&&64&Silica&WBBS\\
5&&225.55&64\\
6&304.87&19&84&69&S-PHM52&L1\\
7&-92.88&8&84&69&S-NBH5&L1\\
8&-482.05&326.60&84&69\\
9&&996.95&&6.8&&P1\\
10&2804.06&15&200&182&SF6&L2\\
11&554.93&3.005&200&182\\
12&577.34&26&200&184&S-LAL18&L2\\
13&-807.01&270&200&184\\
14&&140&200&182&Silica&FPI1\\
15&&200&200&182\\
16&&200&179.8&F2\\
17&&140&200&182&Silica&FPI2\\
18&&270&200&182\\
19&807.01&26&200&184&S-LAL18&L3\\
20&-577.34&3.005&200&184\\
21&-554.93&15&200&182&SF6&L3\\
22&-2408.06&965.89&200&182\\
23&&84.05&&7.4&&P2\\
24&533.43&8&32&24&N-SK2&L4\\
 \hline
  \end{tabular}
    \vspace{1mm}
  \caption{Prescription of the EST-R FPI "mother" system (without camera lens). F1-F3 are focal planes, L1-L4 are cemented doublet lenses, P1-P2 are pupil planes, and WBBS is the wideband beam splitter. The total length from F1 to F3 is ??.}
  \label{table_EST-R}
\end{table}

\begin{table}[h]
  \centering
  \small
  \setlength{\tabcolsep}{3pt}
  \begin{tabular}{rrrrrcc}
    \hline
    \mathstrut
No. & Radius & Thickness & Lens dia & Beam dia & Glass & Label \\
& (mm) & (mm) & (mm) & (mm) & & \\
    \hline
    23&&84.05&&7.4&&P2\\
24&533.43&8&32&24&N-SK2&L4\\
25&-50.58&21.038&32&24\\
26&-23.81&5&28&23&S-FTM16&L4\\
27&84.54&7.092&32&23\\
28&190.39&11&38&30&S-FPM2&L4\\
29&-33.54&20&38&30\\
30&&40&&29&SILICA&PBBS\\
31&&70&&29\\
32&&0&&&&F3\\
 \hline
  \end{tabular}
    \vspace{1mm}
  \caption{Prescription of the camera lens, designed for 6.5~$\mu$m pixel size for the EST-R narrowband FPI system.  P2 is the second pupil stop, F3 the final focal planes, L4 is an air spaced triplet lens, and PBS is the polarising beam splitter.}
  \label{table_EST-R_6.5my_camera_lens}
\end{table}

\begin{table}[h]
  \centering
  \small
  \setlength{\tabcolsep}{3pt}
  \begin{tabular}{rrrrrcc}
    \hline
    \mathstrut
No. & Radius & Thickness & Lens dia & Beam dia & Glass & Label \\
& (mm) & (mm) & (mm) & (mm) & & \\
    \hline
    23&&167.03&P2\\
24&194.07&10&48&39&S-FPM2&L4\\
25&-116.44&43.11&48&39\\
26&-51.13&5&44&35&S-TIM22&L4\\
27&169.73&29.567&44&35\\
28&676.10&12&58&50&S-FPM2&L4\\
29&-71.90&30&58&50\\
30&&60&&48&SILICA&PBBS\\
31&&123.33&&48\\
32&&0&&&&F3\\
 \hline
  \end{tabular}
    \vspace{1mm}
  \caption{Prescription of the camera lens, designed for 12~$\mu$m pixel size for the EST-R narrowband FPI system.  P2 is the second pupil stop, F3 the final focal planes, L4 is an air spaced triplet lens, and PBS is the polarising beam splitter.}
  \label{table_EST-R_12my_camera_lens}
\end{table}

\newpage
\subsection{EST-R wideband design}

\begin{figure}[h]
\center
\includegraphics[angle=0, width=0.8\linewidth,clip]{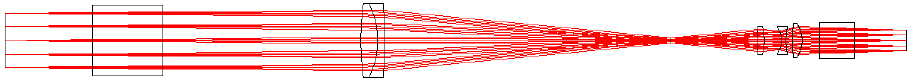}
\includegraphics[angle=0, width=0.99\linewidth,clip]{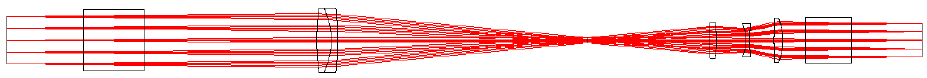}
\caption{Optical layout of EST-R wideband for 6.5~$\mu$m (top) and 12~$\mu$m (bottom) pixel size. The layouts are drawn to roughly the same scale. The first lenses in both layouts are the same and duplicates of L1 in the narrowband system. The second lenses are similar to L4 in the narrowband systems, but the first component of these triplet lenses have slightly different curvatures from those of the narrowband system. Also shown is the phase diversity beam splitter on the output side. The wideband beam splitter cube on the input side is not shown but is included in the layout of the narrowband system.}
\label{fig:EST-R_wideband_layout}
\end{figure}

\begin{table}[h]
  \centering
  \small
  \setlength{\tabcolsep}{2pt}
  \begin{tabular}{rrrrrcc}
    \hline
    \mathstrut
No. & Radius & Thickness & Lens dia & Beam dia & Glass & Label \\
& (mm) & (mm) & (mm) & (mm) & & \\
    \hline
 --&&&&&& \bf{Telescope} \\
 --&&&&&& \bf{+ POP} \\
0&&infinity\\
1&&100&&&&System stop\\
2&&100&&&&Ideal lens\\
-\\
3&&100&&30.5&&F1\\ 
4&&80&&&SILICA&PDBS\\
5&&225.54\\
6&304.87&19&84&&S-PHM52&L1\\
7&-92.88&8&84&&S-NBH5&L1\\
8&-482.05&326.60&84&\\
9&&97.99&&6.8&&P1/P2\\
10&533.43&8&32&25&N-SK2&L4\\
11&-50.58&19.690&32&25\\
12&-23.81&5&28&22&S-FTM16&L4\\
13&84.54&7.28&32&25\\
14&190.39&11&38&29&S-FPM2&L4\\
15&-33.54&20&38&32&\\
16&&40&&30&SILICA&PDBS\\
17&&59.12&&30\\
18&0&F3\\
 \hline
  \end{tabular}
    \vspace{1mm}
  \caption{Prescription of the EST-R wideband system for 6.5~$\mu$m pixel size. F1, F3 are focal planes, L1 is a cemented doublet lens, L4 a triplet lens, P1/P2 is the pupil plane. WBBS is the wideband beam splitter and PDBS is the phase diversity beam splitter. The total length from F1 to F3 is ??.}
  \label{table_EST-R_wideband_6.5mu}
\end{table}

\begin{table}[h]
  \centering
  \small
  \setlength{\tabcolsep}{2pt}
  \begin{tabular}{rrrrrcc}
    \hline
    \mathstrut
No. & Radius & Thickness & Lens dia & Beam dia & Glass & Label \\
& (mm) & (mm) & (mm) & (mm) & & \\
    \hline
9&&159.67&&6.8&&P1/P2\\
10&194.07&10&48&36&S-FPM2&L4\\
11&-116.44&38.076&48&36\\
12&-51.13&5&44&33&S-TIM22&L4\\
13&169.73&31.502&44&33\\
14&676.10&12&58&49&S-FPM2&L4\\
15&-71.90&30&58&49\\
16&&60&&46&SILICA&PDBS\\
17&&92.20&&46\\
18&&0&&&&F3\\
 \hline
  \end{tabular}
    \vspace{1mm}
  \caption{Prescription of the camera lens of the EST-R wideband system for 12~$\mu$m pixel size. The input side of the system is identical to that of the 6.5~$\mu$m wideband system in Table \ref{table_EST-R_wideband_6.5mu}. F3 is the focal plane, L4 is a triplet lens, P1/P2 is the pupil plane. WBBS is the wideband beam splitter and PDBS is the phase diversity beam splitter. The total length from F1 to F3 is ??.}
  \label{table_EST-R_wideband_12mu}
\end{table}
\newpage

\subsection{EST-R narrowband performance} \label{EST-R_narrowband_performance}
\begin{figure}[h]
\center
\includegraphics[angle=0, width=0.99\linewidth,clip]{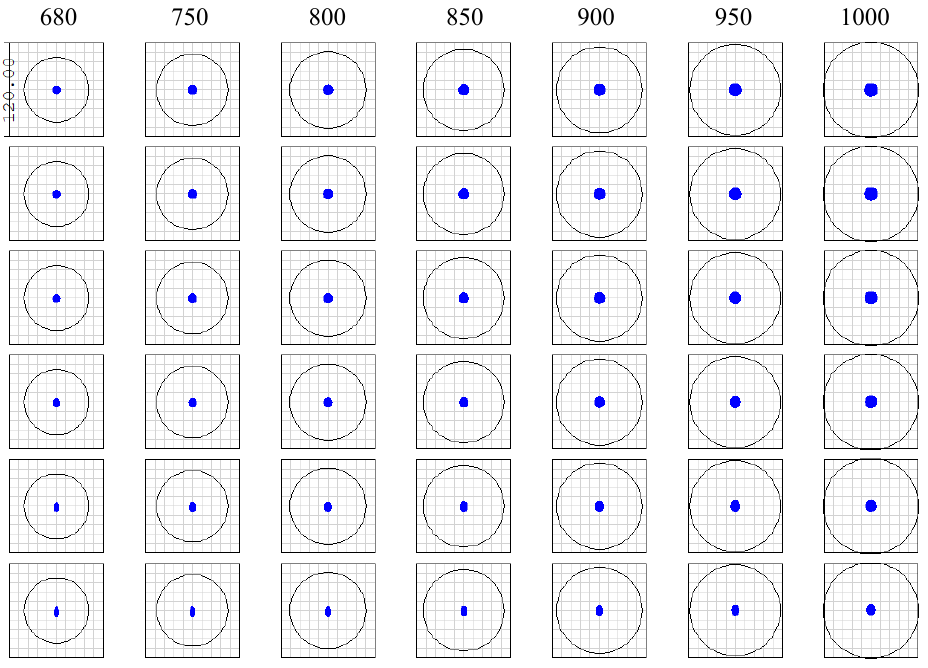}
 \caption{Calculated spot diagrams for POP alone, without FPI system, within the wavelength range of EST-R (680-1000~nm), at a fixed focal plane. POPs focus curve varies by 7.5~mm within this wavelength range.}
\label{fig:EST-R_POP_alone_spot}
\end{figure}
\begin{table}[h]
  \centering
  \small
  \begin{tabular}{cccccccc}
    \hline
    \mathstrut
 $\lambda~(nm)$&680&750&800&850&900&950&1000 \\
FOV dist. \\
\hline
0.0"&1.00&1.00&1.00&0.99&0.99&0.99&0.99\\
6.0"&1.00&1.00&1.00&0.99&0.99&0.99&0.99\\
11.8"&1.00&1.00&1.00&0.99&0.99&0.99&0.99\\
18.0"&1.00&1.00&1.00&1.00&0.99&0.99&0.99\\
23.6"&1.00&1.00&1.00&1.00&1.00&0.99&0.99\\
30.0"&1.00&1.00&1.00&1.00&1.00&1.00&1.00\\
 \hline
  \end{tabular}
    \vspace{1mm}
  \caption{Strehl values for POP alone, without FPI system, within the wavelength range of EST-R (680--1000~nm), at a fixed focal plane. Spot diagrams in Fig. \ref{fig:EST-R_POP_alone_spot}. POPs focus curve varies by 7.5~mm within this wavelength range.}
  \label{table_EST-R_POP_alone_strehl}
\end{table}

\begin{figure}[h]
\center
\includegraphics[angle=0, width=0.99\linewidth,clip]{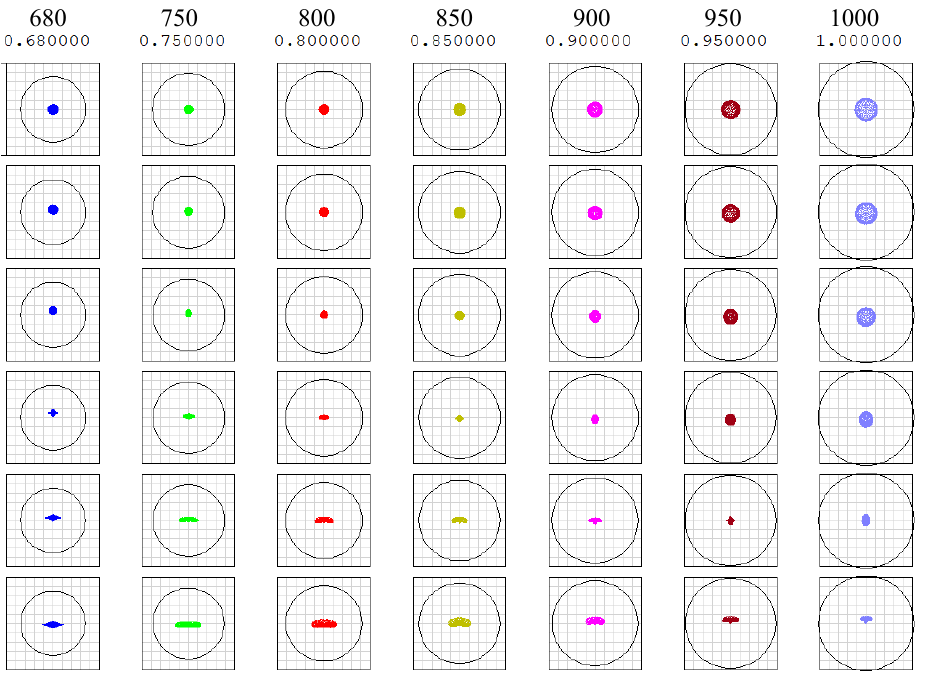}
 \caption{Calculated spot diagrams for EST-R narrowband alone, without POP, within the wavelength range of EST-R (680-1000~nm), for a 6.5~$\mu$m pixel size at a fixed focal plane. The focus curve of EST-R alone is in practice zero.}
\label{fig:EST-R_spot_standalone_6.5mu_pix}
\end{figure}
\begin{table}[h]
  \centering
  \small
  \begin{tabular}{cccccccc}
    \hline
    \mathstrut
 $\lambda~(nm)$&680&750&800&850&900&950&1000\\
FOV dist. \\
\hline
0.0"&0.99&0.99&0.99&0.99&0.98&0.98&0.97\\
6.4"&0.99&0.99&0.99&0.99&0.98&0.98&0.97\\
12.7"&0.99&1.00&1.00&0.99&0.99&0.98&0.98\\
19.1"&0.99&1.00&1.00&1.00&1.00&0.99&0.99\\
25.8"&0.99&1.00&1.00&1.00&1.00&1.00&1.00\\
30.0"&0.99&0.99&0.99&1.00&1.00&1.00&1.00\\
 \hline
  \end{tabular}
    \vspace{1mm}
  \caption{Strehl values for EST-R alone, without POP, within the wavelength range of EST-R (680--1000~nm), for a 6.5~$\mu$m pixel size at a fixed focal plane. Spot diagrams in Fig. \ref{fig:EST-R_spot_standalone_6.5mu_pix}. The focus curve of EST-R alone is in practice zero.}
  \label{table_EST-R_spot_narrowband_FPI_alone_A2.2}
\end{table}

\begin{figure}[h]
\center
\includegraphics[angle=0, width=0.99\linewidth,clip]{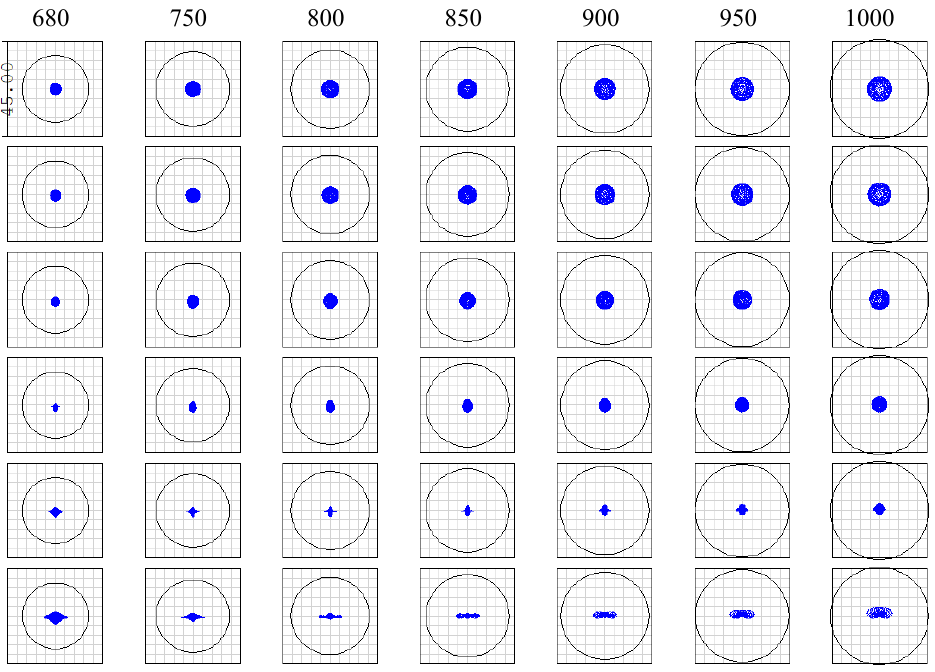}
 \caption{Spot diagrams for EST-R narrowband connected to POP using 6.5 µm pixels, within the 680--1000~nm wavelength range of EST-R. The camera lens needs to move 1.2~mm to compensate the 7.5~mm focus curve of POP.}
\label{fig:EST-R_spot_narrowband_FPI+POP_6.5mu_pix}
\end{figure}
\begin{table}[h]
  \centering
  \small
  \begin{tabular}{cccccccc}
    \hline
    \mathstrut
 $\lambda~(nm)$&680&750&800&850&900&950&1000 \\
FOV dist. \\
\hline
0.0"&0.99&0.98&0.97&0.97&0.97&0.97&0.97\\
6.0"&0.99&0.98&0.98&0.97&0.97&0.97&0.97\\
11.8"&0.99&0.99&0.98&0.98&0.98&0.98&0.98\\
18.0"&1.00&0.99&0.99&0.99&0.99&0.99&0.99\\
23.6"&0.99&1.00&1.00&1.00&1.00&1.00&1.00\\
30.0"&0.98&0.99&0.99&1.00&1.00&1.00&1.00\\
 \hline
  \end{tabular}
    \vspace{1mm}
  \caption{Strehl values for EST-R Narrowband with 6.5~$\mu$m pixels,
when connected to POP with its 7.5 mm focus curve range. The spot
diagrams and Strehl values are obtained by individual focusing of the
camera lens for each wavelength, which requires a total movement of
the camera lens by 1.2~mm. Spot diagrams in \ref{fig:EST-R_spot_narrowband_FPI+POP_6.5mu_pix}.}
\end{table}

\begin{figure}[h]
\center
\includegraphics[angle=0, width=0.99\linewidth,clip]{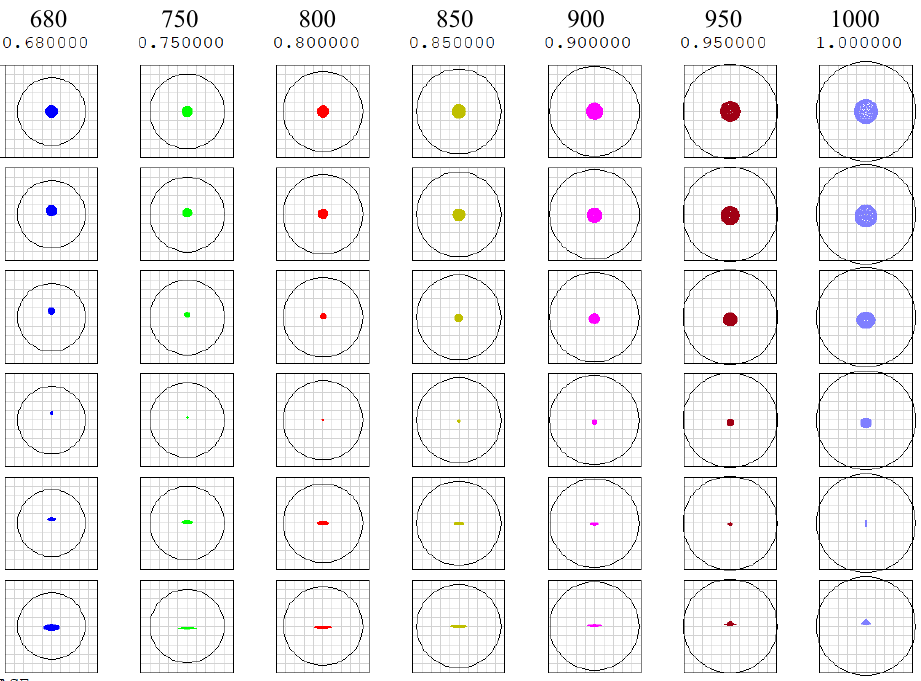}
 \caption{Calculated spot diagrams for EST-R narrowband alone with 12~$\mu$m pixels, without POP, within the wavelength range of EST-V (680--1000~nm), at a fixed focal plane.}
\label{fig:EST-R_spot_standalone_12mu_pix}
\end{figure}
\begin{table}[h]
  \centering
  \small
  \begin{tabular}{cccccccc}
    \hline
    \mathstrut
 $\lambda~(nm)$&680&750&800&850&900&950&1000 \\
FOV dist. \\
\hline
0.0"&0.99&0.99&0.99&0.99&0.99&0.98&0.98\\
6.4"&0.99&0.99&0.99&0.99&0.99&0.99&0.98\\
12.7"&0.99&1.00&1.00&1.00&0.99&0.99&0.99\\
19.1"&1.00&1.00&1.00&1.00&1.00&1.00&1.00\\
25.8"&1.00&1.00&1.00&1.00&1.00&1.00&1.00\\
30.0"&0.99&0.99&0.99&0.99&1.00&1.00&1.00\\
 \hline
  \end{tabular}
    \vspace{1mm}
  \caption{Strehl values for EST-R alone, without POP, with 12~$\mu$m pixels within the wavelength range of EST-R (680--1000~nm), at a fixed focal plane. Spot diagrams in Fig. \ref{fig:EST-R_spot_standalone_12mu_pix}.}
  \label{table_EST-R_spot_FPI_alone_12mu_pix}
\end{table}

\begin{figure}[h]
\center
\includegraphics[angle=0, width=0.99\linewidth,clip]{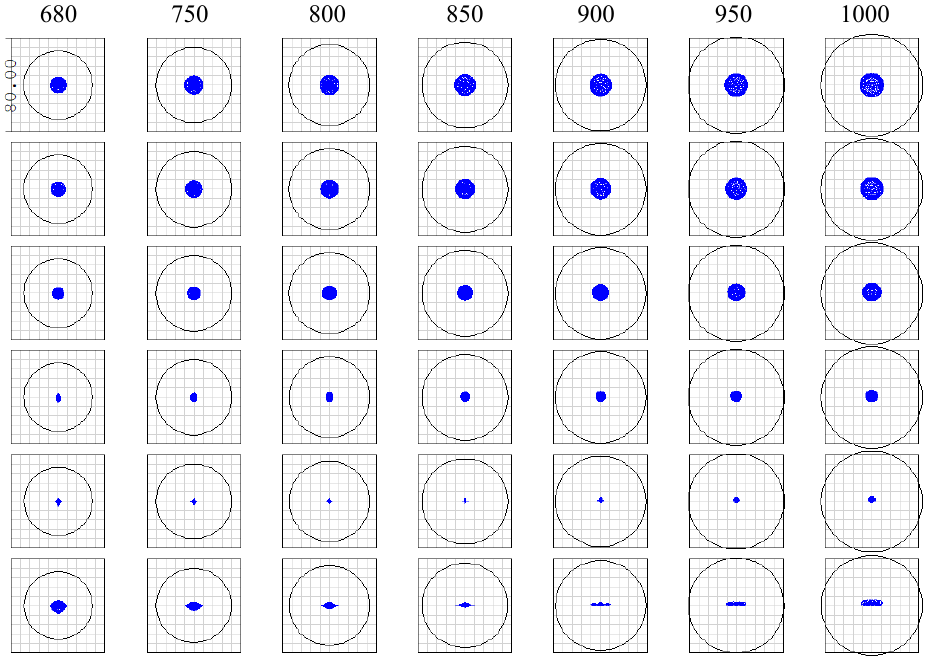}
 \caption{Spot diagrams for EST-R connected to POP using 12~$\mu$m pixels, within the 500-680~nm wavelength range of EST-V.}
\label{fig:EST-R_spot_narrowband_FPI+POP_imscale_12mu_pix}
\end{figure}
\begin{table}[h]
  \centering
  \small
  \begin{tabular}{cccccccc}
    \hline
    \mathstrut
 $\lambda~(nm)$&680&750&800&850&900&950&1000 \\
FOV dist. \\
\hline
0.0"&0.99&0.98&0.98&0.98&0.97&0.97&0.97\\
6.0"&0.99&0.98&0.98&0.98&0.98&0.98&0.98\\
11.8"&0.99&0.99&0.99&0.99&0.99&0.98&0.98\\
18.0"&1.00&1.00&1.00&1.00&0.99&0.99&0.99\\
23.6"&1.00&1.00&1.00&1.00&1.00&1.00&1.00\\
30.0"&0.98&0.99&0.99&0.99&1.00&1.00&1.00\\
 \hline
  \end{tabular}
    \vspace{1mm}
  \caption{Strehl values for EST-R Narrowband with 12~$\mu$m pixels,
when connected to POP with its 7.5mm focus curve range. The spot
diagrams and Strehl values are obtained by individual focusing of the
camera lens for each wavelength, which requires a total movement of
the camera lens by 4~mm. Spot diagrams in \ref{fig:EST-R_spot_narrowband_FPI+POP_imscale_12mu_pix}.}
\label{table_EST-R_spot_narrowband_FPI+POP_imscale_12mu_pix}
\end{table}

\clearpage
\subsection{EST-R wideband performance} \label{EST-R_wideband_performance}

\begin{figure}[h]
\center
\includegraphics[angle=0, width=0.99\linewidth,clip]{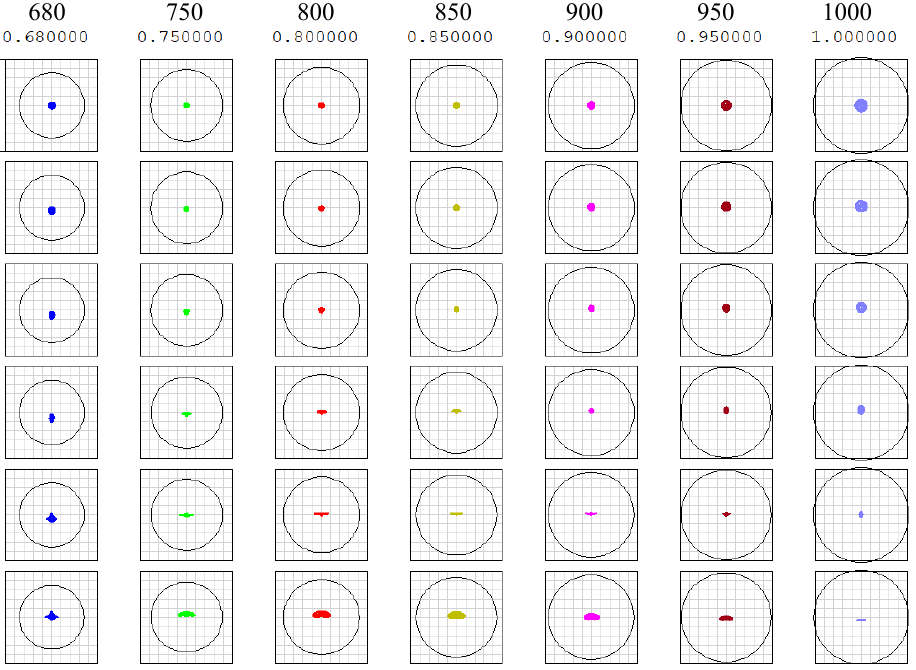}
 \caption{Calculated spot diagrams for EST-R wideband alone, without POP, within the wavelength range of EST-R (680-1000~nm), for a 6.5~$\mu$m pixel size at a fixed focal plane. The focus curve of EST-R alone is in practice zero.}
\label{fig:EST-R_spot_standalone_wideband_6.5mu_pix}
\end{figure}
\begin{table}[h]
  \centering
  \small
  \begin{tabular}{cccccccc}
    \hline
    \mathstrut
 $\lambda~(nm)$&680&750&800&850&900&950&1000\\
FOV dist. \\
\hline
0.0"&0.99&1.00&1.00&1.00&0.99&0.99&0.99\\
6.4"&0.99&1.00&1.00&1.00&1.00&0.99&0.99\\
12.7"&0.99&1.00&1.00&1.00&1.00&1.00&0.99\\
19.1"&0.99&1.00&1.00&1.00&1.00&1.00&1.00\\
25.8"&0.99&1.00&1.00&1.00&1.00&1.00&1.00\\
30.0"&0.99&1.00&1.00&1.00&1.00&1.00&1.00\\
 \hline
  \end{tabular}
    \vspace{1mm}
  \caption{Strehl values for EST-R wideband alone, without POP, within the wavelength range of EST-R (680--1000~nm), for a 6.5~$\mu$m pixel size at a fixed focal plane. Spot diagrams in Fig. \ref{fig:EST-R_spot_standalone_wideband_6.5mu_pix}. The focus curve of EST-R alone is in practice zero.}
  \label{table_EST-R_spot_wideband_FPI_alone_A2.2}
\end{table}

\begin{figure}[h]
\center
\includegraphics[angle=0, width=0.99\linewidth,clip]{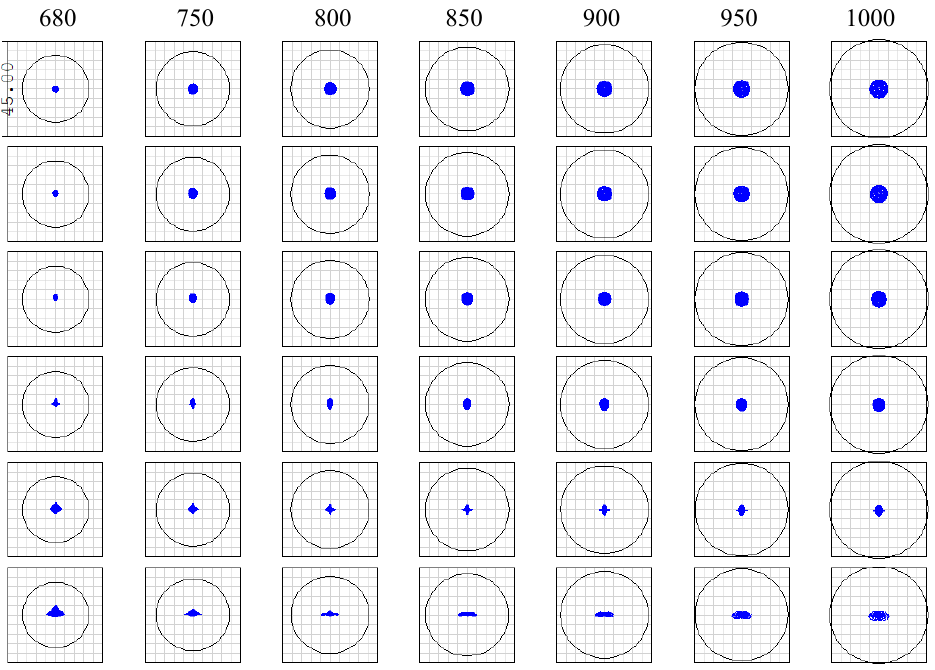}
 \caption{Spot diagrams for EST-R Wideband connected to POP using 6.5 µm pixels, within the 680--1000~nm wavelength range of EST-R. The camera lens needs to move 1.2~mm to compensate the 7.5~mm focus curve of POP.}
\label{fig:EST-R_spot_wideband_FPI+POP_6.5mu_pix}
\end{figure}
\begin{table}[h]
  \centering
  \small
  \begin{tabular}{cccccccc}
    \hline
    \mathstrut
 $\lambda~(nm)$&680&750&800&850&900&950&1000 \\
FOV dist. \\
\hline
0.0"&1.00&0.99&0.99&0.99&0.98&0.98&0.98\\
6.0"&1.00&0.99&0.99&0.99&0.99&0.98&0.98\\
11.8"&1.00&0.99&0.99&0.99&0.99&0.99&0.99\\
18.0"&1.00&1.00&0.99&0.99&0.99&0.99&0.99\\
23.6"&0.99&0.99&1.00&1.00&1.00&1.00&1.00\\
30.0"&0.99&0.99&1.00&1.00&1.00&1.00&1.00\\
 \hline
  \end{tabular}
    \vspace{1mm}
  \caption{Strehl values for EST-R wideband with 6.5~$\mu$m pixels,
when connected to POP with its 7.5 mm focus curve range. The spot
diagrams and Strehl values are obtained by individual focusing of the
camera lens for each wavelength, which requires a total movement of
the camera lens by 1.2~mm. Spot diagrams in \ref{fig:EST-R_spot_wideband_FPI+POP_6.5mu_pix}.}
\end{table}

\begin{figure}[h]
\center
\includegraphics[angle=0, width=0.99\linewidth,clip]{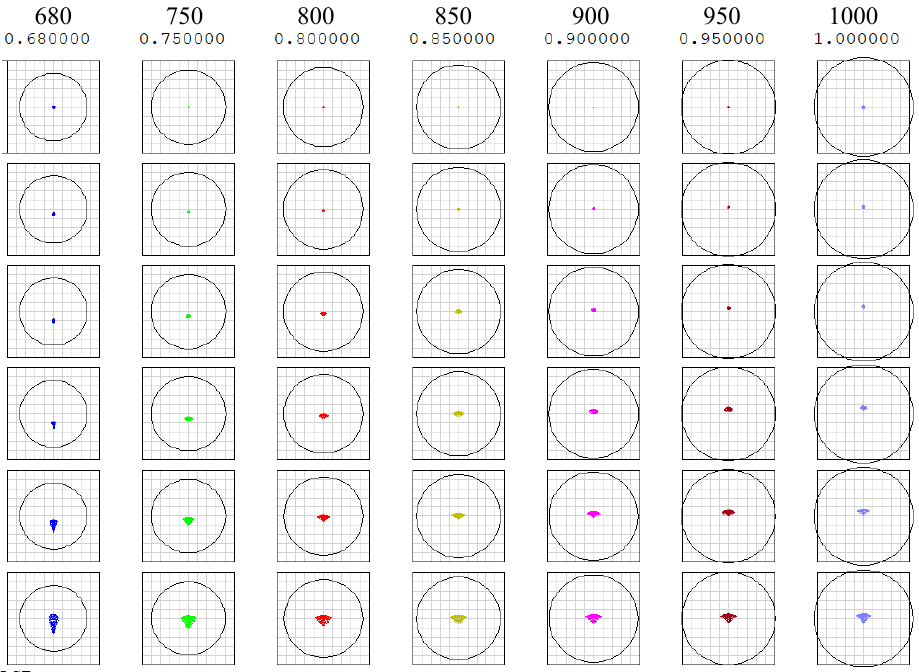}
 \caption{Calculated spot diagrams for EST-R wideband alone with 12~$\mu$m pixels, without POP, within the wavelength range of EST-V (680--1000~nm), at a fixed focal plane.}
\label{fig:EST-R_spot_standalone_wideband_12mu_pix}
\end{figure}
\begin{table}[h]
  \centering
  \small
  \begin{tabular}{cccccccc}
    \hline
    \mathstrut
 $\lambda~(nm)$&680&750&800&850&900&950&1000 \\
FOV dist. \\
\hline
0.0"&1.00&1.00&1.00&1.00&1.00&1.00&1.00\\
6.4"&1.00&1.00&1.00&1.00&1.00&1.00&1.00\\
12.7"&1.00&1.00&1.00&1.00&1.00&1.00&1.00\\
19.1"&1.00&1.00&1.00&1.00&1.00&1.00&1.00\\
25.8"&0.99&0.99&0.99&0.99&1.00&1.00&1.00\\
30.0"&0.97&0.98&0.99&0.99&0.99&0.99&0.99\\
 \hline
  \end{tabular}
    \vspace{1mm}
  \caption{Strehl values for EST-R Wideband alone, without POP, with 12~$\mu$m pixels within the wavelength range of EST-R (680--1000~nm), at a fixed focal plane. Spot diagrams in Fig. \ref{fig:EST-R_spot_standalone_wideband_12mu_pix}.}
  \label{table_EST-R_spot_FPI_alone_wideband_12mu_pix}
\end{table}

\begin{figure}[h]
\center
\includegraphics[angle=0, width=0.99\linewidth,clip]{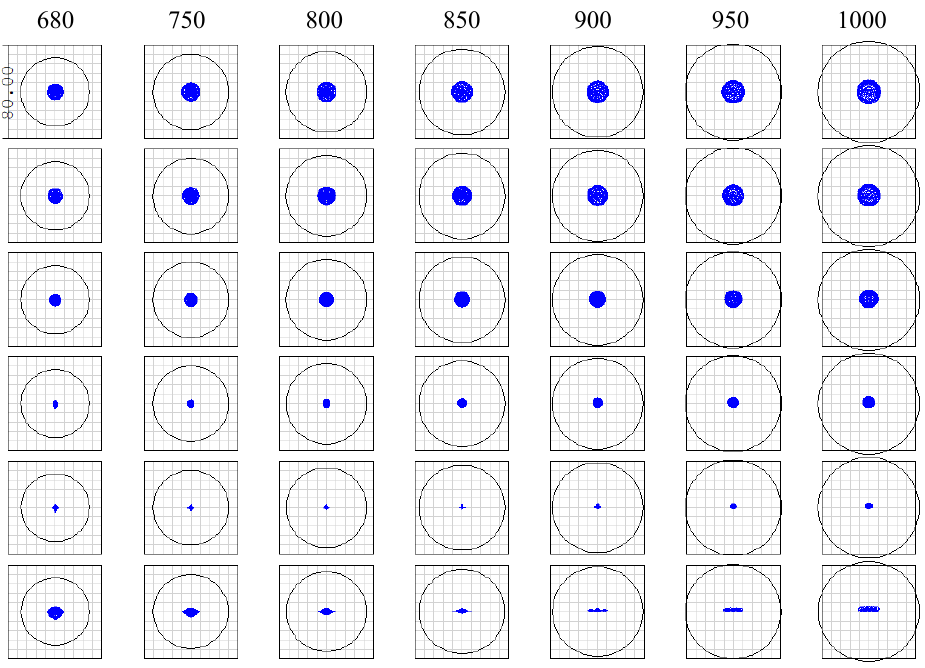}
 \caption{Spot diagrams for EST-R Wideband connected to POP using 12~$\mu$m pixels, within the 500-680~nm wavelength range of EST-V.}
\label{fig:EST-R_spot_wideband_FPI+POP_12mu_pix}
\end{figure}
\begin{table}[h]
  \centering
  \small
  \begin{tabular}{cccccccc}
    \hline
    \mathstrut
 $\lambda~(nm)$&680&750&800&850&900&950&1000 \\
FOV dist. \\
\hline
0.0"&1.00&1.00&1.00&1.00&1.00&0.99&0.99\\
6.0"&1.00&1.00&1.00&1.00&1.00&0.99&0.99\\
11.8"&1.00&1.00&1.00&1.00&1.00&1.00&1.00\\
18.0"&1.00&1.00&1.00&1.00&1.00&1.00&1.00\\
23.6"&0.99&0.99&1.00&1.00&1.00&1.00&1.00\\
30.0"&0.97&0.98&0.99&0.99&0.99&0.99&1.00\\
 \hline
  \end{tabular}
    \vspace{1mm}
  \caption{Strehl values for EST-R Wideband with 12~$\mu$m pixels,
when connected to POP with its 7.5mm focus curve range. The spot
diagrams and Strehl values are obtained by individual focusing of the
camera lens for each wavelength, which requires a total movement of
the camera lens by 4~mm. Spot diagrams in \ref{fig:EST-R_spot_wideband_FPI+POP_12mu_pix}.}
\label{table_EST-R_spot_wideband_FPI+POP_imscale_12mu_pix}
\end{table}

\clearpage

\subsection{EST-R narrowband transmission}

Transmission losses caused by absorption is less than 3\% in all designs for EST-R alone
without POP. For 25 surfaces we get a reflexion loss of 7\% for 25 surfaces, if the loss per surface is 0.3\%. A
more conservative guess of 0.5\% leads to a loss of about 12\%.

\clearpage
\section{Manufacturing and assembly tolerances} \label{tolerances2}
\subsection{EST-B tolerances}
The purpose of this analysis is to ensure that the system can in fact be produced, rather than
optimizing each tolerance for performance/cost. The following discussion of tolerances applies to
EST-B. A final analysis should be made in close cooperation with the manufacturer of the lenses and
mechanical structure.

\subsubsection{Refractive indexes and optical quality}
We assume that the variation of the refractive index with wavelength of the selected blanks is
measured accurately by the manufacturer and that the optical design is iteratively adapted to these
measured values. The requirements on the wavefront errors of the polished surfaces are not tight and
can easily be achieved: ½-1 wave peak to valley is sufficiently good. Therefore, the requirements on
polishing errors and refractive index tolerances are not further discussed.

\subsubsection{Tolerances on curvatures and glass dimensions}
For all lenses, we need to set tolerances on the radii of curvatures of the lens surfaces, the lens center
thicknesses and the wedge angles. Table \ref{table_EST-B_tolerances1} shows the tolerance levels for doublets L1-L3. The
tolerances on the radii correspond to high precision, but should not be a problem to achieve. The
thickness tolerances are standard values. The wedge angle tolerances are based on an edge thickness
runout of 10~$\mu$m. This requires high precision, but tolerance charts from some producers state that
even two times tighter tolerances are achievable with traditional methods.

\begin{table}[h]
  \centering
  \begin{tabular}{|l|l|}
        \hline
    \mathstrut
Parameter & Tolerance \\
\hline
Radius error & $\pm$0.1\% \\
Thickness & $\pm$0.1 mm \\
Wedge angle L2/L3& $\pm$ 0.003\textdegree \\
 Wedge angle L1 & $\pm$0.008\textdegree   \\
 \hline
  \end{tabular}
  \vspace{1mm}
  \caption{Optical tolerances of the EST-B lenses L1 and L2/L3.}
 \label{table_EST-B_tolerances1}  
\end{table}

\subsubsection{Assembly of small and large doublets}
It is assumed that the lenses of the split doublets L1 and L2 in Table \ref{table_EST-B} are produced and centered separately
and then mounted in a lens holder for two lenses in a traditional way (note though, that these doublets
need to be mounted with O-rings on the glass surface facing the FPI etalons). The same is true for the
cemented doublet lens L1 in Table \ref{table_EST-B}. This lens is assumed to be manufactured, cemented, centered
and mounted in its holder, in the order given.

In Table \ref{table_EST-B_tolerances2}  are shown assembly tolerances for doublets L1-L3. The tilt tolerances are the most
critical, but have in fact been achieved with CRISP and CHROMIS.
\begin{table}[h]
  \centering
  \small
  \begin{tabular}{|l|l|}
        \hline
    \mathstrut
Parameter & Tolerance \\
\hline
Decenter of entire FPI system relative to POP & $\pm$0.5 mm \\
Decenter of  L3+L4 relative to L1+L2 & $\pm$0.5 mm \\
Tilt of entire FPI system relative to POP & $\pm$ 0.05 degrees \\
Tilt of L3+L4 relative to L1+L2 & $\pm$ 0.05 degrees \\
Decenter between individual lenses & $\pm$ 0.1 mm \\
Tilt of individual lenses & $\pm$ 0.05 degrees \\ 
 \hline
  \end{tabular}
  \vspace{1mm}
  \caption{Assembly tolerances for lenses L1-L3.}
 \label{table_EST-B_tolerances2}  
\end{table}

The camera lens (L4), which is an air spaced triplet, is assumed to be assembled in a different
way and is discussed in Sect. \ref{assembly_EST-B_cameralens}.

\subsubsection{Assembly of the camera lens}\label{assembly_EST-B_cameralens}
Lensgroup L4 in the camera for EST-B is a triplet lens consisting of three separated singlet lenses, and
is the most demanding lens to manufacture and assemble. This design adds two air-to-glass surfaces
compared to the FBI-V design, but this improves the image quality somewhat and is easier to build, so
it is proposed to be a worthwhile choice. All three lenses are thin and with negligible absorption.

The most demanding tolerance is on surface tilt and wedge angles.

We tentatively propose to mount the two first small lenses in its own holder, and then mount that plus
the rear lens in the final housing, which will probably be a cylinder. This cylinder will be the moving
part when the focal position is changed. The assembly tolerances are strict on the first two lenses, so
we assume they must be built into its sub-housing in an alignment bonding station of some kind. The
requirements on the rear lens are wider.

Below is a possible step by step description of the procedure:

Step 1 The system is re-optimized with the measured index values and the tooling.
Step 2 The lenses are manufactured with the tolerances given in Table \ref{table_EST-B_tolerances3}
\begin{table}[h]
  \centering
  \begin{tabular}{|l|l|}
        \hline
    \mathstrut
Parameter & Tolerance \\
\hline
Radius & $\pm$0.2\% \\
Center thickness & $\pm$0.1 mm \\
Wedge first two lenses & Loose \\
Wedge last lens & $\pm$0.1 degrees \\
 \hline
  \end{tabular}
  \vspace{1mm}
  \caption{EST-B camera lens (L4), Step 2 tolerances.}
   \label{table_EST-B_tolerances3}  
\end{table}

Step 3  The lenses are measured with a higher accuracy than the manufacturing tolerances, and the
objective is re-optimized with the distances between the lenses and the last distance as variables.

Step 4  The two lenses of the first group of the objective are assembled in its own sub-assembly with
the tolerances given in Table \ref{table_EST-B_tolerances4}.

\begin{table}[h]
  \centering
  \begin{tabular}{|l|l|}
        \hline
    \mathstrut
    Parameter & Tolerance \\
\hline
  Tilt of each surface  & $\pm$0.01 degrees\\
  Distance between lenses & $\pm$0.1 mm\\
 \hline
  \end{tabular}
  \vspace{1mm}
  \caption{EST-B camera lens, Step 4 tolerances.}
 \label{table_EST-B_tolerances4}  
\end{table}

Since the tilt tolerances are very tight here we will make the assembly in an alignment-bonding
station. If that is done, then the lenses can be produced with a very loose wedge tolerance, as shown in
Table \ref{table_EST-B_tolerances3} above.

Step 5 The front group assembly and the rear lens are connected in a traditional way in a common
housing with the tolerances relative to the symmetry axis of the housing, given in Table \ref{table_EST-B_tolerances5}.

\begin{table}[h]
  \centering
  \begin{tabular}{|l|l|}
        \hline
    \mathstrut
Parameter & Tolerance \\
\hline
Decenter front group or rear lens & $\pm$0.2 mm \\
Tilt of front group or rear lens  & $\pm$0.05 degrees \\
 \hline
  \end{tabular}
  \vspace{1mm}
  \caption{EST-B camera lens, Step 5 tolerances.}
 \label{table_EST-B_tolerances5}  
\end{table}

Step 6 The objective is positioned relative to the large achromat with the tolerances given in Table  \label{table_EST-B_tolerances6}.
\begin{table}[h]
  \centering
  \begin{tabular}{|l|l|}
        \hline
    \mathstrut
Parameter & Tolerance \\
\hline
Decenter & $\pm$0.3 mm \\
Tilt & $\pm$0.1 degrees \\
 \hline
  \end{tabular}
  \vspace{1mm}
  \caption{EST-B camera lens, Step 6 tolerances.}
 \label{table_EST-B_tolerances6}  
\end{table}
Comparing the above tolerances with those of CHROMIS at SST, the above wedge angles are about 5
times tighter for the large lenses (L2 and L3), and about two times tighter on the cemented L1.
Tolerances on glass thicknesses and curvatures are about the same as for CHROMIS. Other tolerances
are similar to those of CHROMIS.

\clearpage
\subsection{EST-V tolerances}
\subsubsection{Tolerances on small and large doublet lenses}
In the following, we describe a preliminary tolerance analysis. The purpose of this analysis is to ensure
that the system can in fact be produced, rather than optimizing each tolerance for performance/cost.

We assume that the variation of the refractive index with wavelength of the selected blanks is
measured accurately and that the optical design is adapted to these measured values. The requirements
on the wavefront errors of the polished surfaces are not tight and can easily be achieved: ½-1 wave
peak to valley is sufficiently good. Therefore, the requirements on polishing errors and refractive
index tolerances are not further discussed. It is also assumed that the lenses of the split doublets L2 and L3 in
Table \ref{table_EST-V} are produced and centered separately and then mounted in a lens holder for two lenses in a
traditional way (note though, that the doublets L2 and L3 need to be mounted with O-rings on the glass surface
facing the FPI etalons). The same is true for the cemented doublet lens L1 in Tables \ref{table_EST-V}, 
\ref{table_EST-V_6.5my_camera_lens}, \ref{table_EST-V_12my_camera_lens}, 
\ref {table_EST-V_wideband_6.5mu} and \ref {table_EST-V_wideband_12mu}. This
lens is assumed to be manufactured, cemented, centered and mounted in its holder, in the order given.

The camera lens (the last lens group, consisting of singlet  plus cemented doublet in L4 in Tables  \ref{table_EST-V}, 
\ref{table_EST-V_6.5my_camera_lens}, \ref{table_EST-V_12my_camera_lens}, 
\ref {table_EST-V_wideband_6.5mu} and \ref {table_EST-V_wideband_12mu}, in contrast to the other lenses requires tight tolerances that probably cannot be achieved with
traditional methods. It is assumed that this lens is assembled in some kind of alignment - bonding or
alignment turning station. We discuss tolerances on the camera lens separately below.

For all lenses, we need to establish tolerances on the radii of curvatures of the lens surfaces, the lens
center thicknesses, the surface tilt angles, and also on some assembly tolerances. 
Table \ref{table_EST-V_tolerances1}  shows the
tolerance levels for doublets A and B. The tolerances on the radii correspond to high precision, but
should not be a problem to achieve. The thickness tolerances are standard values. The wedge angle
tolerances are based on an edge thickness runout of 10 microns. This a precision tolerance but
tolerance charts from some producers give half of that value as an achievable production limit with
traditional methods.

\begin{table}[h]
  \centering
  \begin{tabular}{|l|l|}
        \hline
    \mathstrut
Parameter & Tolerance \\
\hline
Radius error & $\pm$0.1\% \\
Thickness & $\pm$0.1 mm \\
Wedge angle L2/L3& $\pm$ 0.003\textdegree \\
 Wedge angle L1 & $\pm$0.008\textdegree   \\
 \hline
  \end{tabular}
  \vspace{1mm}
  \caption{Optical tolerances of the EST-V lenses L1 and L2/L3. These are identical to those of  EST-B in \ref{table_EST-B_tolerances1}}
 \label{table_EST-V_tolerances1}  
\end{table}

The next group of tolerances are the assembly tolerances. They are shown for doublets L1-L3 in
Table \ref{table_EST-V_tolerances2}, The tilt tolerances are the most critical, but were in fact achieved with CRISP and
CHROMIS. The internal assembly tolerances of the camera lens are discussed separately in Sect. \ref{EST-V_cameralens_tolerances}.

\begin{table}[h]
  \centering
  \small
  \begin{tabular}{|l|l|}
        \hline
    \mathstrut
Parameter & Tolerance \\
\hline
Decenter of entire FPI system relative to POP & $\pm$0.5 mm \\
Decenter of  L3+L4 relative to L1+L2 & $\pm$0.5 mm \\
Decenter of L4 relative to L3 & $\pm$0.2 mm \\
Tilt of entire FPI system relative to POP & $\pm$ 0.05 degrees \\
Tilt of L3+L4 relative to L1+L2 & $\pm$ 0.05 degrees \\
Tilt L4 relative to L3 & $\pm$ 0.05 degrees \\
Decenter between individual lenses & $\pm$ 0.1 mm \\
Tilt of individual lenses & $\pm$ 0.05 degrees \\ 
 \hline
  \end{tabular}
  \vspace{1mm}
  \caption{Assembly tolerances for lenses L1-L3. These are in principle identical to the corresponding tolerances in \ref{table_EST-B_tolerances2}, but include additional specifications.}
 \label{table_EST-V_tolerances2}  
\end{table}

\subsubsection{Tolerances and assembly of the camera lenses} \label{EST-V_cameralens_tolerances}
The last and most difficult part is the (movable) camera lens (singlet C plus doublet C) in Tables 
\ref{table_EST-V}, 
\ref{table_EST-V_6.5my_camera_lens}, \ref{table_EST-V_12my_camera_lens}, 
\ref {table_EST-V_wideband_6.5mu} and \ref {table_EST-V_wideband_12mu}. Here, the edge ray has steep exit and incidence angles (28\textdegree) inside the air gap
between the singlet and doublet lenses. This and the short radii of curvature make the tolerances very
tight. A possible approach to producing the camera lens is as follows:

\begin{enumerate}
\item Produce the cemented doublet. The tolerances on radius and center thickness for this lens are tight, but
still within the manufacturing limits.
\item Measure the radii and center thicknesses of the manufactured doublet to an accuracy that is much
higher than the manufacturing limits. Then re-design the lens group with the measured values of the
doublet held fixed at these values, while the singlet radii, thickness, distance between that and the
doublet lens, and final distance to the detector are used as free variables and optimized. Then produce
the re-designed singlet lens. The tolerances on the singlet lens are approximately the same as for the
original tolerances of the doublet.
\item Measure the radii and thickness of the singlet lens with high accuracy, and then re-design the lens
group with the distance between the doublet and singlet and the distance from the doublet to the final
focal plane as the only free variables. By implementing this procedure, we achieve tolerances on radii
and thicknesses that are limited only by the measurement accuracy, rather than the production
accuracy of the individual lenses.
\item At this stage, the lens group can be assembled. This must be done using some kind of alignment-
bonding or alignment-turning station, because the tilt angle tolerances are very tight. Table \ref{table_EST-V_tolerances3}  shows the final tolerances.
\end{enumerate}

\begin{table}[h]
  \centering
  \small
  \begin{tabular}{|l|l|l|}
        \hline
            \mathstrut
        Lens & L4 singlet & L4 doublet\\
Parameter&&\\
\hline
Radius & $\pm$0.02\% & $\pm$0.01\%\\
Center thickness & $\pm$0.005~mm & $\pm$0.005~mm\\
Surface tilts & $\pm$0.005\textdegree & $\pm$0.005\textdegree \\
 \hline
  \end{tabular}
  \vspace{1mm}
  \caption{Tolerances for the singlet and doublet of the camera lens.}
 \label{table_EST-V_tolerances3}  
\end{table}

The tolerance on surface tilts is very demanding for such small lenses and this is the most difficult
tolerance for the entire optical system of  EST-V. A ten-year-old article on alignment bonding claim a
tilt accuracy of 0.3 arc min, which corresponds is 0.005 degrees. This point requires more attention and
discussions with one or more potential producers of the camera lens.

\subsubsection{Alternatives to the proposed camera lens}
If the previous tolerances and procedure required for producing the camera lens turn out to be too
demanding or too expensive, there are several possible options. The first option is to redesign the
camera lens and push the 28\textdegree angle to a lower value, which will widen the tolerance a bit. This will
come at a cost in Strehl that perhaps will be lowered by about 0.02. Only the camera lens has to be re-
designed, of course, the remaining optics of EST-V stays the same.

Ongoing work on the design of EST-B indicates that the steep exit and incidence angles of the beam
between the L4 singlet  and L4 doublet of the camera lens can be reduced without any loss of image
quality if one of the surfaces of L4 singlet is made aspheric, which offers a second option to the triplet
lens. Not much work has been done yet to explore this possibility, but this looks like a promising
approach. The downside of this solution is that the back focal distance (distance from last lens surface
to the camera focal plane) is reduced and this might cause problems with space for the cameras, when
the pixel size is small.

A third option is to replace the above triplet camera lens with a cemented doublet: This will widen the
tolerances on this lens to values that are similar to those of CRISP and CHROMIS, which we know
are achievable. The Strehl will probably be reduced to about 0.9 at its worst wavelength and field
point, when tolerances on manufacture and alignment are included, but for most wavelengths and field
points, the Strehl will higher. Probably, the poorest performance will be at the edge of the field-of-
view for a wavelength of 500~nm and perhaps at the center of the field-of-view at a wavelength of 680nm.

A fourth option is obviously to replace the above triplet camera lens with a doublet separated with an
air gap. The performance of this is likely to be better than for the cemented doublet, but we have not
yet explored that design sufficiently well to make more conclusive statements yet.

A fifth option is to use an air spaced triplet, similar to a Cooke triplet. This will add two air-glass
surfaces. This system will be longer than the present system.

\section{files used} \label{files_used}
All files listed below will be made available upon request

\subsection{Files used for EST-B}
BLUE\_POP\_0\\
BLUE\_A\_mother\\
BLUE\_A5\_both\\
BLUE\_A12\_both\\
POP\_BLUE\_A5\_narr\\
POP\_BLUE\_A5\_broad\\
POP\_BLUE\_A12\_narr\\
POP\_BLUE\_A12\_broad\\
BLUE\_AB12\_both\\
POP\_BLUE\_AB12\_narr\\
POP\_BLUE\_AB12\_broad\\

\subsection{Files used for EST-V}
VIS\_POP\_0\\
VIS\_parax\_0\\
VIS65\_0\\
VIS65\_POP\_0\\
VIS65\_BRD\_0\\
VIS65\_BRD\_POP\_0\\
VIS120\_0\\
VIS120\_POP\_0\\
VIS120\_BRD\_0\\
VIS120\_BRD\_POP\_0\\

\subsection{Files used for EST-R}
RED\_POP\_0\\
RED\_narrow\_65\_0\\
RED\_narrow\_65\_POP\_0\\
RED\_broad\_65\_0\\
RED\_broad\_65\_POP\_0\\
RED\_narrow\_120\_0\\
RED\_narrow\_120\_POP\_0\\
RED\_broad\_120\_0\\
RED\_broad\_120\_POP\_0\\

\end{document}